\documentclass[aps,reprint,prl,superscriptaddress, longbibliography]{revtex4-2}
\usepackage{}
\usepackage{amssymb}
\usepackage{amsmath}
\usepackage{dcolumn}
\usepackage{bm}
\usepackage{graphicx}
\usepackage{mathrsfs}
\usepackage[colorlinks,linkcolor=blue,anchorcolor=blue,citecolor=blue,urlcolor=blue]{hyperref}
\usepackage{color}
\begin{document}

\title{Enhanced tripartite interactions in spin-magnon-mechanical hybrid
    systems}

\author{Xin-Lei Hei}
\affiliation{Ministry of Education Key Laboratory for Nonequilibrium Synthesis and Modulation of Condensed Matter,
Shaanxi Province Key Laboratory of Quantum Information and Quantum Optoelectronic Devices,
School of Physics, Xi'an Jiaotong University, Xi'an 710049, China}
\author{Peng-Bo Li}
\email{lipengbo@mail.xjtu.edu.cn}
\affiliation{Ministry of Education Key Laboratory for Nonequilibrium Synthesis and Modulation of Condensed Matter,
Shaanxi Province Key Laboratory of Quantum Information and Quantum Optoelectronic Devices,
School of Physics, Xi'an Jiaotong University, Xi'an 710049, China}
\affiliation{Theoretical Quantum Physics Laboratory, RIKEN Cluster for Pioneering Research, Wako-shi, Saitama 351-0198, Japan}
\author{Xue-Feng Pan}
\affiliation {Ministry of Education Key Laboratory for Nonequilibrium Synthesis and Modulation of Condensed Matter,
Shaanxi Province Key Laboratory of Quantum Information and Quantum Optoelectronic Devices,
School of Physics, Xi'an Jiaotong University, Xi'an 710049, China}
\author{Franco Nori}
\affiliation{Theoretical Quantum Physics Laboratory, RIKEN Cluster for Pioneering Research, Wako-shi, Saitama 351-0198, Japan}
\affiliation{RIKEN Center for Quantum Computing (RQC), 2-1 Hirosawa, Wako-shi, Saitama 351-0198, Japan}
\affiliation{Physics Department, The University of Michigan, Ann Arbor, Michigan 48109-1040, USA }

\begin{abstract}
Coherent tripartite interactions among degrees of freedom of completely different nature are instrumental for quantum information and simulation technologies, but they are generally difficult to realize and remain largely unexplored. Here, we predict a tripartite coupling mechanism in a hybrid setup comprising a single NV center and a
micromagnet. We propose to realize \textit{direct and strong tripartite interactions} among single NV spins, magnons and phonons via modulating the relative motion between the NV center  and the micromagnet. Specifically, by introducing a parametric drive (two-phonon drive) to modulate the mechanical motion (such as the center-of-mass motion of a NV spin in diamond trapped in an electrical trap or a levitated micromagnet in a magnetic trap), we can obtain a tunable and strong spin-magnon-phonon coupling at the single quantum level, with up to two orders of magnitude enhancement for the tripartite coupling strength. This enables, for example, \textit{tripartite entanglement} among solid-state spins, magnons, and mechanical motions in quantum spin-magnonics-mechanics with realistic experimental parameters. This protocol can be readily implemented with the well-developed techniques in ion traps or magnetic traps, and could pave the way for general applications in quantum simulations and information processing based on directly and strongly coupled tripartite systems.
\end{abstract}

\maketitle

\textit{Introduction.---}Coherent interactions between different quantum systems are a fundamental issue in the field of quantum physics and quantum technologies~\cite{RevModPhys.85.623,shandilya2021optomechanical,PhysRevLett.126.203601,PhysRevLett.118.140501,PhysRevLett.124.113602,PhysRevLett.110.156402,PhysRevLett.117.015502,PhysRevApplied.10.024011,PhysRevLett.107.060502,PhysRevLett.106.115305,PhysRevLett.105.210501,PhysRevB.87.144516,PhysRevLett.118.140502,PhysRevLett.124.013602,PhysRevX.7.011030,RevModPhys.91.025005,PhysRevB.103.174106,PhysRevLett.120.093601,PhysRevResearch.2.013121,PhysRevResearch.3.013025,dong2022chiral}.
The Jaynes-Cummings (JC) model~\cite{1443594, doi:10.1080/09500349314551321}, which describes the pairwise coherent interactions between a two-level quantum system and a quantized field, is a textbook example of light-matter interactions in the quantum regime, and lays the foundations of quantum optics \cite{scully1997quantum,KNIGHT198021,STENHOLM19731,haroche1989cavity}. With the fast development of quantum technologies, like quantum information processing \cite{RevModPhys.77.513,RevModPhys.74.347,RevModPhys.82.2313,RevModPhys.84.621} and simulations~\cite{RevModPhys.86.153, RevModPhys.73.33}, the exploration
of interactions beyond the pairwise interactions of the JC model in quantum optics is increasingly appealing, which could enable performing more complex tasks, like generating multipartite entanglement. However, compared to the bipartite interactions of the JC model, the realization of tripartite interactions among completely different degrees of freedom is an
outstanding challenge and remains largely unexplored.

Recently, much attention has been paid to studying hybrid quantum systems based on  nitrogen-vacancy (NV) centers in
diamond~\cite{Aharonovich2011, DOHERTY20131, PhysRevLett.112.047601, RevModPhys.92.015004, Bar-Gill2013, Abobeih2018One,Casola2018,PhysRevLett.124.210502,PhysRevB.104.014109,PhysRevLett.129.093605}, magnons in microscopic magnets~\cite{PhysRevA.99.021801, Lachance_Quirion_2019, PhysRevLett.128.013602, TABUCHI2016729, PhysRevLett.120.057202, PhysRevLett.123.127202, doi:10.1126/science.aaz9236, doi:10.1063/5.0020277, Andrich2017, huillery2020spin, Zhang2015, PhysRevLett.123.107702, PhysRevLett.123.107701, PhysRevB.101.125404, PhysRevLett.128.013602, PhysRevLett.124.093602, 2018LiP203601203601, Zhange1501286}, and mechanical motions~\cite{RN750, RN276, RN415, cha2021superconducting,RN302, RN679, RN542, RN678, Ovartchaiyapong2014, PhysRevLett.121.123604, PhysRevLett.126.193602, PhysRevLett.113.020503, Arcizet2011, Zhange1501286}. Recent theoretical and experimental advances have revealed the coupling of NV spins to phonons in nanomechanical oscillators \cite{PhysRevLett.124.163604,rusconi2022spin,PhysRevLett.126.193602, Lee_2017, doi:10.1126/science.1216821, Arcizet2011, Whiteley2019, PhysRevLett.121.123604, PhysRevX.8.041027, PhysRevLett.111.227602, PhysRevApplied.5.034010, PhysRevLett.113.020503, PhysRevB.79.041302, PhysRevApplied.4.044003, doi:10.1021/nl300775c, doi:10.1126/science.1216821, Barfuss2015, Ovartchaiyapong2014, Cai2014, Rabl2010, MacQuarrie2017, Arcizet2011, RN1054} and magnons of micromagnets~\cite{PhysRevLett.125.247702, PhysRevA.103.043706, Candido_2020, PRXQuantum.2.040314,RN382}, in addition to the interactions between magnons and phonons~\cite{PhysRevLett.124.093602, 2018LiP203601203601, Zhange1501286, PhysRevB.101.125404} or photons~\cite{RN623, RN691, RN472, RN433, PhysRevLett.113.083603, RN463, RN343, RN448, PhysRevLett.113.156401,kani2022intensive}.
However, previous studies mostly focus on pairwise interactions between completely different
physical systems to construct hybrid quantum setups; it seems to us that the tripartite coupling among spins, magnons as well as mechanical motions, which is fundamentally different from spin-magnon, spin-phonon, and magnon-phonon couplings, is still lacking.

In this work, we theoretically show how it is possible to achieve the tripartite interaction among single spins, magnons, and phonons
in a hybrid setup comprising a single NV center in diamond and a micromagnet.
We show that when the relative motion between the spin and the micromagnet is modulated, it will change
the magnetic field of the magnons felt by the nearby spin, which thus leads to direct coherent couplings among these three degrees of freedom at the single quantum level. To control and enhance this tripartite coupling, we propose to make use of a parametric drive to
amplify the mechanical zero-point fluctuations of the vibration mode ~\cite{PhysRevLett.67.699, PhysRevLett.107.213603, PhysRevLett.125.153602, Szorkovszky_2014, Lemonde2016, PhysRevA.95.053861,PhysRevLett.122.030501,Burd2021}, which can exponentially enhance the spin-magnon-phonon coupling. Specifically, here the mechanical motion could be either the center of mass motion of a NV spin in diamond or that of a levitated micromagnet. For the former, it can be implemented in a setup with a nano-diamond sphere containing a single NV spin in a Paul trap~\cite{RN742, RN738, RN735, RN737,RN752, doi:10.1063/1.4893575, RN734} or a diamond cantilever with embedded NV centers, while for the latter it can be realized with a levitated micromagnet (such as a Yttrium iron garnet (YIG) sphere) in a magnetic trap \cite{PhysRevApplied.13.064027, RN427, PhysRevLett.124.163604, doi:10.1063/1.5129145}. For both cases, we only need a time-dependent electrical driving to manipulate the effective spring constant of the harmonic motion, thus remarkably simplifying the experimental implementation with only minor modifications of existing experimental setups. But our proposal differs
fundamentally from these experimental works  with a markedly
different kind of spin-magnon-phonon tripartite interaction.
As intriguing applications,
we also show the appearance and enlargement of the tripartite entanglement via the enhanced interaction of the spin-magnon-phonon coupling system, which could find useful applications in modern quantum  technologies.

\begin{figure}[t]
	\includegraphics[width=8.5cm]{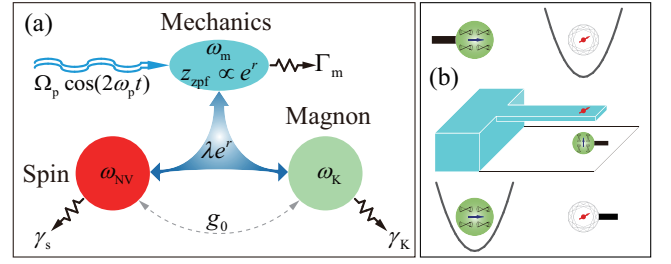}
	\caption{\label{fig:f1}(color online). (a) Schematic of the physical model. The spin qubit (red circle), the phonon mode (cyan ellipse), and the Kittel magnon mode (green circle) are simultaneously coupled, with the enhanced coupling rate $ \lambda e^{r} $ (blue trichotomous arrow) via a two-phonon driving (blue wavy arrow). (b) Schematic illustration of this proposal: a diamond particle with single NV spins in an electrical trap (top); an NV center embedded in a cantilever (middle); a YIG microsphere levitated in a magnetic trap (bottom). }
\end{figure}

\textit{The model.---}As illustrated in Fig.~\ref{fig:f1}, we consider a hybrid system comprising a
single NV center in diamond and a micromagnet of radius \(R\) (such as a YIG sphere), but with three degrees of freedom, including  the single NV spin ($ \hat{\sigma}_{i} $),  the magnon mode ($ \hat{a} $), and the mechanical mode ($ \hat{b} $). Here, the mechanical mode is the relative motion between the spin and the micromagnet, which is subject to a two-phonon (parametric) drive $ \Omega(t) = \Omega_{\mathrm{p}}\cos(2\omega_{\mathrm{p}} t) $. The spin operators are defined as the Pauli operators $ \hat{\sigma_i} $ (with $ i=x,y,z $) in the two-level-energy basis $\{|g\rangle,|e\rangle\}$.
The interaction among the NV spin, the mechanical mode, and the magnon can be described by the Hamiltonian (let $ \hbar=1 $)
\begin{equation}\label{Eq1}
\hat{H}_{\mathrm{Trip}} = \lambda(\hat{b}+\hat{b}^{\dagger})(\hat{a}^{\dagger}\hat{\sigma}^{-} + \hat{a}\hat{\sigma}^{+}),
\end{equation}
with tripartite coupling strength $ \lambda $. Here, the spin operators satisfy $\hat{\sigma}^{\pm} = (\hat{\sigma}_x \pm i\hat{\sigma}_y)/2$.

We then present more details regarding the above interactions. The tripartite spin-magnon-phonon coupling results from the magnetic coupling between
the spin and the magnon mode of the YIG sphere.
First, we focus on the Kittel mode that is supported by the magnetic microsphere~\cite{SupplementalMaterial}. For this mode, all spins in the micromagnet precess in phase and with the same amplitude~\cite{TABUCHI2016729}.
The free Hamiltonian of the magnon can be $
\hat{H}_{\mathrm{K}} = \omega_{\mathrm{K}} \hat{a}^{\dagger}\hat{a}$.
Here, $ \omega_{\mathrm{K}}=|\gamma|B_{z,\mathrm{K}} $, with a large external magnetic field $ B_{z,\mathrm{K}} $ resulting in saturation magnetization of the spherical magnet and the gyromagnetic ratio $ \gamma $. Then, a quantized magnetic field $ \hat{\vec{B}} $ is generated by the Kittel mode. The nearby NV center  as a magnetic dipole, with the free Hamiltonian $ \hat{H}_{NV} = \omega_{NV}\hat{\sigma}_{z}/2 $, experiences the magnetic field of the magnons. The interaction can be naturally described as the Hamiltonian $ \hat{H}_{\mathrm{int}} = -({g_{\mathrm{e}} \mu_{\mathrm{B}}}/{\hbar}) \hat{\vec {B}}\cdot \hat{\vec{S}} $, with the Land\'{e} factor $g_{\mathrm{e}}$, Bohr magneton $ \mu_{\mathrm{B}} $ and spin operators $ \hat{\vec{S}} = (\hbar/2)(\hat{\sigma}_x,\hat{\sigma}_y,\hat{\sigma}_z) $.
To be much clearer, the interaction Hamiltonian can be written as $ \hat{H}_{\mathrm{int}} = g(r)(\hat{a}^{\dagger}\hat{\sigma}^{-} + \hat{a}\hat{\sigma}^{+}) $, with the coupling strength $g(r)$ dependent on the distance between the NV spin and the micromagnet $ r=r_0 + z $. Here, $ z $ ($r_0$) denotes the modulated (static) part of the distance relative to the equilibrium, respectively.
Then, by quantizing the modulated motion $z$, it is possible to introduce a mechanical mode with the vibration frequency $\omega_m$.
Up to first order on the quantized coordinate $ \hat{z} = z_{\mathrm{zpf}}(\hat{b}+\hat{b}^{\dagger}) $, with the zero-point fluctuation $ z_{\mathrm{zpf}}=\sqrt{\hbar/(2M\omega_{\mathrm{m}})} $, the tripartite interaction appears with the coupling rate~\cite{SupplementalMaterial}
\begin{equation}\label{Eq2}
\lambda = \frac{3{g_e}{\mu _0}{\mu _B}} {{8\pi r_0^4}}\sqrt {\frac{|\gamma |{M_s}V}{ M \omega_{\mathrm{m}}}},
\end{equation}
where $ \mu_0 $ is the permeability of vacuum, $ M_{\mathrm{s}} $ is the saturation magnetization, and $ V $ is the volume of YIG sphere.

For the mechanical mode, we propose three probable schemes that can generate the relative motion between the NV center and the YIG sphere:  an NV center in a trapped diamond nanoparticle or embedded in a cantilever~\cite{RN679, RN678} coupled to the magnon mode of a YIG sphere, and a single NV spin interacting with the magnon mode of a levitated micromagnet~\cite{PhysRevApplied.8.034002, doi:10.1063/1.5129145}, as shown in Fig.~\ref{fig:f1}(b). Here, we focus on the setup where the diamond nanoparticle containing a single NV center is trapped in a Paul trap [top part of Fig.~\ref{fig:f1}(b)]. An additional oscillating electrical potential~\cite{PhysRevA.42.2977, PhysRevLett.122.030501} supplies the approach to modulate and drive the center-of-mass motion of the trapped diamond particle, which gives rise to an added potential $ \hat{V}_{\mathrm{dr}} = - 2 q U_{\mathrm{T}} (\hat{z}/d_{\mathrm{T}})^2 \cos(2\omega_{\mathrm{p}}t) $, with the diamond particle charge $q$, the voltage amplitude $ U_\mathrm{T} $, and the characteristic trap dimension $ d_\mathrm{T} $. Hence, the center-of-mass motion of the diamond particle can be described by the Hamiltomian $ \hat{H}_{\mathrm{m}} = \frac{\hat{p}_{z}^{2}}{2M} + \frac{1}{2}M\omega_{\mathrm{m}}^2\hat{z}^2 + \frac{1}{2}k_{\mathrm{e}}(t)\hat{z}^2 $, with momentum operator $\hat{p}_{z}$. Here, the effective mass $M$ of the mechanical mode is the mass of the diamond particle, while the frequency $ \omega_{\mathrm{m}} $ is relevant to the electrical trap and the charge to mass ratio of the diamond particle. The rotation mode of the trapped diamond particle can be safely neglected, since its frequency is vanished with a spherical diamond~\cite{PhysRevA.96.063810}.
The last term is the parametric drive with the time-dependent tunable stiffness coefficient~\cite{SupplementalMaterial}
\begin{equation}\label{Eq3}
k_{\mathrm{e}}(t) = -\frac{4 q U_{\mathrm{T}}}{d_{\mathrm{T}}^2}\cos(2\omega_{\mathrm{p}}t).
\end{equation}
Employing the transformation $\hat{b} = \hat{z}/(2z_{\mathrm{zpf}}) + i z_{\mathrm{zpf}}\hat{p}_z/\hbar $, the Hamiltonian of the mechanical mode can be written as
\begin{equation}\label{Eq4}
\hat{H}_{\mathrm{m}} = \omega_{\mathrm{m}} \hat{b}^{\dagger} \hat{b}-\Omega_{\mathrm{p}}\cos(2\omega_{\mathrm{p}}t)(\hat{b}+\hat{b}^{\dagger})^2,
\end{equation}
with the parametric-drive amplitude $ \Omega_{\mathrm{p}} = 2qU_{\mathrm{T}} z_{\mathrm{zpf}}^2/\hbar d_{\mathrm{T}}^2 $. As alternatives in the middle and bottom parts of Fig.~\ref{fig:f1}(b), similar results can be obtained for both cases~\cite{SupplementalMaterial}.

In a suitable rotation framework, dropping the high-frequency oscillation and the constant terms as well, the total Hamitonian of the system can be obtained as
\begin{eqnarray}\label{Eq5}
\hat{H}_{\mathrm{Tot}} = &&\delta_{\mathrm{K}}\hat{a}^{\dagger}\hat{a} + \delta_{\mathrm{m}} \hat{b}^{\dagger} \hat{b} + \frac{\delta_{\mathrm{NV}}}{2} \hat{\sigma}_{z} - \frac{\Omega_{\mathrm{p}}}{2}( {\hat b}^{\dagger 2} + {\hat b}^2 )\nonumber\\
&&+\lambda(\hat{b}+\hat{b}^{\dagger})(\hat{a}^{\dagger}\hat{\sigma}^{-} + \hat{a}\hat{\sigma}^{+}) + \hat{H}_{\mathrm{JC}},
\end{eqnarray}
with the detunings ${\delta _{\mathrm{K}}} = {\omega _{\mathrm{K}}} - {\omega _{\mathrm{p}}}$, ${\delta _{\mathrm{m}}} = {\omega _{\mathrm{m}}} - {\omega _{\mathrm{p}}}$, and ${\delta _{\mathrm{NV}}} = {\omega _{\mathrm{NV}}} - {\omega _{\mathrm{p}}}$. Here, we have included the spin-magnon coupling term $ \hat{H}_{\mathrm{JC}} = g_0(\hat{a}^{\dagger}\hat{\sigma}^{-} + \hat{a}\hat{\sigma}^{+}) $, with the coupling rate $ g_0 = r_0\lambda/(3z_{\mathrm{zpf}}) $.
.

\textit{Enhanced tripartite interactions.---}For the Hamiltonian (\ref{Eq5}), we can apply the unitary transformation $ \hat{U}_{\mathrm{S}}(r) = \exp [r(\hat{b}^2 - \hat{b}^{\dagger 2})/2] $ to diagonalize the center-of-mass mechanical mode.
Here, the squeezing parameter $r$ is defined as $ \tanh 2r = \Omega_{\mathrm{p}}/\delta_{\mathrm{m}} $. In this squeezed frame, the total Hamiltonian can be written as
\begin{eqnarray}{\label{Eq6}}
\hat H_{\mathrm{Tot}}^{\mathrm{S}} = &&{\delta_{\mathrm{K}}}\hat {a}^{\dagger}\hat {a} + {\Delta_{\mathrm{m}}}\hat {b}^{\dagger}\hat b + \frac{\delta_{\mathrm{NV}}}{2}\hat {\sigma}_z \nonumber\\
&&+ \lambda _{\mathrm{eff}}(\hat{b}+\hat{b}^{\dagger})(\hat{a}^{\dagger}\hat{\sigma}^{-} + \hat{a}\hat{\sigma}^{+}) + \hat{H}_{\mathrm{JC}},
\end{eqnarray}
where $ \Delta_{\mathrm{m}} = \delta_{\mathrm{m}}/\cosh 2r $ and $ \lambda_{\mathrm{eff}} = \lambda e^{r}$.
The eigenstates of the free Hmiltionian $ \{|g,m,k\rangle, |e,m\pm 1,k-1\rangle\} $ can be applied to clarify the process of the tripartite interaction. Here, $ g $ ($e$) denotes the $ |0\rangle $ ($ |+1\rangle $) state of the NV spin. The particle numbers of the phonons and magnons are denoted by $\{m, m\pm 1\}$ and $ \{k, k-1\} $.
The condition for red (blue) detuning, $ \delta_{\mathrm{K}} \sim \delta_{\mathrm{NV}} - \Delta_{\mathrm{m}} $ ($ \delta_{\mathrm{K}} \sim \delta_{\mathrm{NV}} + \Delta_{\mathrm{m}} $), allows for the interaction $ \hat{a}\hat{b}\hat{\sigma}^{+} + H.c. $ ($ \hat{a}^{\dagger}\hat{b}\hat{\sigma}^{-} + H.c. $) in Eq.~(\ref{Eq1}) with the transition between $ \{|g,m,k\rangle $ and $ |e, m-1, k-1 \rangle \} $ ($ |e, m+1, k-1 \rangle \} $), which describes the spin and phonon annihilation upon magnon excitation (the spin annihilation with magnon and phonon excitation) and the inverse process.

\begin{figure}[t]
	\includegraphics[width=8.5cm]{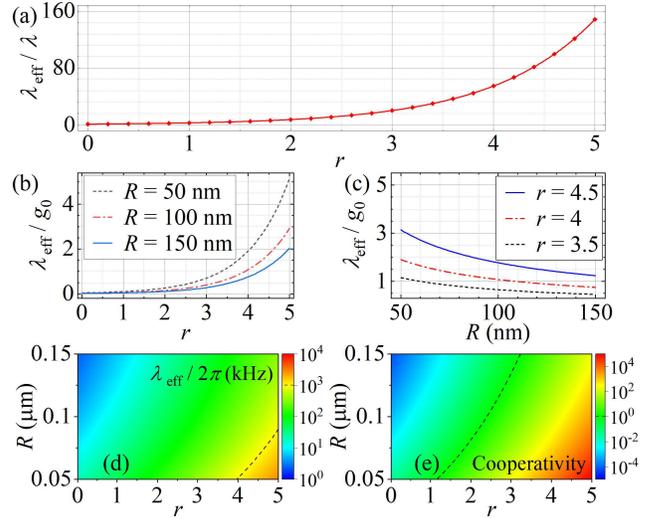}
	\caption{\label{fig:f2}(color online). (a) Tripartite coupling enhancement $ \lambda_{\mathrm{eff}} /\lambda$ versus the squeezing parameter $r$. (b) The ratio $ \lambda_{\mathrm{eff}}/g_0 $ versus $r$ with different radius of the YIG sphere. (c) The ratio $ \lambda_{\mathrm{eff}}/g_0 $ versus $ R $ for different $ r $. (d) and (e) Contour maps of $ \lambda_{\mathrm{eff}} $ and the tripartite cooperativity $ \mathcal{C} $ versus $R$ and $r$. The dashed line in (d) indicates the value of 1 MHz. The dashed line in (e) indicates the value of 1.}
\end{figure}
Remarkably, we find that the tripartite coupling strength $\lambda_{\mathrm{eff}}$ can be exponentially enhanced due to the amplification of the mechanical fluctuation caused by the phonon squeezing [Fig.~\ref{fig:f2}(a)].
For the scheme of the trapped diamond nanoparticle, the tripartite interaction can have the same magnitude as the bipartite interaction.
To make this clear, we define the ratio $ \lambda_{\mathrm{eff}}/g_0 = 3e^rz_{\mathrm{zpf}}/(d+R+R_s) $ with the diamond particle radius $R_s$ and the surface spacing $d$. With a proper choice of $r$ and $ R $, this ratio exceeds 1, indicating the coexistence of the two different interactions [see Fig.~\ref{fig:f2}(b) and ~\ref{fig:f2}(c)].
Naturally, as shown in Fig.~\ref{fig:f2}(d), the effective tripartite coupling strength $ \lambda_{\mathrm{eff}} $ exponentially increases with the squeezing parameter $ r $, and is inversely proportional to $ R^{5/2} $.

We now consider this tripartite coupling system in a realistic situation.
Here, we take into account the dephasing of the NV center spin ($\gamma_{\mathrm{s}}$), the decay of the Kittel mode ($ \gamma_{\mathrm{K}} $), and the effective mechanical phonon ($ \Gamma_{\mathrm{m}} $).
Though the effective mechanical decay rate is exponentially enlarged as well, one can define a generalized cooperativity $ \mathcal{C} = \lambda_{\mathrm{eff}}^3/(\Gamma_m \gamma_K \gamma_s) $ to quantify the coupling regime.
As shown in Fig.~\ref{fig:f2}(e), the system can reach the strong coupling regime ($ \mathcal{C}>1 $) with a large range of $R$ and $ r $. The result shows that increasing $ r $ and decreasing $R$ enable a large enhancement of the cooperativity.
Note that the results displayed in Fig.~\ref{fig:f2} are obtained with the surface spacing $ d=5 $ nm and the diamond radius $ R_s = 10 $ nm.

\begin{figure}[t]
	\includegraphics[width=8.5cm]{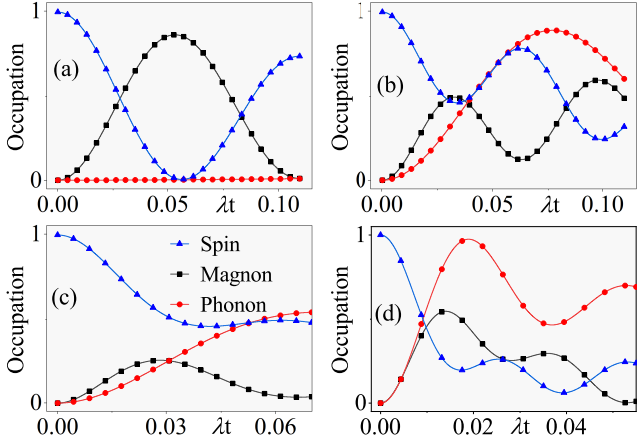}
	\caption{\label{fig:f3}(color online). Quantum dynamics of the NV spin, the Kittel magnon, and the center-of-mass motion (a) without mechanical amplification ($ r=0 $), (b) and (c) with $ r=3 $, and (d) with $ r=4.5 $. The magnon decay rate is $ \gamma_{\mathrm{K}}\sim 5\lambda $ in (a) and (b) , while it is $ \gamma_{\mathrm{K}}\sim 50\lambda $ in (c) and (d). The other parameters are $ g_0 \sim 30\lambda $, $ \gamma_{\mathrm{s}} \sim 0.05\lambda $, $ \Gamma_{\mathrm{m}}\sim 1.1\lambda $, $ R_s = 10 $ nm, $ R = 50 $ nm, and $ d = 5 $ nm. The results are obtained with the red detuning $ \delta_{\mathrm{K}} \sim \delta_{\mathrm{NV}} - \Delta_{\mathrm{m}} $ and initial state $ |e,0,0\rangle $.}
\end{figure}

To give more insight into this proposal, we numerically simulate the time-dependent occupation evolution of the spin qubit, the Kittel magnon, and the mechanical motion, as shown in the Fig.~\ref{fig:f3}.
Fig.~\ref{fig:f3} (a) shows, without mechanical amplification, the population of the mechanical mode can be neglected, and the dominant term is the spin-mangon coupling. As the center-of-mass motion is
modulated, the tripartite spin-magnon-phonon interaction needs to be considered.
In the intermediate squeezed regime, e.g. $r = 3$, tripartite
and dual interactions coexist, as shown in Fig.~\ref{fig:f3}(b) and (c), with different decays.
The direct tripartite coupling dominates the pairwise interaction  when the squeezing parameter is large enough, despite the large decay of the magnon, as illustrated in Fig.~\ref{fig:f3}(d). Therefore, by properly
choosing the experimental parameters, strong spin-magnon-phonon coupling at the single quantum level can
be obtained.

\begin{figure}[t]
	\includegraphics[width=8.5cm]{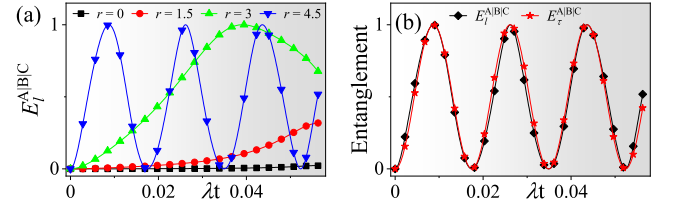}
	\caption{\label{fig:f4}(color online). (a) Entanglement versus time for different squeezing parameters $ r $, quantified by the minimum residual contangle $E_{l}^{A|B|C}$ without decay. (b) Entanglement versus time with two entanglement measures, minimum residual contangle $E_{l}^{A|B|C}$ and three-tangle $ E_{\tau}^{A|B|C} $. The initial state of the system and the parameters \{$ g_0 $, $ r_0 $\} are consistent with those in Fig.~\ref{fig:f3}. The squeezing parameter is $r = 4.5$.}
\end{figure}

\textit{Applications.}---We now consider generating tripartite entanglement among the spin qubit, the Kittel magnon and the mechanical phonon via the enhanced tripartite coupling.
Here, we employ the measure of genuine tripartite entanglement, \emph{minimum residual contangle} ranging from 0 to 1, defined as $ E_{l}^{A|B|C} = \min_{(A,B,C)}[E_{l}^{A|(BC)} - E_{l}^{A|B} - E_{l}^{A|C}] $, where (A,B,C) denotes all the permutations of the tripartite system~\cite{doi:10.1142/S0219749906001852}. The contangles $ \{E_{l}^{A|(BC)},E_{l}^{A|B},E_{l}^{A|C}\} $ are defined as the quadratic logarithm of $ \{||\rho^{T_A}||, ||\rho_{AB}^{T_A}||, ||\rho_{AC}^{T_A}||\} $ with the trace norm ($ ||\cdot|| $), partial transpose (superscript), and partial trace (subscript). We consider that the whole system starts from the state $ |e,0,0\rangle $ under the Hamiltonian Eq.~(\ref{Eq6}), with different squeezing parameters $r = \{0,1.5,3,4.5\} $, as shown in Fig.~\ref{fig:f4}(a). The minimum residual contangle vanishes without mechanical amplification, implying that the tripartite interaction is insignificant. When the center-of-mass motion is modulated by an applied electrical potential, the quality of the produced tripartite entangled state and the speed with which it is generated can be greatly improved. Using another entanglement measure three-tangle extended from the concurrence~\cite{PhysRevA.61.052306}, we can obtain the same result as the one of minimum residual contangle[Fig.~\ref{fig:f4}(b)]. The true tripartite entanglement between the degrees of freedom of the system can be widely used to execute tasks in the field of quantum information, such as quantum teleportation~\cite{1998KarlssonP43944400,2004YonezawaP430433}, dense coding~\cite{2006YeoP6050260502}, quantum computation~\cite{2009BriegelP1926}, and quantum secure sharing~\cite{1999CleveP648651,2004LanceP177903177903}.

We proceed to discuss how to detect the tripartite entanglement. The above measure of genuine tripartite entanglement is calculated from the density matrix of the whole system, which indicates that a possible approach can be the measurement of the density matrix using quantum state tomography ~\cite{PhysRevLett.94.070402,PhysRevLett.109.120403,PhysRevX.5.041006} or direct measurement ~\cite{RN758,RN755,RN751}. The readout of magnons can be realized by single-shot detection with a superconducting qubit~\cite{doi:10.1126/science.aaz9236,2017LachanceQuirionP16031501603150}. For the NV center, the state can be detected by cycling optical transition~\cite{RN591}, or photoelectrical detection of magnetic resonance~\cite{RN566}. The motion of nanodiamond particles can be detected by optical detection~\cite{RN734}.

\textit{Experimental feasibility.}---To examine the feasibility of this proposal for experiments, the center-of-mass vibration of a diamond particle can be obtained by levitating it in a quadratic potential. The paul trap~\cite{RN742, RN738, RN735, RN737,RN752, doi:10.1063/1.4893575, RN734, RN714} is a proper electric potential to realize this scheme. At the equilibrium location, the electric field can operate as a force to oppose gravity.
The levitated regime has been accomplished experimentally with a large mechanical factor $ Q\sim 10^8 $~\cite{RN743,PhysRevLett.124.093602}.
In this setup, for the spin qubit, we select the transition between the state $ |0\rangle $ and $ |+1\rangle $ in the ground states of NV center with frequency $ \omega_{\mathrm{NV}}=D_0 + |\gamma|B_{z,\mathrm{s}} $. Here, $D_0/2\pi = 2.87$ GHz is the electronic zero-field splitting. Applying variable external static magnetic field $ B_{z,\mathrm{s}} $ and $ B_{z,\mathrm{K}} $, hence, the detunings of the spin ($ \delta_{\mathrm{NV}} $) and the Kittel mode ($ \delta_{\mathrm{K}} $) can be tunable at the order of magnitudes of $10$ GHz.
To enlarge the direct tripartite interaction, we assume the driving amplitude $ \Omega_{\mathrm{p}}/2\pi\sim\omega_{\mathrm{p}}/2\pi\sim 200 $ MHz with the voltage amplitude $ U_{\mathrm{T}} = 12.6 $ V and the characteristic trap dimension $ d_{\mathrm{T}} = 100 $ $ \mu $m~\cite{SupplementalMaterial}. Here, we estimate the charge to mass ratios on the order of $ \mathrm{mC/kg} $~\cite{RN737}. At the same time, we estimate the mechanical frequency $ \omega_{\mathrm{m}}/2\pi\sim 1 $ kHz~\cite{PhysRevLett.124.093602}. Then the squeezing parameter satisfies $ r\in [0,5] $ to allow for the effective tripartite coupling $ \lambda_{\mathrm{eff}}\sim 100 \lambda$.
Given that $ r = 4.5 $, the enhanced coupling strength is $ \lambda_{\mathrm{eff}}/2\pi\sim 1.7 $ MHz.
Note that the frequencies $ {\omega_{\mathrm{K}},\omega_{\mathrm{NV}}} $ are on the order of 10 GHz, far larger than the mechanical frequency $ \omega_{\mathrm{m}} $.
At low temperature $ T\sim 10 $ mK, the thermal magnon number can be ignored with $ \bar{n}_{\mathrm{K}} \ll 1 $ for the case of $ \omega_{\mathrm{K}}/2\pi\sim 10 $ GHz. For practical considerations with saturation magnetization, we assume the decay of Kittel mode as $ \gamma_{\mathrm{K}}/2\pi\sim 1 $ MHz~\cite{PhysRevLett.113.083603, PhysRevLett.124.093602}.
For the mechanical mode, the thermal decay rate is $ \gamma_{\mathrm{th}}/2\pi = k_B T/(2\pi\hbar Q)\sim 2 $ Hz, which comes from the heating due to collisions with gas molecules~\cite{RN246}. Here, the gas damping satisfies $ \gamma_{\mathrm{gas}} = \omega_{\mathrm{m}}/Q $ with a ultra-low pressure $ P_{\mathrm{gas}} \sim 10^{-9} $ mBar~\cite{SupplementalMaterial,RN714}.
The mechanical amplification also leads to a magnification of the phonon decay by $e^{2r}$. The effective decay of mechanical mode can be obtained as $ \Gamma_{\mathrm{m}}/2\pi = e^{2r} \gamma_{\mathrm{th}}/2\pi \sim 21 $ kHz.
For a single NV center spin in diamond, the dephasing rate is about $ \gamma_{\mathrm{s}}/2\pi\sim 1 $ kHz~\cite{Bar-Gill2013}. Therefore, we can naturally estimate the tripartite cooperativity $ \mathcal{C}\sim 10^5\gg 1 $, which definitely indicates the strong coupling regime.

\textit{Conclusion.}---In this work, we propose an experimentally
feasible method for realizing direct and strong tripartite interactions among single NV spins, the Kittel magnon mode, and the phonon by introducing the relative motion between a single NV  center and a nearby micromagnet.
We show that the direct tripartite coupling strength can be exponentially enhanced by up to two orders
of magnitude via modulating the mechanical motion via parametric amplification.
We have shown the presence of tripartite entanglement via the enhanced spin-magnon-phonon coupling, and the possibility to actively control the tripartite coupling for realistic experimental parameters.
This is a promising platform for quantum science and technology based on spin-magnon-phonon tripartite strongly coupled systems.
\begin{acknowledgments}
P.B.L. is supported by the National Natural Science
Foundation of China under Grant No. 92065105, and the Natural Science Basic Research Program
of Shaanxi (Program No. 2020JC-02).
F.N. is supported in part by Nippon
Telegraph and Telephone Corporation (NTT) Research,
Japan Science, and Technology Agency (JST) (via the
Quantum Leap Flagship Program (Q-LEAP), Moonshot
R\&D Grant No. JPMJMS2061, Japan Society for
the Promotion of Science (JSPS) (via the Grants-inAid for Scientific Research (KAKENHI) Grant
No. JP20H00134, Army Research Office (ARO)
(Grant No. W911NF-18-1-0358), the Asian Office of
Aerospace Research and Development (AOARD) (via
Grant No. FA2386-20-1-4069), and the Foundational
Questions Institute (FQXi) (via Grant No. FQXiIAF19-06)
The simulations are obtained using QuTiP~\cite{JOHANSSON20121760,JOHANSSON20131234}.
\end{acknowledgments}

\begin{thebibliography}{148}%
\makeatletter
\providecommand \@ifxundefined [1]{%
 \@ifx{#1\undefined}
}%
\providecommand \@ifnum [1]{%
 \ifnum #1\expandafter \@firstoftwo
 \else \expandafter \@secondoftwo
 \fi
}%
\providecommand \@ifx [1]{%
 \ifx #1\expandafter \@firstoftwo
 \else \expandafter \@secondoftwo
 \fi
}%
\providecommand \natexlab [1]{#1}%
\providecommand \enquote  [1]{``#1''}%
\providecommand \bibnamefont  [1]{#1}%
\providecommand \bibfnamefont [1]{#1}%
\providecommand \citenamefont [1]{#1}%
\providecommand \href@noop [0]{\@secondoftwo}%
\providecommand \href [0]{\begingroup \@sanitize@url \@href}%
\providecommand \@href[1]{\@@startlink{#1}\@@href}%
\providecommand \@@href[1]{\endgroup#1\@@endlink}%
\providecommand \@sanitize@url [0]{\catcode `\\12\catcode `\$12\catcode
  `\&12\catcode `\#12\catcode `\^12\catcode `\_12\catcode `\%12\relax}%
\providecommand \@@startlink[1]{}%
\providecommand \@@endlink[0]{}%
\providecommand \url  [0]{\begingroup\@sanitize@url \@url }%
\providecommand \@url [1]{\endgroup\@href {#1}{\urlprefix }}%
\providecommand \urlprefix  [0]{URL }%
\providecommand \Eprint [0]{\href }%
\providecommand \doibase [0]{https://doi.org/}%
\providecommand \selectlanguage [0]{\@gobble}%
\providecommand \bibinfo  [0]{\@secondoftwo}%
\providecommand \bibfield  [0]{\@secondoftwo}%
\providecommand \translation [1]{[#1]}%
\providecommand \BibitemOpen [0]{}%
\providecommand \bibitemStop [0]{}%
\providecommand \bibitemNoStop [0]{.\EOS\space}%
\providecommand \EOS [0]{\spacefactor3000\relax}%
\providecommand \BibitemShut  [1]{\csname bibitem#1\endcsname}%
\let\auto@bib@innerbib\@empty
\bibitem [{\citenamefont {Xiang}\ \emph
  {et~al.}(2013{\natexlab{a}})\citenamefont {Xiang}, \citenamefont {Ashhab},
  \citenamefont {You},\ and\ \citenamefont {Nori}}]{RevModPhys.85.623}%
  \BibitemOpen
  \bibfield  {author} {\bibinfo {author} {\bibfnamefont {Z.-L.}\ \bibnamefont
  {Xiang}}, \bibinfo {author} {\bibfnamefont {S.}~\bibnamefont {Ashhab}},
  \bibinfo {author} {\bibfnamefont {J.~Q.}\ \bibnamefont {You}},\ and\ \bibinfo
  {author} {\bibfnamefont {F.}~\bibnamefont {Nori}},\ }\bibfield  {title}
  {\bibinfo {title} {Hybrid quantum circuits: Superconducting circuits
  interacting with other quantum systems},\ }\href
  {https://doi.org/10.1103/RevModPhys.85.623} {\bibfield  {journal} {\bibinfo
  {journal} {Rev. Mod. Phys.}\ }\textbf {\bibinfo {volume} {85}},\ \bibinfo
  {pages} {623} (\bibinfo {year} {2013}{\natexlab{a}})}\BibitemShut {NoStop}%
\bibitem [{\citenamefont {Shandilya}\ \emph
  {et~al.}(2021{\natexlab{a}})\citenamefont {Shandilya}, \citenamefont {Lake},
  \citenamefont {Mitchell}, \citenamefont {Sukachev},\ and\ \citenamefont
  {Barclay}}]{shandilya2021optomechanical}%
  \BibitemOpen
  \bibfield  {author} {\bibinfo {author} {\bibfnamefont {P.~K.}\ \bibnamefont
  {Shandilya}}, \bibinfo {author} {\bibfnamefont {D.~P.}\ \bibnamefont {Lake}},
  \bibinfo {author} {\bibfnamefont {M.~J.}\ \bibnamefont {Mitchell}}, \bibinfo
  {author} {\bibfnamefont {D.~D.}\ \bibnamefont {Sukachev}},\ and\ \bibinfo
  {author} {\bibfnamefont {P.~E.}\ \bibnamefont {Barclay}},\ }\bibfield
  {title} {\bibinfo {title} {Optomechanical interface between telecom photons
  and spin quantum memory},\ }\href
  {https://wwwnature.53yu.com/articles/s41567-021-01237-9} {\bibfield
  {journal} {\bibinfo  {journal} {Nat. Phys.}\ }\textbf {\bibinfo {volume}
  {17}},\ \bibinfo {pages} {1420} (\bibinfo {year}
  {2021}{\natexlab{a}})}\BibitemShut {NoStop}%
\bibitem [{\citenamefont {Dong}\ \emph {et~al.}(2021)\citenamefont {Dong},
  \citenamefont {Li}, \citenamefont {Liu},\ and\ \citenamefont
  {Nori}}]{PhysRevLett.126.203601}%
  \BibitemOpen
  \bibfield  {author} {\bibinfo {author} {\bibfnamefont {X.-L.}\ \bibnamefont
  {Dong}}, \bibinfo {author} {\bibfnamefont {P.-B.}\ \bibnamefont {Li}},
  \bibinfo {author} {\bibfnamefont {T.}~\bibnamefont {Liu}},\ and\ \bibinfo
  {author} {\bibfnamefont {F.}~\bibnamefont {Nori}},\ }\bibfield  {title}
  {\bibinfo {title} {Unconventional quantum sound-matter interactions in
  spin-optomechanical-crystal hybrid systems},\ }\href
  {https://doi.org/10.1103/PhysRevLett.126.203601} {\bibfield  {journal}
  {\bibinfo  {journal} {Phys. Rev. Lett.}\ }\textbf {\bibinfo {volume} {126}},\
  \bibinfo {pages} {203601} (\bibinfo {year} {2021})}\BibitemShut {NoStop}%
\bibitem [{\citenamefont {Das}\ \emph {et~al.}(2017)\citenamefont {Das},
  \citenamefont {Elfving}, \citenamefont {Faez},\ and\ \citenamefont
  {S\o{}rensen}}]{PhysRevLett.118.140501}%
  \BibitemOpen
  \bibfield  {author} {\bibinfo {author} {\bibfnamefont {S.}~\bibnamefont
  {Das}}, \bibinfo {author} {\bibfnamefont {V.~E.}\ \bibnamefont {Elfving}},
  \bibinfo {author} {\bibfnamefont {S.}~\bibnamefont {Faez}},\ and\ \bibinfo
  {author} {\bibfnamefont {A.~S.}\ \bibnamefont {S\o{}rensen}},\ }\bibfield
  {title} {\bibinfo {title} {Interfacing superconducting qubits and single
  optical photons using molecules in waveguides},\ }\href
  {https://doi.org/10.1103/PhysRevLett.118.140501} {\bibfield  {journal}
  {\bibinfo  {journal} {Phys. Rev. Lett.}\ }\textbf {\bibinfo {volume} {118}},\
  \bibinfo {pages} {140501} (\bibinfo {year} {2017})}\BibitemShut {NoStop}%
\bibitem [{\citenamefont {Sch\"utz}\ \emph {et~al.}(2020)\citenamefont
  {Sch\"utz}, \citenamefont {Schachenmayer}, \citenamefont {Hagenm\"uller},
  \citenamefont {Brennen}, \citenamefont {Volz}, \citenamefont {Sandoghdar},
  \citenamefont {Ebbesen}, \citenamefont {Genes},\ and\ \citenamefont
  {Pupillo}}]{PhysRevLett.124.113602}%
  \BibitemOpen
  \bibfield  {author} {\bibinfo {author} {\bibfnamefont {S.}~\bibnamefont
  {Sch\"utz}}, \bibinfo {author} {\bibfnamefont {J.}~\bibnamefont
  {Schachenmayer}}, \bibinfo {author} {\bibfnamefont {D.}~\bibnamefont
  {Hagenm\"uller}}, \bibinfo {author} {\bibfnamefont {G.~K.}\ \bibnamefont
  {Brennen}}, \bibinfo {author} {\bibfnamefont {T.}~\bibnamefont {Volz}},
  \bibinfo {author} {\bibfnamefont {V.}~\bibnamefont {Sandoghdar}}, \bibinfo
  {author} {\bibfnamefont {T.~W.}\ \bibnamefont {Ebbesen}}, \bibinfo {author}
  {\bibfnamefont {C.}~\bibnamefont {Genes}},\ and\ \bibinfo {author}
  {\bibfnamefont {G.}~\bibnamefont {Pupillo}},\ }\bibfield  {title} {\bibinfo
  {title} {Ensemble-induced strong light-matter coupling of a single quantum
  emitter},\ }\href {https://doi.org/10.1103/PhysRevLett.124.113602} {\bibfield
   {journal} {\bibinfo  {journal} {Phys. Rev. Lett.}\ }\textbf {\bibinfo
  {volume} {124}},\ \bibinfo {pages} {113602} (\bibinfo {year}
  {2020})}\BibitemShut {NoStop}%
\bibitem [{\citenamefont {Bennett}\ \emph {et~al.}(2013)\citenamefont
  {Bennett}, \citenamefont {Yao}, \citenamefont {Otterbach}, \citenamefont
  {Zoller}, \citenamefont {Rabl},\ and\ \citenamefont
  {Lukin}}]{PhysRevLett.110.156402}%
  \BibitemOpen
  \bibfield  {author} {\bibinfo {author} {\bibfnamefont {S.~D.}\ \bibnamefont
  {Bennett}}, \bibinfo {author} {\bibfnamefont {N.~Y.}\ \bibnamefont {Yao}},
  \bibinfo {author} {\bibfnamefont {J.}~\bibnamefont {Otterbach}}, \bibinfo
  {author} {\bibfnamefont {P.}~\bibnamefont {Zoller}}, \bibinfo {author}
  {\bibfnamefont {P.}~\bibnamefont {Rabl}},\ and\ \bibinfo {author}
  {\bibfnamefont {M.~D.}\ \bibnamefont {Lukin}},\ }\bibfield  {title} {\bibinfo
  {title} {Phonon-induced spin-spin interactions in diamond nanostructures:
  Application to spin squeezing},\ }\href
  {https://doi.org/10.1103/PhysRevLett.110.156402} {\bibfield  {journal}
  {\bibinfo  {journal} {Phys. Rev. Lett.}\ }\textbf {\bibinfo {volume} {110}},\
  \bibinfo {pages} {156402} (\bibinfo {year} {2013})}\BibitemShut {NoStop}%
\bibitem [{\citenamefont {Li}\ \emph {et~al.}(2016)\citenamefont {Li},
  \citenamefont {Xiang}, \citenamefont {Rabl},\ and\ \citenamefont
  {Nori}}]{PhysRevLett.117.015502}%
  \BibitemOpen
  \bibfield  {author} {\bibinfo {author} {\bibfnamefont {P.-B.}\ \bibnamefont
  {Li}}, \bibinfo {author} {\bibfnamefont {Z.-L.}\ \bibnamefont {Xiang}},
  \bibinfo {author} {\bibfnamefont {P.}~\bibnamefont {Rabl}},\ and\ \bibinfo
  {author} {\bibfnamefont {F.}~\bibnamefont {Nori}},\ }\bibfield  {title}
  {\bibinfo {title} {Hybrid quantum device with nitrogen-vacancy centers in
  diamond coupled to carbon nanotubes},\ }\href
  {https://doi.org/10.1103/PhysRevLett.117.015502} {\bibfield  {journal}
  {\bibinfo  {journal} {Phys. Rev. Lett.}\ }\textbf {\bibinfo {volume} {117}},\
  \bibinfo {pages} {015502} (\bibinfo {year} {2016})}\BibitemShut {NoStop}%
\bibitem [{\citenamefont {Li}\ and\ \citenamefont
  {Nori}(2018)}]{PhysRevApplied.10.024011}%
  \BibitemOpen
  \bibfield  {author} {\bibinfo {author} {\bibfnamefont {P.-B.}\ \bibnamefont
  {Li}}\ and\ \bibinfo {author} {\bibfnamefont {F.}~\bibnamefont {Nori}},\
  }\bibfield  {title} {\bibinfo {title} {Hybrid quantum system with
  nitrogen-vacancy centers in diamond coupled to surface-phonon polaritons in
  piezomagnetic superlattices},\ }\href
  {https://doi.org/10.1103/PhysRevApplied.10.024011} {\bibfield  {journal}
  {\bibinfo  {journal} {Phys. Rev. Applied}\ }\textbf {\bibinfo {volume}
  {10}},\ \bibinfo {pages} {024011} (\bibinfo {year} {2018})}\BibitemShut
  {NoStop}%
\bibitem [{\citenamefont {Ams\"uss}\ \emph {et~al.}(2011)\citenamefont
  {Ams\"uss}, \citenamefont {Koller}, \citenamefont {N\"obauer}, \citenamefont
  {Putz}, \citenamefont {Rotter}, \citenamefont {Sandner}, \citenamefont
  {Schneider}, \citenamefont {Schramb\"ock}, \citenamefont {Steinhauser},
  \citenamefont {Ritsch}, \citenamefont {Schmiedmayer},\ and\ \citenamefont
  {Majer}}]{PhysRevLett.107.060502}%
  \BibitemOpen
  \bibfield  {author} {\bibinfo {author} {\bibfnamefont {R.}~\bibnamefont
  {Ams\"uss}}, \bibinfo {author} {\bibfnamefont {C.}~\bibnamefont {Koller}},
  \bibinfo {author} {\bibfnamefont {T.}~\bibnamefont {N\"obauer}}, \bibinfo
  {author} {\bibfnamefont {S.}~\bibnamefont {Putz}}, \bibinfo {author}
  {\bibfnamefont {S.}~\bibnamefont {Rotter}}, \bibinfo {author} {\bibfnamefont
  {K.}~\bibnamefont {Sandner}}, \bibinfo {author} {\bibfnamefont
  {S.}~\bibnamefont {Schneider}}, \bibinfo {author} {\bibfnamefont
  {M.}~\bibnamefont {Schramb\"ock}}, \bibinfo {author} {\bibfnamefont
  {G.}~\bibnamefont {Steinhauser}}, \bibinfo {author} {\bibfnamefont
  {H.}~\bibnamefont {Ritsch}}, \bibinfo {author} {\bibfnamefont
  {J.}~\bibnamefont {Schmiedmayer}},\ and\ \bibinfo {author} {\bibfnamefont
  {J.}~\bibnamefont {Majer}},\ }\bibfield  {title} {\bibinfo {title} {Cavity
  {QED} with magnetically coupled collective spin states},\ }\href
  {https://doi.org/10.1103/PhysRevLett.107.060502} {\bibfield  {journal}
  {\bibinfo  {journal} {Phys. Rev. Lett.}\ }\textbf {\bibinfo {volume} {107}},\
  \bibinfo {pages} {060502} (\bibinfo {year} {2011})}\BibitemShut {NoStop}%
\bibitem [{\citenamefont {Will}\ \emph {et~al.}(2011)\citenamefont {Will},
  \citenamefont {Best}, \citenamefont {Braun}, \citenamefont {Schneider},\ and\
  \citenamefont {Bloch}}]{PhysRevLett.106.115305}%
  \BibitemOpen
  \bibfield  {author} {\bibinfo {author} {\bibfnamefont {S.}~\bibnamefont
  {Will}}, \bibinfo {author} {\bibfnamefont {T.}~\bibnamefont {Best}}, \bibinfo
  {author} {\bibfnamefont {S.}~\bibnamefont {Braun}}, \bibinfo {author}
  {\bibfnamefont {U.}~\bibnamefont {Schneider}},\ and\ \bibinfo {author}
  {\bibfnamefont {I.}~\bibnamefont {Bloch}},\ }\bibfield  {title} {\bibinfo
  {title} {Coherent interaction of a single fermion with a small bosonic
  field},\ }\href {https://doi.org/10.1103/PhysRevLett.106.115305} {\bibfield
  {journal} {\bibinfo  {journal} {Phys. Rev. Lett.}\ }\textbf {\bibinfo
  {volume} {106}},\ \bibinfo {pages} {115305} (\bibinfo {year}
  {2011})}\BibitemShut {NoStop}%
\bibitem [{\citenamefont {Marcos}\ \emph {et~al.}(2010)\citenamefont {Marcos},
  \citenamefont {Wubs}, \citenamefont {Taylor}, \citenamefont {Aguado},
  \citenamefont {Lukin},\ and\ \citenamefont
  {S\o{}rensen}}]{PhysRevLett.105.210501}%
  \BibitemOpen
  \bibfield  {author} {\bibinfo {author} {\bibfnamefont {D.}~\bibnamefont
  {Marcos}}, \bibinfo {author} {\bibfnamefont {M.}~\bibnamefont {Wubs}},
  \bibinfo {author} {\bibfnamefont {J.~M.}\ \bibnamefont {Taylor}}, \bibinfo
  {author} {\bibfnamefont {R.}~\bibnamefont {Aguado}}, \bibinfo {author}
  {\bibfnamefont {M.~D.}\ \bibnamefont {Lukin}},\ and\ \bibinfo {author}
  {\bibfnamefont {A.~S.}\ \bibnamefont {S\o{}rensen}},\ }\bibfield  {title}
  {\bibinfo {title} {Coupling nitrogen-vacancy centers in diamond to
  superconducting flux qubits},\ }\href
  {https://doi.org/10.1103/PhysRevLett.105.210501} {\bibfield  {journal}
  {\bibinfo  {journal} {Phys. Rev. Lett.}\ }\textbf {\bibinfo {volume} {105}},\
  \bibinfo {pages} {210501} (\bibinfo {year} {2010})}\BibitemShut {NoStop}%
\bibitem [{\citenamefont {Xiang}\ \emph
  {et~al.}(2013{\natexlab{b}})\citenamefont {Xiang}, \citenamefont {L\"u},
  \citenamefont {Li}, \citenamefont {You},\ and\ \citenamefont
  {Nori}}]{PhysRevB.87.144516}%
  \BibitemOpen
  \bibfield  {author} {\bibinfo {author} {\bibfnamefont {Z.-L.}\ \bibnamefont
  {Xiang}}, \bibinfo {author} {\bibfnamefont {X.-Y.}\ \bibnamefont {L\"u}},
  \bibinfo {author} {\bibfnamefont {T.-F.}\ \bibnamefont {Li}}, \bibinfo
  {author} {\bibfnamefont {J.~Q.}\ \bibnamefont {You}},\ and\ \bibinfo {author}
  {\bibfnamefont {F.}~\bibnamefont {Nori}},\ }\bibfield  {title} {\bibinfo
  {title} {Hybrid quantum circuit consisting of a superconducting flux qubit
  coupled to a spin ensemble and a transmission-line resonator},\ }\href
  {https://doi.org/10.1103/PhysRevB.87.144516} {\bibfield  {journal} {\bibinfo
  {journal} {Phys. Rev. B}\ }\textbf {\bibinfo {volume} {87}},\ \bibinfo
  {pages} {144516} (\bibinfo {year} {2013}{\natexlab{b}})}\BibitemShut
  {NoStop}%
\bibitem [{\citenamefont {Astner}\ \emph {et~al.}(2017)\citenamefont {Astner},
  \citenamefont {Nevlacsil}, \citenamefont {Peterschofsky}, \citenamefont
  {Angerer}, \citenamefont {Rotter}, \citenamefont {Putz}, \citenamefont
  {Schmiedmayer},\ and\ \citenamefont {Majer}}]{PhysRevLett.118.140502}%
  \BibitemOpen
  \bibfield  {author} {\bibinfo {author} {\bibfnamefont {T.}~\bibnamefont
  {Astner}}, \bibinfo {author} {\bibfnamefont {S.}~\bibnamefont {Nevlacsil}},
  \bibinfo {author} {\bibfnamefont {N.}~\bibnamefont {Peterschofsky}}, \bibinfo
  {author} {\bibfnamefont {A.}~\bibnamefont {Angerer}}, \bibinfo {author}
  {\bibfnamefont {S.}~\bibnamefont {Rotter}}, \bibinfo {author} {\bibfnamefont
  {S.}~\bibnamefont {Putz}}, \bibinfo {author} {\bibfnamefont {J.}~\bibnamefont
  {Schmiedmayer}},\ and\ \bibinfo {author} {\bibfnamefont {J.}~\bibnamefont
  {Majer}},\ }\bibfield  {title} {\bibinfo {title} {Coherent coupling of remote
  spin ensembles via a cavity bus},\ }\href
  {https://doi.org/10.1103/PhysRevLett.118.140502} {\bibfield  {journal}
  {\bibinfo  {journal} {Phys. Rev. Lett.}\ }\textbf {\bibinfo {volume} {118}},\
  \bibinfo {pages} {140502} (\bibinfo {year} {2017})}\BibitemShut {NoStop}%
\bibitem [{\citenamefont {Takahashi}\ \emph {et~al.}(2020)\citenamefont
  {Takahashi}, \citenamefont {Kassa}, \citenamefont {Christoforou},\ and\
  \citenamefont {Keller}}]{PhysRevLett.124.013602}%
  \BibitemOpen
  \bibfield  {author} {\bibinfo {author} {\bibfnamefont {H.}~\bibnamefont
  {Takahashi}}, \bibinfo {author} {\bibfnamefont {E.}~\bibnamefont {Kassa}},
  \bibinfo {author} {\bibfnamefont {C.}~\bibnamefont {Christoforou}},\ and\
  \bibinfo {author} {\bibfnamefont {M.}~\bibnamefont {Keller}},\ }\bibfield
  {title} {\bibinfo {title} {Strong coupling of a single ion to an optical
  cavity},\ }\href {https://doi.org/10.1103/PhysRevLett.124.013602} {\bibfield
  {journal} {\bibinfo  {journal} {Phys. Rev. Lett.}\ }\textbf {\bibinfo
  {volume} {124}},\ \bibinfo {pages} {013602} (\bibinfo {year}
  {2020})}\BibitemShut {NoStop}%
\bibitem [{\citenamefont {Stockklauser}\ \emph {et~al.}(2017)\citenamefont
  {Stockklauser}, \citenamefont {Scarlino}, \citenamefont {Koski},
  \citenamefont {Gasparinetti}, \citenamefont {Andersen}, \citenamefont
  {Reichl}, \citenamefont {Wegscheider}, \citenamefont {Ihn}, \citenamefont
  {Ensslin},\ and\ \citenamefont {Wallraff}}]{PhysRevX.7.011030}%
  \BibitemOpen
  \bibfield  {author} {\bibinfo {author} {\bibfnamefont {A.}~\bibnamefont
  {Stockklauser}}, \bibinfo {author} {\bibfnamefont {P.}~\bibnamefont
  {Scarlino}}, \bibinfo {author} {\bibfnamefont {J.~V.}\ \bibnamefont {Koski}},
  \bibinfo {author} {\bibfnamefont {S.}~\bibnamefont {Gasparinetti}}, \bibinfo
  {author} {\bibfnamefont {C.~K.}\ \bibnamefont {Andersen}}, \bibinfo {author}
  {\bibfnamefont {C.}~\bibnamefont {Reichl}}, \bibinfo {author} {\bibfnamefont
  {W.}~\bibnamefont {Wegscheider}}, \bibinfo {author} {\bibfnamefont
  {T.}~\bibnamefont {Ihn}}, \bibinfo {author} {\bibfnamefont {K.}~\bibnamefont
  {Ensslin}},\ and\ \bibinfo {author} {\bibfnamefont {A.}~\bibnamefont
  {Wallraff}},\ }\bibfield  {title} {\bibinfo {title} {Strong coupling cavity
  {QED} with gate-defined double quantum dots enabled by a high impedance
  resonator},\ }\href {https://doi.org/10.1103/PhysRevX.7.011030} {\bibfield
  {journal} {\bibinfo  {journal} {Phys. Rev. X}\ }\textbf {\bibinfo {volume}
  {7}},\ \bibinfo {pages} {011030} (\bibinfo {year} {2017})}\BibitemShut
  {NoStop}%
\bibitem [{\citenamefont {Forn-D\'{\i}az}\ \emph {et~al.}(2019)\citenamefont
  {Forn-D\'{\i}az}, \citenamefont {Lamata}, \citenamefont {Rico}, \citenamefont
  {Kono},\ and\ \citenamefont {Solano}}]{RevModPhys.91.025005}%
  \BibitemOpen
  \bibfield  {author} {\bibinfo {author} {\bibfnamefont {P.}~\bibnamefont
  {Forn-D\'{\i}az}}, \bibinfo {author} {\bibfnamefont {L.}~\bibnamefont
  {Lamata}}, \bibinfo {author} {\bibfnamefont {E.}~\bibnamefont {Rico}},
  \bibinfo {author} {\bibfnamefont {J.}~\bibnamefont {Kono}},\ and\ \bibinfo
  {author} {\bibfnamefont {E.}~\bibnamefont {Solano}},\ }\bibfield  {title}
  {\bibinfo {title} {Ultrastrong coupling regimes of light-matter
  interaction},\ }\href {https://doi.org/10.1103/RevModPhys.91.025005}
  {\bibfield  {journal} {\bibinfo  {journal} {Rev. Mod. Phys.}\ }\textbf
  {\bibinfo {volume} {91}},\ \bibinfo {pages} {025005} (\bibinfo {year}
  {2019})}\BibitemShut {NoStop}%
\bibitem [{\citenamefont {Xiong}\ \emph {et~al.}(2021)\citenamefont {Xiong},
  \citenamefont {Chen}, \citenamefont {Fang}, \citenamefont {Wang},
  \citenamefont {Ye},\ and\ \citenamefont {You}}]{PhysRevB.103.174106}%
  \BibitemOpen
  \bibfield  {author} {\bibinfo {author} {\bibfnamefont {W.}~\bibnamefont
  {Xiong}}, \bibinfo {author} {\bibfnamefont {J.}~\bibnamefont {Chen}},
  \bibinfo {author} {\bibfnamefont {B.}~\bibnamefont {Fang}}, \bibinfo {author}
  {\bibfnamefont {M.}~\bibnamefont {Wang}}, \bibinfo {author} {\bibfnamefont
  {L.}~\bibnamefont {Ye}},\ and\ \bibinfo {author} {\bibfnamefont {J.~Q.}\
  \bibnamefont {You}},\ }\bibfield  {title} {\bibinfo {title} {Strong tunable
  spin-spin interaction in a weakly coupled nitrogen vacancy spin-cavity
  electromechanical system},\ }\href
  {https://doi.org/10.1103/PhysRevB.103.174106} {\bibfield  {journal} {\bibinfo
   {journal} {Phys. Rev. B}\ }\textbf {\bibinfo {volume} {103}},\ \bibinfo
  {pages} {174106} (\bibinfo {year} {2021})}\BibitemShut {NoStop}%
\bibitem [{\citenamefont {Qin}\ \emph {et~al.}(2018)\citenamefont {Qin},
  \citenamefont {Miranowicz}, \citenamefont {Li}, \citenamefont {L\"u},
  \citenamefont {You},\ and\ \citenamefont {Nori}}]{PhysRevLett.120.093601}%
  \BibitemOpen
  \bibfield  {author} {\bibinfo {author} {\bibfnamefont {W.}~\bibnamefont
  {Qin}}, \bibinfo {author} {\bibfnamefont {A.}~\bibnamefont {Miranowicz}},
  \bibinfo {author} {\bibfnamefont {P.-B.}\ \bibnamefont {Li}}, \bibinfo
  {author} {\bibfnamefont {X.-Y.}\ \bibnamefont {L\"u}}, \bibinfo {author}
  {\bibfnamefont {J.~Q.}\ \bibnamefont {You}},\ and\ \bibinfo {author}
  {\bibfnamefont {F.}~\bibnamefont {Nori}},\ }\bibfield  {title} {\bibinfo
  {title} {Exponentially enhanced light-matter interaction, cooperativities,
  and steady-state entanglement using parametric amplification},\ }\href
  {https://doi.org/10.1103/PhysRevLett.120.093601} {\bibfield  {journal}
  {\bibinfo  {journal} {Phys. Rev. Lett.}\ }\textbf {\bibinfo {volume} {120}},\
  \bibinfo {pages} {093601} (\bibinfo {year} {2018})}\BibitemShut {NoStop}%
\bibitem [{\citenamefont {Li}\ \emph {et~al.}(2020{\natexlab{a}})\citenamefont
  {Li}, \citenamefont {Li},\ and\ \citenamefont
  {Li}}]{PhysRevResearch.2.013121}%
  \BibitemOpen
  \bibfield  {author} {\bibinfo {author} {\bibfnamefont {X.-X.}\ \bibnamefont
  {Li}}, \bibinfo {author} {\bibfnamefont {B.}~\bibnamefont {Li}},\ and\
  \bibinfo {author} {\bibfnamefont {P.-B.}\ \bibnamefont {Li}},\ }\bibfield
  {title} {\bibinfo {title} {Simulation of topological phases with color center
  arrays in phononic crystals},\ }\href
  {https://doi.org/10.1103/PhysRevResearch.2.013121} {\bibfield  {journal}
  {\bibinfo  {journal} {Phys. Rev. Research}\ }\textbf {\bibinfo {volume}
  {2}},\ \bibinfo {pages} {013121} (\bibinfo {year}
  {2020}{\natexlab{a}})}\BibitemShut {NoStop}%
\bibitem [{\citenamefont {Li}\ \emph {et~al.}(2021)\citenamefont {Li},
  \citenamefont {Li}, \citenamefont {Li}, \citenamefont {Gao},\ and\
  \citenamefont {Li}}]{PhysRevResearch.3.013025}%
  \BibitemOpen
  \bibfield  {author} {\bibinfo {author} {\bibfnamefont {X.-X.}\ \bibnamefont
  {Li}}, \bibinfo {author} {\bibfnamefont {P.-B.}\ \bibnamefont {Li}}, \bibinfo
  {author} {\bibfnamefont {H.-R.}\ \bibnamefont {Li}}, \bibinfo {author}
  {\bibfnamefont {H.}~\bibnamefont {Gao}},\ and\ \bibinfo {author}
  {\bibfnamefont {F.-L.}\ \bibnamefont {Li}},\ }\bibfield  {title} {\bibinfo
  {title} {Simulation of topological {Zak} phase in spin-phononic crystal
  networks},\ }\href {https://doi.org/10.1103/PhysRevResearch.3.013025}
  {\bibfield  {journal} {\bibinfo  {journal} {Phys. Rev. Research}\ }\textbf
  {\bibinfo {volume} {3}},\ \bibinfo {pages} {013025} (\bibinfo {year}
  {2021})}\BibitemShut {NoStop}%
\bibitem [{\citenamefont {Dong}\ \emph {et~al.}(2022)\citenamefont {Dong},
  \citenamefont {Shen}, \citenamefont {Gao}, \citenamefont {Li}, \citenamefont
  {Gao}, \citenamefont {Li},\ and\ \citenamefont {Li}}]{dong2022chiral}%
  \BibitemOpen
  \bibfield  {author} {\bibinfo {author} {\bibfnamefont {X.-L.}\ \bibnamefont
  {Dong}}, \bibinfo {author} {\bibfnamefont {C.-P.}\ \bibnamefont {Shen}},
  \bibinfo {author} {\bibfnamefont {S.-Y.}\ \bibnamefont {Gao}}, \bibinfo
  {author} {\bibfnamefont {H.-R.}\ \bibnamefont {Li}}, \bibinfo {author}
  {\bibfnamefont {H.}~\bibnamefont {Gao}}, \bibinfo {author} {\bibfnamefont
  {F.-L.}\ \bibnamefont {Li}},\ and\ \bibinfo {author} {\bibfnamefont {P.-B.}\
  \bibnamefont {Li}},\ }\bibfield  {title} {\bibinfo {title} {Chiral
  spin-phonon bound states and spin-spin interactions with phononic lattices},\
  }\href
  {https://journals.aps.org/prresearch/abstract/10.1103/PhysRevResearch.4.023077}
  {\bibfield  {journal} {\bibinfo  {journal} {Phys. Rev. Research}\ }\textbf
  {\bibinfo {volume} {4}},\ \bibinfo {pages} {023077} (\bibinfo {year}
  {2022})}\BibitemShut {NoStop}%
\bibitem [{\citenamefont {Jaynes}\ and\ \citenamefont
  {Cummings}(1963)}]{1443594}%
  \BibitemOpen
  \bibfield  {author} {\bibinfo {author} {\bibfnamefont {E.}~\bibnamefont
  {Jaynes}}\ and\ \bibinfo {author} {\bibfnamefont {F.}~\bibnamefont
  {Cummings}},\ }\bibfield  {title} {\bibinfo {title} {Comparison of quantum
  and semiclassical radiation theories with application to the beam maser},\
  }\href {https://doi.org/10.1109/PROC.1963.1664} {\bibfield  {journal}
  {\bibinfo  {journal} {Pro. IEEE.}\ }\textbf {\bibinfo {volume} {51}},\
  \bibinfo {pages} {89} (\bibinfo {year} {1963})}\BibitemShut {NoStop}%
\bibitem [{\citenamefont {Shore}\ and\ \citenamefont
  {Knight}(1993)}]{doi:10.1080/09500349314551321}%
  \BibitemOpen
  \bibfield  {author} {\bibinfo {author} {\bibfnamefont {B.~W.}\ \bibnamefont
  {Shore}}\ and\ \bibinfo {author} {\bibfnamefont {P.~L.}\ \bibnamefont
  {Knight}},\ }\bibfield  {title} {\bibinfo {title} {The {Jaynes-Cummings}
  model},\ }\href {https://doi.org/10.1080/09500349314551321} {\bibfield
  {journal} {\bibinfo  {journal} {J. Mod. Opt.}\ }\textbf {\bibinfo {volume}
  {40}},\ \bibinfo {pages} {1195} (\bibinfo {year} {1993})}\BibitemShut
  {NoStop}%
\bibitem [{\citenamefont {Scully}\ and\ \citenamefont
  {Zubairy}(1997)}]{scully1997quantum}%
  \BibitemOpen
  \bibfield  {author} {\bibinfo {author} {\bibfnamefont {M.~O.}\ \bibnamefont
  {Scully}}\ and\ \bibinfo {author} {\bibfnamefont {M.~S.}\ \bibnamefont
  {Zubairy}},\ }\href@noop {} {\emph {\bibinfo {title} {Quantum Optics}}}\
  (\bibinfo  {publisher} {Cambridge University Press},\ \bibinfo {year}
  {1997})\BibitemShut {NoStop}%
\bibitem [{\citenamefont {Knight}\ and\ \citenamefont
  {Milonni}(1980)}]{KNIGHT198021}%
  \BibitemOpen
  \bibfield  {author} {\bibinfo {author} {\bibfnamefont {P.}~\bibnamefont
  {Knight}}\ and\ \bibinfo {author} {\bibfnamefont {P.}~\bibnamefont
  {Milonni}},\ }\bibfield  {title} {\bibinfo {title} {The {Rabi} frequency in
  optical spectra},\ }\href
  {https://doi.org/https://doi.org/10.1016/0370-1573(80)90119-2} {\bibfield
  {journal} {\bibinfo  {journal} {Phys. Rep.}\ }\textbf {\bibinfo {volume}
  {66}},\ \bibinfo {pages} {21} (\bibinfo {year} {1980})}\BibitemShut {NoStop}%
\bibitem [{\citenamefont {Stenholm}(1973)}]{STENHOLM19731}%
  \BibitemOpen
  \bibfield  {author} {\bibinfo {author} {\bibfnamefont {S.}~\bibnamefont
  {Stenholm}},\ }\bibfield  {title} {\bibinfo {title} {Quantum theory of
  electromagnetic fields interacting with atoms and molecules},\ }\href
  {https://doi.org/https://doi.org/10.1016/0370-1573(73)90011-2} {\bibfield
  {journal} {\bibinfo  {journal} {Phys. Rep.}\ }\textbf {\bibinfo {volume}
  {6}},\ \bibinfo {pages} {1} (\bibinfo {year} {1973})}\BibitemShut {NoStop}%
\bibitem [{\citenamefont {Haroche}\ and\ \citenamefont
  {Kleppner}(1989)}]{haroche1989cavity}%
  \BibitemOpen
  \bibfield  {author} {\bibinfo {author} {\bibfnamefont {S.}~\bibnamefont
  {Haroche}}\ and\ \bibinfo {author} {\bibfnamefont {D.}~\bibnamefont
  {Kleppner}},\ }\bibfield  {title} {\bibinfo {title} {Cavity quantum
  electrodynamics},\ }\href {https://doi.org/10.1063/1.881201} {\bibfield
  {journal} {\bibinfo  {journal} {Phys. Today}\ }\textbf {\bibinfo {volume}
  {42}},\ \bibinfo {pages} {24} (\bibinfo {year} {1989})}\BibitemShut {NoStop}%
\bibitem [{\citenamefont {Braunstein}\ and\ \citenamefont {van
  Loock}(2005)}]{RevModPhys.77.513}%
  \BibitemOpen
  \bibfield  {author} {\bibinfo {author} {\bibfnamefont {S.~L.}\ \bibnamefont
  {Braunstein}}\ and\ \bibinfo {author} {\bibfnamefont {P.}~\bibnamefont {van
  Loock}},\ }\bibfield  {title} {\bibinfo {title} {Quantum information with
  continuous variables},\ }\href {https://doi.org/10.1103/RevModPhys.77.513}
  {\bibfield  {journal} {\bibinfo  {journal} {Rev. Mod. Phys.}\ }\textbf
  {\bibinfo {volume} {77}},\ \bibinfo {pages} {513} (\bibinfo {year}
  {2005})}\BibitemShut {NoStop}%
\bibitem [{\citenamefont {Galindo}\ and\ \citenamefont
  {Mart\'{\i}n-Delgado}(2002)}]{RevModPhys.74.347}%
  \BibitemOpen
  \bibfield  {author} {\bibinfo {author} {\bibfnamefont {A.}~\bibnamefont
  {Galindo}}\ and\ \bibinfo {author} {\bibfnamefont {M.~A.}\ \bibnamefont
  {Mart\'{\i}n-Delgado}},\ }\bibfield  {title} {\bibinfo {title} {Information
  and computation: Classical and quantum aspects},\ }\href
  {https://doi.org/10.1103/RevModPhys.74.347} {\bibfield  {journal} {\bibinfo
  {journal} {Rev. Mod. Phys.}\ }\textbf {\bibinfo {volume} {74}},\ \bibinfo
  {pages} {347} (\bibinfo {year} {2002})}\BibitemShut {NoStop}%
\bibitem [{\citenamefont {Saffman}\ \emph {et~al.}(2010)\citenamefont
  {Saffman}, \citenamefont {Walker},\ and\ \citenamefont
  {M\o{}lmer}}]{RevModPhys.82.2313}%
  \BibitemOpen
  \bibfield  {author} {\bibinfo {author} {\bibfnamefont {M.}~\bibnamefont
  {Saffman}}, \bibinfo {author} {\bibfnamefont {T.~G.}\ \bibnamefont
  {Walker}},\ and\ \bibinfo {author} {\bibfnamefont {K.}~\bibnamefont
  {M\o{}lmer}},\ }\bibfield  {title} {\bibinfo {title} {Quantum information
  with {Rydberg} atoms},\ }\href {https://doi.org/10.1103/RevModPhys.82.2313}
  {\bibfield  {journal} {\bibinfo  {journal} {Rev. Mod. Phys.}\ }\textbf
  {\bibinfo {volume} {82}},\ \bibinfo {pages} {2313} (\bibinfo {year}
  {2010})}\BibitemShut {NoStop}%
\bibitem [{\citenamefont {Weedbrook}\ \emph {et~al.}(2012)\citenamefont
  {Weedbrook}, \citenamefont {Pirandola}, \citenamefont {Garc\'{\i}a-Patr\'on},
  \citenamefont {Cerf}, \citenamefont {Ralph}, \citenamefont {Shapiro},\ and\
  \citenamefont {Lloyd}}]{RevModPhys.84.621}%
  \BibitemOpen
  \bibfield  {author} {\bibinfo {author} {\bibfnamefont {C.}~\bibnamefont
  {Weedbrook}}, \bibinfo {author} {\bibfnamefont {S.}~\bibnamefont
  {Pirandola}}, \bibinfo {author} {\bibfnamefont {R.}~\bibnamefont
  {Garc\'{\i}a-Patr\'on}}, \bibinfo {author} {\bibfnamefont {N.~J.}\
  \bibnamefont {Cerf}}, \bibinfo {author} {\bibfnamefont {T.~C.}\ \bibnamefont
  {Ralph}}, \bibinfo {author} {\bibfnamefont {J.~H.}\ \bibnamefont {Shapiro}},\
  and\ \bibinfo {author} {\bibfnamefont {S.}~\bibnamefont {Lloyd}},\ }\bibfield
   {title} {\bibinfo {title} {Gaussian quantum information},\ }\href
  {https://doi.org/10.1103/RevModPhys.84.621} {\bibfield  {journal} {\bibinfo
  {journal} {Rev. Mod. Phys.}\ }\textbf {\bibinfo {volume} {84}},\ \bibinfo
  {pages} {621} (\bibinfo {year} {2012})}\BibitemShut {NoStop}%
\bibitem [{\citenamefont {Georgescu}\ \emph {et~al.}(2014)\citenamefont
  {Georgescu}, \citenamefont {Ashhab},\ and\ \citenamefont
  {Nori}}]{RevModPhys.86.153}%
  \BibitemOpen
  \bibfield  {author} {\bibinfo {author} {\bibfnamefont {I.~M.}\ \bibnamefont
  {Georgescu}}, \bibinfo {author} {\bibfnamefont {S.}~\bibnamefont {Ashhab}},\
  and\ \bibinfo {author} {\bibfnamefont {F.}~\bibnamefont {Nori}},\ }\bibfield
  {title} {\bibinfo {title} {Quantum simulation},\ }\href
  {https://doi.org/10.1103/RevModPhys.86.153} {\bibfield  {journal} {\bibinfo
  {journal} {Rev. Mod. Phys.}\ }\textbf {\bibinfo {volume} {86}},\ \bibinfo
  {pages} {153} (\bibinfo {year} {2014})}\BibitemShut {NoStop}%
\bibitem [{\citenamefont {Foulkes}\ \emph {et~al.}(2001)\citenamefont
  {Foulkes}, \citenamefont {Mitas}, \citenamefont {Needs},\ and\ \citenamefont
  {Rajagopal}}]{RevModPhys.73.33}%
  \BibitemOpen
  \bibfield  {author} {\bibinfo {author} {\bibfnamefont {W.~M.~C.}\
  \bibnamefont {Foulkes}}, \bibinfo {author} {\bibfnamefont {L.}~\bibnamefont
  {Mitas}}, \bibinfo {author} {\bibfnamefont {R.~J.}\ \bibnamefont {Needs}},\
  and\ \bibinfo {author} {\bibfnamefont {G.}~\bibnamefont {Rajagopal}},\
  }\bibfield  {title} {\bibinfo {title} {Quantum {Monte Carlo} simulations of
  solids},\ }\href {https://doi.org/10.1103/RevModPhys.73.33} {\bibfield
  {journal} {\bibinfo  {journal} {Rev. Mod. Phys.}\ }\textbf {\bibinfo {volume}
  {73}},\ \bibinfo {pages} {33} (\bibinfo {year} {2001})}\BibitemShut {NoStop}%
\bibitem [{\citenamefont {Aharonovich}\ \emph {et~al.}(2011)\citenamefont
  {Aharonovich}, \citenamefont {Greentree},\ and\ \citenamefont
  {Prawer}}]{Aharonovich2011}%
  \BibitemOpen
  \bibfield  {author} {\bibinfo {author} {\bibfnamefont {I.}~\bibnamefont
  {Aharonovich}}, \bibinfo {author} {\bibfnamefont {A.~D.}\ \bibnamefont
  {Greentree}},\ and\ \bibinfo {author} {\bibfnamefont {S.}~\bibnamefont
  {Prawer}},\ }\bibfield  {title} {\bibinfo {title} {Diamond photonics},\
  }\href {https://doi.org/10.1038/nphoton.2011.54} {\bibfield  {journal}
  {\bibinfo  {journal} {Nat. Photonics}\ }\textbf {\bibinfo {volume} {5}},\
  \bibinfo {pages} {397} (\bibinfo {year} {2011})}\BibitemShut {NoStop}%
\bibitem [{\citenamefont {Marcus}\ \emph {et~al.}(2013)\citenamefont {Marcus},
  \citenamefont {Neil}, \citenamefont {Paul}, \citenamefont {Fedor},
  \citenamefont {Jörg},\ and\ \citenamefont {Lloyd}}]{DOHERTY20131}%
  \BibitemOpen
  \bibfield  {author} {\bibinfo {author} {\bibfnamefont {W.~D.}\ \bibnamefont
  {Marcus}}, \bibinfo {author} {\bibfnamefont {B.~M.}\ \bibnamefont {Neil}},
  \bibinfo {author} {\bibfnamefont {D.}~\bibnamefont {Paul}}, \bibinfo {author}
  {\bibfnamefont {J.}~\bibnamefont {Fedor}}, \bibinfo {author} {\bibfnamefont
  {W.}~\bibnamefont {Jörg}},\ and\ \bibinfo {author} {\bibfnamefont {C.~H.}\
  \bibnamefont {Lloyd}},\ }\bibfield  {title} {\bibinfo {title} {The
  nitrogen-vacancy colour centre in diamond},\ }\href
  {https://doi.org/https://doi.org/10.1016/j.physrep.2013.02.001} {\bibfield
  {journal} {\bibinfo  {journal} {Phys. Rep.}\ }\textbf {\bibinfo {volume}
  {528}},\ \bibinfo {pages} {1} (\bibinfo {year} {2013})}\BibitemShut {NoStop}%
\bibitem [{\citenamefont {Doherty}\ \emph {et~al.}(2014)\citenamefont
  {Doherty}, \citenamefont {Struzhkin}, \citenamefont {Simpson}, \citenamefont
  {McGuinness}, \citenamefont {Meng}, \citenamefont {Stacey}, \citenamefont
  {Karle}, \citenamefont {Hemley}, \citenamefont {Manson}, \citenamefont
  {Hollenberg},\ and\ \citenamefont {Prawer}}]{PhysRevLett.112.047601}%
  \BibitemOpen
  \bibfield  {author} {\bibinfo {author} {\bibfnamefont {M.~W.}\ \bibnamefont
  {Doherty}}, \bibinfo {author} {\bibfnamefont {V.~V.}\ \bibnamefont
  {Struzhkin}}, \bibinfo {author} {\bibfnamefont {D.~A.}\ \bibnamefont
  {Simpson}}, \bibinfo {author} {\bibfnamefont {L.~P.}\ \bibnamefont
  {McGuinness}}, \bibinfo {author} {\bibfnamefont {Y.}~\bibnamefont {Meng}},
  \bibinfo {author} {\bibfnamefont {A.}~\bibnamefont {Stacey}}, \bibinfo
  {author} {\bibfnamefont {T.~J.}\ \bibnamefont {Karle}}, \bibinfo {author}
  {\bibfnamefont {R.~J.}\ \bibnamefont {Hemley}}, \bibinfo {author}
  {\bibfnamefont {N.~B.}\ \bibnamefont {Manson}}, \bibinfo {author}
  {\bibfnamefont {L.~C.~L.}\ \bibnamefont {Hollenberg}},\ and\ \bibinfo
  {author} {\bibfnamefont {S.}~\bibnamefont {Prawer}},\ }\bibfield  {title}
  {\bibinfo {title} {Electronic properties and metrology applications of the
  diamond {${\mathrm{NV}}^{\ensuremath{-}}$} center under pressure},\ }\href
  {https://doi.org/10.1103/PhysRevLett.112.047601} {\bibfield  {journal}
  {\bibinfo  {journal} {Phys. Rev. Lett.}\ }\textbf {\bibinfo {volume} {112}},\
  \bibinfo {pages} {047601} (\bibinfo {year} {2014})}\BibitemShut {NoStop}%
\bibitem [{\citenamefont {Barry}\ \emph {et~al.}(2020)\citenamefont {Barry},
  \citenamefont {Schloss}, \citenamefont {Bauch}, \citenamefont {Turner},
  \citenamefont {Hart}, \citenamefont {Pham},\ and\ \citenamefont
  {Walsworth}}]{RevModPhys.92.015004}%
  \BibitemOpen
  \bibfield  {author} {\bibinfo {author} {\bibfnamefont {J.~F.}\ \bibnamefont
  {Barry}}, \bibinfo {author} {\bibfnamefont {J.~M.}\ \bibnamefont {Schloss}},
  \bibinfo {author} {\bibfnamefont {E.}~\bibnamefont {Bauch}}, \bibinfo
  {author} {\bibfnamefont {M.~J.}\ \bibnamefont {Turner}}, \bibinfo {author}
  {\bibfnamefont {C.~A.}\ \bibnamefont {Hart}}, \bibinfo {author}
  {\bibfnamefont {L.~M.}\ \bibnamefont {Pham}},\ and\ \bibinfo {author}
  {\bibfnamefont {R.~L.}\ \bibnamefont {Walsworth}},\ }\bibfield  {title}
  {\bibinfo {title} {Sensitivity optimization for {NV-diamond} magnetometry},\
  }\href {https://doi.org/10.1103/RevModPhys.92.015004} {\bibfield  {journal}
  {\bibinfo  {journal} {Rev. Mod. Phys.}\ }\textbf {\bibinfo {volume} {92}},\
  \bibinfo {pages} {015004} (\bibinfo {year} {2020})}\BibitemShut {NoStop}%
\bibitem [{\citenamefont {Bar-Gill}\ \emph {et~al.}(2013)\citenamefont
  {Bar-Gill}, \citenamefont {Pham}, \citenamefont {Jarmola}, \citenamefont
  {Budker},\ and\ \citenamefont {Walsworth}}]{Bar-Gill2013}%
  \BibitemOpen
  \bibfield  {author} {\bibinfo {author} {\bibfnamefont {N.}~\bibnamefont
  {Bar-Gill}}, \bibinfo {author} {\bibfnamefont {L.}~\bibnamefont {Pham}},
  \bibinfo {author} {\bibfnamefont {A.}~\bibnamefont {Jarmola}}, \bibinfo
  {author} {\bibfnamefont {D.}~\bibnamefont {Budker}},\ and\ \bibinfo {author}
  {\bibfnamefont {R.}~\bibnamefont {Walsworth}},\ }\bibfield  {title} {\bibinfo
  {title} {Solid-state electronic spin coherence time approaching one second},\
  }\href {https://doi.org/10.1038/ncomms2771} {\bibfield  {journal} {\bibinfo
  {journal} {Nat. Commun.}\ }\textbf {\bibinfo {volume} {4}},\ \bibinfo {pages}
  {1743} (\bibinfo {year} {2013})}\BibitemShut {NoStop}%
\bibitem [{\citenamefont {Abobeih}\ \emph {et~al.}(2018)\citenamefont
  {Abobeih}, \citenamefont {Cramer}, \citenamefont {Bakker}, \citenamefont
  {Kalb}, \citenamefont {Markham}, \citenamefont {Twitchen},\ and\
  \citenamefont {Taminiau}}]{Abobeih2018One}%
  \BibitemOpen
  \bibfield  {author} {\bibinfo {author} {\bibfnamefont {M.~H.}\ \bibnamefont
  {Abobeih}}, \bibinfo {author} {\bibfnamefont {J.}~\bibnamefont {Cramer}},
  \bibinfo {author} {\bibfnamefont {M.~A.}\ \bibnamefont {Bakker}}, \bibinfo
  {author} {\bibfnamefont {N.}~\bibnamefont {Kalb}}, \bibinfo {author}
  {\bibfnamefont {M.}~\bibnamefont {Markham}}, \bibinfo {author} {\bibfnamefont
  {D.~J.}\ \bibnamefont {Twitchen}},\ and\ \bibinfo {author} {\bibfnamefont
  {T.~H.}\ \bibnamefont {Taminiau}},\ }\bibfield  {title} {\bibinfo {title}
  {One-second coherence for a single electron spin coupled to a multi-qubit
  nuclear-spin environment},\ }\href
  {https://doi.org/10.1038/s41467-018-04916-z} {\bibfield  {journal} {\bibinfo
  {journal} {Nat. Commun.}\ }\textbf {\bibinfo {volume} {9}},\ \bibinfo {pages}
  {2552} (\bibinfo {year} {2018})}\BibitemShut {NoStop}%
\bibitem [{\citenamefont {Casola}\ \emph {et~al.}(2018)\citenamefont {Casola},
  \citenamefont {van~der Sar},\ and\ \citenamefont {Yacoby}}]{Casola2018}%
  \BibitemOpen
  \bibfield  {author} {\bibinfo {author} {\bibfnamefont {F.}~\bibnamefont
  {Casola}}, \bibinfo {author} {\bibfnamefont {T.}~\bibnamefont {van~der
  Sar}},\ and\ \bibinfo {author} {\bibfnamefont {A.}~\bibnamefont {Yacoby}},\
  }\bibfield  {title} {\bibinfo {title} {Probing condensed matter physics with
  magnetometry based on nitrogen-vacancy centres in diamond},\ }\href
  {https://doi.org/10.1038/natrevmats.2017.88} {\bibfield  {journal} {\bibinfo
  {journal} {Nat. Rev. Mat.}\ }\textbf {\bibinfo {volume} {3}},\ \bibinfo
  {pages} {17088} (\bibinfo {year} {2018})}\BibitemShut {NoStop}%
\bibitem [{\citenamefont {Lu}\ \emph {et~al.}(2020)\citenamefont {Lu},
  \citenamefont {Zhang}, \citenamefont {Liu}, \citenamefont {Nori},
  \citenamefont {Fan},\ and\ \citenamefont {Pan}}]{PhysRevLett.124.210502}%
  \BibitemOpen
  \bibfield  {author} {\bibinfo {author} {\bibfnamefont {Y.-N.}\ \bibnamefont
  {Lu}}, \bibinfo {author} {\bibfnamefont {Y.-R.}\ \bibnamefont {Zhang}},
  \bibinfo {author} {\bibfnamefont {G.-Q.}\ \bibnamefont {Liu}}, \bibinfo
  {author} {\bibfnamefont {F.}~\bibnamefont {Nori}}, \bibinfo {author}
  {\bibfnamefont {H.}~\bibnamefont {Fan}},\ and\ \bibinfo {author}
  {\bibfnamefont {X.-Y.}\ \bibnamefont {Pan}},\ }\bibfield  {title} {\bibinfo
  {title} {Observing information backflow from controllable non-{Markovian}
  multichannels in diamond},\ }\href
  {https://doi.org/10.1103/PhysRevLett.124.210502} {\bibfield  {journal}
  {\bibinfo  {journal} {Phys. Rev. Lett.}\ }\textbf {\bibinfo {volume} {124}},\
  \bibinfo {pages} {210502} (\bibinfo {year} {2020})}\BibitemShut {NoStop}%
\bibitem [{\citenamefont {Ai}\ \emph {et~al.}(2021)\citenamefont {Ai},
  \citenamefont {Li}, \citenamefont {Qin}, \citenamefont {Zhao}, \citenamefont
  {Sun},\ and\ \citenamefont {Nori}}]{PhysRevB.104.014109}%
  \BibitemOpen
  \bibfield  {author} {\bibinfo {author} {\bibfnamefont {Q.}~\bibnamefont
  {Ai}}, \bibinfo {author} {\bibfnamefont {P.-B.}\ \bibnamefont {Li}}, \bibinfo
  {author} {\bibfnamefont {W.}~\bibnamefont {Qin}}, \bibinfo {author}
  {\bibfnamefont {J.-X.}\ \bibnamefont {Zhao}}, \bibinfo {author}
  {\bibfnamefont {C.~P.}\ \bibnamefont {Sun}},\ and\ \bibinfo {author}
  {\bibfnamefont {F.}~\bibnamefont {Nori}},\ }\bibfield  {title} {\bibinfo
  {title} {The {NV} metamaterial: Tunable quantum hyperbolic metamaterial using
  nitrogen vacancy centers in diamond},\ }\href
  {https://doi.org/10.1103/PhysRevB.104.014109} {\bibfield  {journal} {\bibinfo
   {journal} {Phys. Rev. B}\ }\textbf {\bibinfo {volume} {104}},\ \bibinfo
  {pages} {014109} (\bibinfo {year} {2021})}\BibitemShut {NoStop}%
\bibitem [{\citenamefont {Rusconi}\ \emph
  {et~al.}(2022{\natexlab{a}})\citenamefont {Rusconi}, \citenamefont
  {Perdriat}, \citenamefont {H\'etet}, \citenamefont {Romero-Isart},\ and\
  \citenamefont {Stickler}}]{PhysRevLett.129.093605}%
  \BibitemOpen
  \bibfield  {author} {\bibinfo {author} {\bibfnamefont {C.~C.}\ \bibnamefont
  {Rusconi}}, \bibinfo {author} {\bibfnamefont {M.}~\bibnamefont {Perdriat}},
  \bibinfo {author} {\bibfnamefont {G.}~\bibnamefont {H\'etet}}, \bibinfo
  {author} {\bibfnamefont {O.}~\bibnamefont {Romero-Isart}},\ and\ \bibinfo
  {author} {\bibfnamefont {B.~A.}\ \bibnamefont {Stickler}},\ }\bibfield
  {title} {\bibinfo {title} {Spin-controlled quantum interference of levitated
  nanorotors},\ }\href {https://doi.org/10.1103/PhysRevLett.129.093605}
  {\bibfield  {journal} {\bibinfo  {journal} {Phys. Rev. Lett.}\ }\textbf
  {\bibinfo {volume} {129}},\ \bibinfo {pages} {093605} (\bibinfo {year}
  {2022}{\natexlab{a}})}\BibitemShut {NoStop}%
\bibitem [{\citenamefont {Li}\ \emph {et~al.}(2019{\natexlab{a}})\citenamefont
  {Li}, \citenamefont {Zhu},\ and\ \citenamefont
  {Agarwal}}]{PhysRevA.99.021801}%
  \BibitemOpen
  \bibfield  {author} {\bibinfo {author} {\bibfnamefont {J.}~\bibnamefont
  {Li}}, \bibinfo {author} {\bibfnamefont {S.-Y.}\ \bibnamefont {Zhu}},\ and\
  \bibinfo {author} {\bibfnamefont {G.~S.}\ \bibnamefont {Agarwal}},\
  }\bibfield  {title} {\bibinfo {title} {Squeezed states of magnons and phonons
  in cavity magnomechanics},\ }\href
  {https://doi.org/10.1103/PhysRevA.99.021801} {\bibfield  {journal} {\bibinfo
  {journal} {Phys. Rev. A}\ }\textbf {\bibinfo {volume} {99}},\ \bibinfo
  {pages} {021801} (\bibinfo {year} {2019}{\natexlab{a}})}\BibitemShut
  {NoStop}%
\bibitem [{\citenamefont {Lachance-Quirion}\ \emph {et~al.}(2019)\citenamefont
  {Lachance-Quirion}, \citenamefont {Tabuchi}, \citenamefont {Gloppe},
  \citenamefont {Usami},\ and\ \citenamefont
  {Nakamura}}]{Lachance_Quirion_2019}%
  \BibitemOpen
  \bibfield  {author} {\bibinfo {author} {\bibfnamefont {D.}~\bibnamefont
  {Lachance-Quirion}}, \bibinfo {author} {\bibfnamefont {Y.}~\bibnamefont
  {Tabuchi}}, \bibinfo {author} {\bibfnamefont {A.}~\bibnamefont {Gloppe}},
  \bibinfo {author} {\bibfnamefont {K.}~\bibnamefont {Usami}},\ and\ \bibinfo
  {author} {\bibfnamefont {Y.}~\bibnamefont {Nakamura}},\ }\bibfield  {title}
  {\bibinfo {title} {Hybrid quantum systems based on magnonics},\ }\href
  {https://doi.org/10.7567/1882-0786/ab248d} {\bibfield  {journal} {\bibinfo
  {journal} {Appl. Phys. Express}\ }\textbf {\bibinfo {volume} {12}},\ \bibinfo
  {pages} {070101} (\bibinfo {year} {2019})}\BibitemShut {NoStop}%
\bibitem [{\citenamefont {Kani}\ \emph
  {et~al.}(2022{\natexlab{a}})\citenamefont {Kani}, \citenamefont {Sarma},\
  and\ \citenamefont {Twamley}}]{PhysRevLett.128.013602}%
  \BibitemOpen
  \bibfield  {author} {\bibinfo {author} {\bibfnamefont {A.}~\bibnamefont
  {Kani}}, \bibinfo {author} {\bibfnamefont {B.}~\bibnamefont {Sarma}},\ and\
  \bibinfo {author} {\bibfnamefont {J.}~\bibnamefont {Twamley}},\ }\bibfield
  {title} {\bibinfo {title} {Intensive cavity-magnomechanical cooling of a
  levitated macromagnet},\ }\href
  {https://doi.org/10.1103/PhysRevLett.128.013602} {\bibfield  {journal}
  {\bibinfo  {journal} {Phys. Rev. Lett.}\ }\textbf {\bibinfo {volume} {128}},\
  \bibinfo {pages} {013602} (\bibinfo {year} {2022}{\natexlab{a}})}\BibitemShut
  {NoStop}%
\bibitem [{\citenamefont {Tabuchi}\ \emph {et~al.}(2016)\citenamefont
  {Tabuchi}, \citenamefont {Ishino}, \citenamefont {Noguchi}, \citenamefont
  {Ishikawa}, \citenamefont {Yamazaki}, \citenamefont {Usami},\ and\
  \citenamefont {Nakamura}}]{TABUCHI2016729}%
  \BibitemOpen
  \bibfield  {author} {\bibinfo {author} {\bibfnamefont {Y.}~\bibnamefont
  {Tabuchi}}, \bibinfo {author} {\bibfnamefont {S.}~\bibnamefont {Ishino}},
  \bibinfo {author} {\bibfnamefont {A.}~\bibnamefont {Noguchi}}, \bibinfo
  {author} {\bibfnamefont {T.}~\bibnamefont {Ishikawa}}, \bibinfo {author}
  {\bibfnamefont {R.}~\bibnamefont {Yamazaki}}, \bibinfo {author}
  {\bibfnamefont {K.}~\bibnamefont {Usami}},\ and\ \bibinfo {author}
  {\bibfnamefont {Y.}~\bibnamefont {Nakamura}},\ }\bibfield  {title} {\bibinfo
  {title} {Quantum magnonics: The magnon meets the superconducting qubit},\
  }\href {https://doi.org/https://doi.org/10.1016/j.crhy.2016.07.009}
  {\bibfield  {journal} {\bibinfo  {journal} {C. R. Phys.}\ }\textbf {\bibinfo
  {volume} {17}},\ \bibinfo {pages} {729} (\bibinfo {year} {2016})}\BibitemShut
  {NoStop}%
\bibitem [{\citenamefont {Wang}\ \emph {et~al.}(2018)\citenamefont {Wang},
  \citenamefont {Zhang}, \citenamefont {Zhang}, \citenamefont {Li},
  \citenamefont {Hu},\ and\ \citenamefont {You}}]{PhysRevLett.120.057202}%
  \BibitemOpen
  \bibfield  {author} {\bibinfo {author} {\bibfnamefont {Y.-P.}\ \bibnamefont
  {Wang}}, \bibinfo {author} {\bibfnamefont {G.-Q.}\ \bibnamefont {Zhang}},
  \bibinfo {author} {\bibfnamefont {D.}~\bibnamefont {Zhang}}, \bibinfo
  {author} {\bibfnamefont {T.-F.}\ \bibnamefont {Li}}, \bibinfo {author}
  {\bibfnamefont {C.-M.}\ \bibnamefont {Hu}},\ and\ \bibinfo {author}
  {\bibfnamefont {J.~Q.}\ \bibnamefont {You}},\ }\bibfield  {title} {\bibinfo
  {title} {Bistability of cavity magnon polaritons},\ }\href
  {https://doi.org/10.1103/PhysRevLett.120.057202} {\bibfield  {journal}
  {\bibinfo  {journal} {Phys. Rev. Lett.}\ }\textbf {\bibinfo {volume} {120}},\
  \bibinfo {pages} {057202} (\bibinfo {year} {2018})}\BibitemShut {NoStop}%
\bibitem [{\citenamefont {Wang}\ \emph {et~al.}(2019)\citenamefont {Wang},
  \citenamefont {Rao}, \citenamefont {Yang}, \citenamefont {Xu}, \citenamefont
  {Gui}, \citenamefont {Yao}, \citenamefont {You},\ and\ \citenamefont
  {Hu}}]{PhysRevLett.123.127202}%
  \BibitemOpen
  \bibfield  {author} {\bibinfo {author} {\bibfnamefont {Y.-P.}\ \bibnamefont
  {Wang}}, \bibinfo {author} {\bibfnamefont {J.~W.}\ \bibnamefont {Rao}},
  \bibinfo {author} {\bibfnamefont {Y.}~\bibnamefont {Yang}}, \bibinfo {author}
  {\bibfnamefont {P.-C.}\ \bibnamefont {Xu}}, \bibinfo {author} {\bibfnamefont
  {Y.~S.}\ \bibnamefont {Gui}}, \bibinfo {author} {\bibfnamefont {B.~M.}\
  \bibnamefont {Yao}}, \bibinfo {author} {\bibfnamefont {J.~Q.}\ \bibnamefont
  {You}},\ and\ \bibinfo {author} {\bibfnamefont {C.-M.}\ \bibnamefont {Hu}},\
  }\bibfield  {title} {\bibinfo {title} {Nonreciprocity and unidirectional
  invisibility in cavity magnonics},\ }\href
  {https://doi.org/10.1103/PhysRevLett.123.127202} {\bibfield  {journal}
  {\bibinfo  {journal} {Phys. Rev. Lett.}\ }\textbf {\bibinfo {volume} {123}},\
  \bibinfo {pages} {127202} (\bibinfo {year} {2019})}\BibitemShut {NoStop}%
\bibitem [{\citenamefont {Lachance-Quirion}\ \emph {et~al.}(2020)\citenamefont
  {Lachance-Quirion}, \citenamefont {Wolski}, \citenamefont {Tabuchi},
  \citenamefont {Kono}, \citenamefont {Usami},\ and\ \citenamefont
  {Nakamura}}]{doi:10.1126/science.aaz9236}%
  \BibitemOpen
  \bibfield  {author} {\bibinfo {author} {\bibfnamefont {D.}~\bibnamefont
  {Lachance-Quirion}}, \bibinfo {author} {\bibfnamefont {S.~P.}\ \bibnamefont
  {Wolski}}, \bibinfo {author} {\bibfnamefont {Y.}~\bibnamefont {Tabuchi}},
  \bibinfo {author} {\bibfnamefont {S.}~\bibnamefont {Kono}}, \bibinfo {author}
  {\bibfnamefont {K.}~\bibnamefont {Usami}},\ and\ \bibinfo {author}
  {\bibfnamefont {Y.}~\bibnamefont {Nakamura}},\ }\bibfield  {title} {\bibinfo
  {title} {Entanglement-based single-shot detection of a single magnon with a
  superconducting qubit},\ }\href {https://doi.org/10.1126/science.aaz9236}
  {\bibfield  {journal} {\bibinfo  {journal} {Science}\ }\textbf {\bibinfo
  {volume} {367}},\ \bibinfo {pages} {425} (\bibinfo {year}
  {2020})}\BibitemShut {NoStop}%
\bibitem [{\citenamefont {Li}\ \emph {et~al.}(2020{\natexlab{b}})\citenamefont
  {Li}, \citenamefont {Zhang}, \citenamefont {Tyberkevych}, \citenamefont
  {Kwok}, \citenamefont {Hoffmann},\ and\ \citenamefont
  {Novosad}}]{doi:10.1063/5.0020277}%
  \BibitemOpen
  \bibfield  {author} {\bibinfo {author} {\bibfnamefont {Y.}~\bibnamefont
  {Li}}, \bibinfo {author} {\bibfnamefont {W.}~\bibnamefont {Zhang}}, \bibinfo
  {author} {\bibfnamefont {V.}~\bibnamefont {Tyberkevych}}, \bibinfo {author}
  {\bibfnamefont {W.-K.}\ \bibnamefont {Kwok}}, \bibinfo {author}
  {\bibfnamefont {A.}~\bibnamefont {Hoffmann}},\ and\ \bibinfo {author}
  {\bibfnamefont {V.}~\bibnamefont {Novosad}},\ }\bibfield  {title} {\bibinfo
  {title} {Hybrid magnonics: Physics, circuits, and applications for coherent
  information processing},\ }\href {https://doi.org/10.1063/5.0020277}
  {\bibfield  {journal} {\bibinfo  {journal} {J. Applied Phys.}\ }\textbf
  {\bibinfo {volume} {128}},\ \bibinfo {pages} {130902} (\bibinfo {year}
  {2020}{\natexlab{b}})}\BibitemShut {NoStop}%
\bibitem [{\citenamefont {Andrich}\ \emph {et~al.}(2017)\citenamefont
  {Andrich}, \citenamefont {las Casas}, \citenamefont {Liu}, \citenamefont
  {Bretscher}, \citenamefont {Berman}, \citenamefont {Heremans}, \citenamefont
  {Nealey},\ and\ \citenamefont {Awschalom}}]{Andrich2017}%
  \BibitemOpen
  \bibfield  {author} {\bibinfo {author} {\bibfnamefont {P.}~\bibnamefont
  {Andrich}}, \bibinfo {author} {\bibfnamefont {C.~F.}\ \bibnamefont {las
  Casas}}, \bibinfo {author} {\bibfnamefont {X.}~\bibnamefont {Liu}}, \bibinfo
  {author} {\bibfnamefont {H.~L.}\ \bibnamefont {Bretscher}}, \bibinfo {author}
  {\bibfnamefont {J.~R.}\ \bibnamefont {Berman}}, \bibinfo {author}
  {\bibfnamefont {F.~J.}\ \bibnamefont {Heremans}}, \bibinfo {author}
  {\bibfnamefont {P.~F.}\ \bibnamefont {Nealey}},\ and\ \bibinfo {author}
  {\bibfnamefont {D.~D.}\ \bibnamefont {Awschalom}},\ }\bibfield  {title}
  {\bibinfo {title} {Long-range spin wave mediated control of defect qubits in
  nanodiamonds},\ }\href {https://doi.org/10.1038/s41534-017-0029-z} {\bibfield
   {journal} {\bibinfo  {journal} {npj Quantum Inf.}\ }\textbf {\bibinfo
  {volume} {3}},\ \bibinfo {pages} {28} (\bibinfo {year} {2017})}\BibitemShut
  {NoStop}%
\bibitem [{\citenamefont {Huillery}\ \emph {et~al.}(2020)\citenamefont
  {Huillery}, \citenamefont {Delord}, \citenamefont {Nicolas}, \citenamefont
  {Van Den~Bossche}, \citenamefont {Perdriat},\ and\ \citenamefont
  {Hetet}}]{huillery2020spin}%
  \BibitemOpen
  \bibfield  {author} {\bibinfo {author} {\bibfnamefont {P.}~\bibnamefont
  {Huillery}}, \bibinfo {author} {\bibfnamefont {T.}~\bibnamefont {Delord}},
  \bibinfo {author} {\bibfnamefont {L.}~\bibnamefont {Nicolas}}, \bibinfo
  {author} {\bibfnamefont {M.}~\bibnamefont {Van Den~Bossche}}, \bibinfo
  {author} {\bibfnamefont {M.}~\bibnamefont {Perdriat}},\ and\ \bibinfo
  {author} {\bibfnamefont {G.}~\bibnamefont {Hetet}},\ }\bibfield  {title}
  {\bibinfo {title} {Spin mechanics with levitating ferromagnetic particles},\
  }\href {https://doi.org/0} {\bibfield  {journal} {\bibinfo  {journal} {Phys.
  Rev. B}\ }\textbf {\bibinfo {volume} {101}},\ \bibinfo {pages} {134415}
  (\bibinfo {year} {2020})}\BibitemShut {NoStop}%
\bibitem [{\citenamefont {Zhang}\ \emph {et~al.}(2015)\citenamefont {Zhang},
  \citenamefont {Wang}, \citenamefont {Li}, \citenamefont {Luo}, \citenamefont
  {Wu}, \citenamefont {Nori},\ and\ \citenamefont {You}}]{Zhang2015}%
  \BibitemOpen
  \bibfield  {author} {\bibinfo {author} {\bibfnamefont {D.}~\bibnamefont
  {Zhang}}, \bibinfo {author} {\bibfnamefont {X.-M.}\ \bibnamefont {Wang}},
  \bibinfo {author} {\bibfnamefont {T.-F.}\ \bibnamefont {Li}}, \bibinfo
  {author} {\bibfnamefont {X.-Q.}\ \bibnamefont {Luo}}, \bibinfo {author}
  {\bibfnamefont {W.}~\bibnamefont {Wu}}, \bibinfo {author} {\bibfnamefont
  {F.}~\bibnamefont {Nori}},\ and\ \bibinfo {author} {\bibfnamefont {J.~Q.}\
  \bibnamefont {You}},\ }\bibfield  {title} {\bibinfo {title} {Cavity quantum
  electrodynamics with ferromagnetic magnons in a small yttrium-iron-garnet
  sphere},\ }\href {https://doi.org/10.1038/npjqi.2015.14} {\bibfield
  {journal} {\bibinfo  {journal} {npj Quantum Inf.}\ }\textbf {\bibinfo
  {volume} {1}},\ \bibinfo {pages} {15014} (\bibinfo {year}
  {2015})}\BibitemShut {NoStop}%
\bibitem [{\citenamefont {Hou}\ and\ \citenamefont
  {Liu}(2019)}]{PhysRevLett.123.107702}%
  \BibitemOpen
  \bibfield  {author} {\bibinfo {author} {\bibfnamefont {J.~T.}\ \bibnamefont
  {Hou}}\ and\ \bibinfo {author} {\bibfnamefont {L.}~\bibnamefont {Liu}},\
  }\bibfield  {title} {\bibinfo {title} {Strong coupling between microwave
  photons and nanomagnet magnons},\ }\href
  {https://doi.org/10.1103/PhysRevLett.123.107702} {\bibfield  {journal}
  {\bibinfo  {journal} {Phys. Rev. Lett.}\ }\textbf {\bibinfo {volume} {123}},\
  \bibinfo {pages} {107702} (\bibinfo {year} {2019})}\BibitemShut {NoStop}%
\bibitem [{\citenamefont {Li}\ \emph {et~al.}(2019{\natexlab{b}})\citenamefont
  {Li}, \citenamefont {Polakovic}, \citenamefont {Wang}, \citenamefont {Xu},
  \citenamefont {Lendinez}, \citenamefont {Zhang}, \citenamefont {Ding},
  \citenamefont {Khaire}, \citenamefont {Saglam}, \citenamefont {Divan},
  \citenamefont {Pearson}, \citenamefont {Kwok}, \citenamefont {Xiao},
  \citenamefont {Novosad}, \citenamefont {Hoffmann},\ and\ \citenamefont
  {Zhang}}]{PhysRevLett.123.107701}%
  \BibitemOpen
  \bibfield  {author} {\bibinfo {author} {\bibfnamefont {Y.}~\bibnamefont
  {Li}}, \bibinfo {author} {\bibfnamefont {T.}~\bibnamefont {Polakovic}},
  \bibinfo {author} {\bibfnamefont {Y.-L.}\ \bibnamefont {Wang}}, \bibinfo
  {author} {\bibfnamefont {J.}~\bibnamefont {Xu}}, \bibinfo {author}
  {\bibfnamefont {S.}~\bibnamefont {Lendinez}}, \bibinfo {author}
  {\bibfnamefont {Z.}~\bibnamefont {Zhang}}, \bibinfo {author} {\bibfnamefont
  {J.}~\bibnamefont {Ding}}, \bibinfo {author} {\bibfnamefont {T.}~\bibnamefont
  {Khaire}}, \bibinfo {author} {\bibfnamefont {H.}~\bibnamefont {Saglam}},
  \bibinfo {author} {\bibfnamefont {R.}~\bibnamefont {Divan}}, \bibinfo
  {author} {\bibfnamefont {J.}~\bibnamefont {Pearson}}, \bibinfo {author}
  {\bibfnamefont {W.-K.}\ \bibnamefont {Kwok}}, \bibinfo {author}
  {\bibfnamefont {Z.}~\bibnamefont {Xiao}}, \bibinfo {author} {\bibfnamefont
  {V.}~\bibnamefont {Novosad}}, \bibinfo {author} {\bibfnamefont
  {A.}~\bibnamefont {Hoffmann}},\ and\ \bibinfo {author} {\bibfnamefont
  {W.}~\bibnamefont {Zhang}},\ }\bibfield  {title} {\bibinfo {title} {Strong
  coupling between magnons and microwave photons in on-chip
  ferromagnet-superconductor thin-film devices},\ }\href
  {https://doi.org/10.1103/PhysRevLett.123.107701} {\bibfield  {journal}
  {\bibinfo  {journal} {Phys. Rev. Lett.}\ }\textbf {\bibinfo {volume} {123}},\
  \bibinfo {pages} {107701} (\bibinfo {year} {2019}{\natexlab{b}})}\BibitemShut
  {NoStop}%
\bibitem [{\citenamefont {Gonzalez-Ballestero}\ \emph
  {et~al.}(2020{\natexlab{a}})\citenamefont {Gonzalez-Ballestero},
  \citenamefont {H\"ummer}, \citenamefont {Gieseler},\ and\ \citenamefont
  {Romero-Isart}}]{PhysRevB.101.125404}%
  \BibitemOpen
  \bibfield  {author} {\bibinfo {author} {\bibfnamefont {C.}~\bibnamefont
  {Gonzalez-Ballestero}}, \bibinfo {author} {\bibfnamefont {D.}~\bibnamefont
  {H\"ummer}}, \bibinfo {author} {\bibfnamefont {J.}~\bibnamefont {Gieseler}},\
  and\ \bibinfo {author} {\bibfnamefont {O.}~\bibnamefont {Romero-Isart}},\
  }\bibfield  {title} {\bibinfo {title} {Theory of quantum acoustomagnonics and
  acoustomechanics with a micromagnet},\ }\href
  {https://doi.org/10.1103/PhysRevB.101.125404} {\bibfield  {journal} {\bibinfo
   {journal} {Phys. Rev. B}\ }\textbf {\bibinfo {volume} {101}},\ \bibinfo
  {pages} {125404} (\bibinfo {year} {2020}{\natexlab{a}})}\BibitemShut
  {NoStop}%
\bibitem [{\citenamefont {Gonzalez-Ballestero}\ \emph
  {et~al.}(2020{\natexlab{b}})\citenamefont {Gonzalez-Ballestero},
  \citenamefont {Gieseler},\ and\ \citenamefont
  {Romero-Isart}}]{PhysRevLett.124.093602}%
  \BibitemOpen
  \bibfield  {author} {\bibinfo {author} {\bibfnamefont {C.}~\bibnamefont
  {Gonzalez-Ballestero}}, \bibinfo {author} {\bibfnamefont {J.}~\bibnamefont
  {Gieseler}},\ and\ \bibinfo {author} {\bibfnamefont {O.}~\bibnamefont
  {Romero-Isart}},\ }\bibfield  {title} {\bibinfo {title} {Quantum
  acoustomechanics with a micromagnet},\ }\href
  {https://doi.org/10.1103/PhysRevLett.124.093602} {\bibfield  {journal}
  {\bibinfo  {journal} {Phys. Rev. Lett.}\ }\textbf {\bibinfo {volume} {124}},\
  \bibinfo {pages} {093602} (\bibinfo {year} {2020}{\natexlab{b}})}\BibitemShut
  {NoStop}%
\bibitem [{\citenamefont {Li}\ \emph {et~al.}(2018)\citenamefont {Li},
  \citenamefont {Zhu},\ and\ \citenamefont {Agarwal}}]{2018LiP203601203601}%
  \BibitemOpen
  \bibfield  {author} {\bibinfo {author} {\bibfnamefont {J.}~\bibnamefont
  {Li}}, \bibinfo {author} {\bibfnamefont {S.-Y.}\ \bibnamefont {Zhu}},\ and\
  \bibinfo {author} {\bibfnamefont {G.~S.}\ \bibnamefont {Agarwal}},\
  }\bibfield  {title} {\bibinfo {title} {Magnon-photon-phonon entanglement in
  cavity magnomechanics},\ }\href
  {https://doi.org/10.1103/PhysRevLett.121.203601} {\bibfield  {journal}
  {\bibinfo  {journal} {Phys. Rev. Lett.}\ }\textbf {\bibinfo {volume} {121}},\
  \bibinfo {pages} {203601} (\bibinfo {year} {2018})}\BibitemShut {NoStop}%
\bibitem [{\citenamefont {Zhang}\ \emph {et~al.}(2016)\citenamefont {Zhang},
  \citenamefont {Zou}, \citenamefont {Jiang},\ and\ \citenamefont
  {Tang}}]{Zhange1501286}%
  \BibitemOpen
  \bibfield  {author} {\bibinfo {author} {\bibfnamefont {X.}~\bibnamefont
  {Zhang}}, \bibinfo {author} {\bibfnamefont {C.-L.}\ \bibnamefont {Zou}},
  \bibinfo {author} {\bibfnamefont {L.}~\bibnamefont {Jiang}},\ and\ \bibinfo
  {author} {\bibfnamefont {H.~X.}\ \bibnamefont {Tang}},\ }\bibfield  {title}
  {\bibinfo {title} {Cavity magnomechanics},\ }\href
  {https://doi.org/10.1126/sciadv.1501286} {\bibfield  {journal} {\bibinfo
  {journal} {Sci. Adv.}\ }\textbf {\bibinfo {volume} {2}},\ \bibinfo {pages}
  {e1501286} (\bibinfo {year} {2016})}\BibitemShut {NoStop}%
\bibitem [{\citenamefont {Riedinger}\ \emph {et~al.}(2016)\citenamefont
  {Riedinger}, \citenamefont {Hong}, \citenamefont {Norte}, \citenamefont
  {Slater}, \citenamefont {Shang}, \citenamefont {Krause}, \citenamefont
  {Anant}, \citenamefont {Aspelmeyer},\ and\ \citenamefont
  {Groblacher}}]{RN750}%
  \BibitemOpen
  \bibfield  {author} {\bibinfo {author} {\bibfnamefont {R.}~\bibnamefont
  {Riedinger}}, \bibinfo {author} {\bibfnamefont {S.}~\bibnamefont {Hong}},
  \bibinfo {author} {\bibfnamefont {R.~A.}\ \bibnamefont {Norte}}, \bibinfo
  {author} {\bibfnamefont {J.~A.}\ \bibnamefont {Slater}}, \bibinfo {author}
  {\bibfnamefont {J.}~\bibnamefont {Shang}}, \bibinfo {author} {\bibfnamefont
  {A.~G.}\ \bibnamefont {Krause}}, \bibinfo {author} {\bibfnamefont
  {V.}~\bibnamefont {Anant}}, \bibinfo {author} {\bibfnamefont
  {M.}~\bibnamefont {Aspelmeyer}},\ and\ \bibinfo {author} {\bibfnamefont
  {S.}~\bibnamefont {Groblacher}},\ }\bibfield  {title} {\bibinfo {title}
  {Non-classical correlations between single photons and phonons from a
  mechanical oscillator},\ }\href {https://doi.org/10.1038/nature16536}
  {\bibfield  {journal} {\bibinfo  {journal} {Nature}\ }\textbf {\bibinfo
  {volume} {530}},\ \bibinfo {pages} {313} (\bibinfo {year}
  {2016})}\BibitemShut {NoStop}%
\bibitem [{\citenamefont {Kim}\ \emph {et~al.}(2016)\citenamefont {Kim},
  \citenamefont {Kwon}, \citenamefont {Kim}, \citenamefont {Do}, \citenamefont
  {Lee},\ and\ \citenamefont {Han}}]{RN276}%
  \BibitemOpen
  \bibfield  {author} {\bibinfo {author} {\bibfnamefont {J.}~\bibnamefont
  {Kim}}, \bibinfo {author} {\bibfnamefont {J.}~\bibnamefont {Kwon}}, \bibinfo
  {author} {\bibfnamefont {M.}~\bibnamefont {Kim}}, \bibinfo {author}
  {\bibfnamefont {J.}~\bibnamefont {Do}}, \bibinfo {author} {\bibfnamefont
  {D.}~\bibnamefont {Lee}},\ and\ \bibinfo {author} {\bibfnamefont
  {H.}~\bibnamefont {Han}},\ }\bibfield  {title} {\bibinfo {title}
  {Low-dielectric-constant polyimide aerogel composite films with low water
  uptake},\ }\href {https://doi.org/10.1038/pj.2016.37} {\bibfield  {journal}
  {\bibinfo  {journal} {Nat. Commun.}\ }\textbf {\bibinfo {volume} {48}},\
  \bibinfo {pages} {829} (\bibinfo {year} {2016})}\BibitemShut {NoStop}%
\bibitem [{\citenamefont {Kovalev}\ \emph {et~al.}(2014)\citenamefont
  {Kovalev}, \citenamefont {De},\ and\ \citenamefont {Shtengel}}]{RN415}%
  \BibitemOpen
  \bibfield  {author} {\bibinfo {author} {\bibfnamefont {A.~A.}\ \bibnamefont
  {Kovalev}}, \bibinfo {author} {\bibfnamefont {A.}~\bibnamefont {De}},\ and\
  \bibinfo {author} {\bibfnamefont {K.}~\bibnamefont {Shtengel}},\ }\bibfield
  {title} {\bibinfo {title} {Spin transfer of quantum information between
  majorana modes and a resonator},\ }\href
  {https://doi.org/10.1103/PhysRevLett.112.106402} {\bibfield  {journal}
  {\bibinfo  {journal} {Phys. Rev. Lett.}\ }\textbf {\bibinfo {volume} {112}},\
  \bibinfo {pages} {106402} (\bibinfo {year} {2014})}\BibitemShut {NoStop}%
\bibitem [{\citenamefont {Cha}\ \emph {et~al.}(2021)\citenamefont {Cha},
  \citenamefont {Kim}, \citenamefont {Kim}, \citenamefont {Shim},\ and\
  \citenamefont {Suh}}]{cha2021superconducting}%
  \BibitemOpen
  \bibfield  {author} {\bibinfo {author} {\bibfnamefont {J.}~\bibnamefont
  {Cha}}, \bibinfo {author} {\bibfnamefont {H.}~\bibnamefont {Kim}}, \bibinfo
  {author} {\bibfnamefont {J.}~\bibnamefont {Kim}}, \bibinfo {author}
  {\bibfnamefont {S.-B.}\ \bibnamefont {Shim}},\ and\ \bibinfo {author}
  {\bibfnamefont {J.}~\bibnamefont {Suh}},\ }\bibfield  {title} {\bibinfo
  {title} {Superconducting nanoelectromechanical transducer resilient to
  magnetic fields},\ }\href
  {https://pubs.acs.org/doi/abs/10.1021/acs.nanolett.0c04845} {\bibfield
  {journal} {\bibinfo  {journal} {Nano Letters}\ }\textbf {\bibinfo {volume}
  {21}},\ \bibinfo {pages} {1800} (\bibinfo {year} {2021})}\BibitemShut
  {NoStop}%
\bibitem [{\citenamefont {Qin}\ \emph {et~al.}(2019)\citenamefont {Qin},
  \citenamefont {Miranowicz}, \citenamefont {Long}, \citenamefont {You},\ and\
  \citenamefont {Nori}}]{RN302}%
  \BibitemOpen
  \bibfield  {author} {\bibinfo {author} {\bibfnamefont {W.}~\bibnamefont
  {Qin}}, \bibinfo {author} {\bibfnamefont {A.}~\bibnamefont {Miranowicz}},
  \bibinfo {author} {\bibfnamefont {G.}~\bibnamefont {Long}}, \bibinfo {author}
  {\bibfnamefont {J.~Q.}\ \bibnamefont {You}},\ and\ \bibinfo {author}
  {\bibfnamefont {F.}~\bibnamefont {Nori}},\ }\bibfield  {title} {\bibinfo
  {title} {Proposal to test quantum wave-particle superposition on massive
  mechanical resonators},\ }\href {https://doi.org/10.1038/s41534-019-0172-9}
  {\bibfield  {journal} {\bibinfo  {journal} {npj Quantum Inf.}\ }\textbf
  {\bibinfo {volume} {5}},\ \bibinfo {pages} {58} (\bibinfo {year}
  {2019})}\BibitemShut {NoStop}%
\bibitem [{\citenamefont {Rao}\ \emph {et~al.}(2016)\citenamefont {Rao},
  \citenamefont {Momenzadeh},\ and\ \citenamefont {Wrachtrup}}]{RN679}%
  \BibitemOpen
  \bibfield  {author} {\bibinfo {author} {\bibfnamefont {D.~D.}\ \bibnamefont
  {Rao}}, \bibinfo {author} {\bibfnamefont {S.~A.}\ \bibnamefont
  {Momenzadeh}},\ and\ \bibinfo {author} {\bibfnamefont {J.}~\bibnamefont
  {Wrachtrup}},\ }\bibfield  {title} {\bibinfo {title} {Heralded control of
  mechanical motion by single spins},\ }\href
  {https://doi.org/10.1103/PhysRevLett.117.077203} {\bibfield  {journal}
  {\bibinfo  {journal} {Phys. Rev. Lett.}\ }\textbf {\bibinfo {volume} {117}},\
  \bibinfo {pages} {077203} (\bibinfo {year} {2016})}\BibitemShut {NoStop}%
\bibitem [{\citenamefont {Shandilya}\ \emph
  {et~al.}(2021{\natexlab{b}})\citenamefont {Shandilya}, \citenamefont {Lake},
  \citenamefont {Mitchell}, \citenamefont {Sukachev},\ and\ \citenamefont
  {Barclay}}]{RN542}%
  \BibitemOpen
  \bibfield  {author} {\bibinfo {author} {\bibfnamefont {P.~K.}\ \bibnamefont
  {Shandilya}}, \bibinfo {author} {\bibfnamefont {D.~P.}\ \bibnamefont {Lake}},
  \bibinfo {author} {\bibfnamefont {M.~J.}\ \bibnamefont {Mitchell}}, \bibinfo
  {author} {\bibfnamefont {D.~D.}\ \bibnamefont {Sukachev}},\ and\ \bibinfo
  {author} {\bibfnamefont {P.~E.}\ \bibnamefont {Barclay}},\ }\bibfield
  {title} {\bibinfo {title} {Optomechanical interface between telecom photons
  and spin quantum memory},\ }\href
  {https://doi.org/10.1038/s41567-021-01364-3} {\bibfield  {journal} {\bibinfo
  {journal} {Nat. Phys.}\ }\textbf {\bibinfo {volume} {17}},\ \bibinfo {pages}
  {1420} (\bibinfo {year} {2021}{\natexlab{b}})}\BibitemShut {NoStop}%
\bibitem [{\citenamefont {Wang}\ \emph {et~al.}(2022)\citenamefont {Wang},
  \citenamefont {Wu}, \citenamefont {Han}, \citenamefont {Xia}, \citenamefont
  {Jiang},\ and\ \citenamefont {Song}}]{RN678}%
  \BibitemOpen
  \bibfield  {author} {\bibinfo {author} {\bibfnamefont {Y.}~\bibnamefont
  {Wang}}, \bibinfo {author} {\bibfnamefont {J.-L.}\ \bibnamefont {Wu}},
  \bibinfo {author} {\bibfnamefont {J.-X.}\ \bibnamefont {Han}}, \bibinfo
  {author} {\bibfnamefont {Y.}~\bibnamefont {Xia}}, \bibinfo {author}
  {\bibfnamefont {Y.-Y.}\ \bibnamefont {Jiang}},\ and\ \bibinfo {author}
  {\bibfnamefont {J.}~\bibnamefont {Song}},\ }\bibfield  {title} {\bibinfo
  {title} {Enhanced phonon blockade in a weakly coupled hybrid system via
  mechanical parametric amplification},\ }\href
  {https://doi.org/10.1103/PhysRevApplied.17.024009} {\bibfield  {journal}
  {\bibinfo  {journal} {Phys. Rev. Applied}\ }\textbf {\bibinfo {volume}
  {17}},\ \bibinfo {pages} {024009} (\bibinfo {year} {2022})}\BibitemShut
  {NoStop}%
\bibitem [{\citenamefont {Ovartchaiyapong}\ \emph {et~al.}(2014)\citenamefont
  {Ovartchaiyapong}, \citenamefont {Lee}, \citenamefont {Myers},\ and\
  \citenamefont {Jayich}}]{Ovartchaiyapong2014}%
  \BibitemOpen
  \bibfield  {author} {\bibinfo {author} {\bibfnamefont {P.}~\bibnamefont
  {Ovartchaiyapong}}, \bibinfo {author} {\bibfnamefont {K.~W.}\ \bibnamefont
  {Lee}}, \bibinfo {author} {\bibfnamefont {B.~A.}\ \bibnamefont {Myers}},\
  and\ \bibinfo {author} {\bibfnamefont {A.~C.~B.}\ \bibnamefont {Jayich}},\
  }\bibfield  {title} {\bibinfo {title} {Dynamic strain-mediated coupling of a
  single diamond spin to a mechanical resonator},\ }\href
  {https://doi.org/10.1038/ncomms5429} {\bibfield  {journal} {\bibinfo
  {journal} {Nat. Commun.}\ }\textbf {\bibinfo {volume} {5}},\ \bibinfo {pages}
  {4429} (\bibinfo {year} {2014})}\BibitemShut {NoStop}%
\bibitem [{\citenamefont {S\'anchez Mu\~noz}\ \emph {et~al.}(2018)\citenamefont
  {S\'anchez Mu\~noz}, \citenamefont {Lara}, \citenamefont {Puebla},\ and\
  \citenamefont {Nori}}]{PhysRevLett.121.123604}%
  \BibitemOpen
  \bibfield  {author} {\bibinfo {author} {\bibfnamefont {C.}~\bibnamefont
  {S\'anchez Mu\~noz}}, \bibinfo {author} {\bibfnamefont {A.}~\bibnamefont
  {Lara}}, \bibinfo {author} {\bibfnamefont {J.}~\bibnamefont {Puebla}},\ and\
  \bibinfo {author} {\bibfnamefont {F.}~\bibnamefont {Nori}},\ }\bibfield
  {title} {\bibinfo {title} {Hybrid systems for the generation of nonclassical
  mechanical states via quadratic interactions},\ }\href
  {https://doi.org/10.1103/PhysRevLett.121.123604} {\bibfield  {journal}
  {\bibinfo  {journal} {Phys. Rev. Lett.}\ }\textbf {\bibinfo {volume} {121}},\
  \bibinfo {pages} {123604} (\bibinfo {year} {2018})}\BibitemShut {NoStop}%
\bibitem [{\citenamefont {Streltsov}\ \emph {et~al.}(2021)\citenamefont
  {Streltsov}, \citenamefont {Pedernales},\ and\ \citenamefont
  {Plenio}}]{PhysRevLett.126.193602}%
  \BibitemOpen
  \bibfield  {author} {\bibinfo {author} {\bibfnamefont {K.}~\bibnamefont
  {Streltsov}}, \bibinfo {author} {\bibfnamefont {J.~S.}\ \bibnamefont
  {Pedernales}},\ and\ \bibinfo {author} {\bibfnamefont {M.~B.}\ \bibnamefont
  {Plenio}},\ }\bibfield  {title} {\bibinfo {title} {Ground-state cooling of
  levitated magnets in low-frequency traps},\ }\href
  {https://doi.org/10.1103/PhysRevLett.126.193602} {\bibfield  {journal}
  {\bibinfo  {journal} {Phys. Rev. Lett.}\ }\textbf {\bibinfo {volume} {126}},\
  \bibinfo {pages} {193602} (\bibinfo {year} {2021})}\BibitemShut {NoStop}%
\bibitem [{\citenamefont {Teissier}\ \emph {et~al.}(2014)\citenamefont
  {Teissier}, \citenamefont {Barfuss}, \citenamefont {Appel}, \citenamefont
  {Neu},\ and\ \citenamefont {Maletinsky}}]{PhysRevLett.113.020503}%
  \BibitemOpen
  \bibfield  {author} {\bibinfo {author} {\bibfnamefont {J.}~\bibnamefont
  {Teissier}}, \bibinfo {author} {\bibfnamefont {A.}~\bibnamefont {Barfuss}},
  \bibinfo {author} {\bibfnamefont {P.}~\bibnamefont {Appel}}, \bibinfo
  {author} {\bibfnamefont {E.}~\bibnamefont {Neu}},\ and\ \bibinfo {author}
  {\bibfnamefont {P.}~\bibnamefont {Maletinsky}},\ }\bibfield  {title}
  {\bibinfo {title} {Strain coupling of a nitrogen-vacancy center spin to a
  diamond mechanical oscillator},\ }\href
  {https://doi.org/10.1103/PhysRevLett.113.020503} {\bibfield  {journal}
  {\bibinfo  {journal} {Phys. Rev. Lett.}\ }\textbf {\bibinfo {volume} {113}},\
  \bibinfo {pages} {020503} (\bibinfo {year} {2014})}\BibitemShut {NoStop}%
\bibitem [{\citenamefont {Arcizet}\ \emph {et~al.}(2011)\citenamefont
  {Arcizet}, \citenamefont {Jacques}, \citenamefont {Siria}, \citenamefont
  {Poncharal}, \citenamefont {Vincent},\ and\ \citenamefont
  {Seidelin}}]{Arcizet2011}%
  \BibitemOpen
  \bibfield  {author} {\bibinfo {author} {\bibfnamefont {O.}~\bibnamefont
  {Arcizet}}, \bibinfo {author} {\bibfnamefont {V.}~\bibnamefont {Jacques}},
  \bibinfo {author} {\bibfnamefont {A.}~\bibnamefont {Siria}}, \bibinfo
  {author} {\bibfnamefont {P.}~\bibnamefont {Poncharal}}, \bibinfo {author}
  {\bibfnamefont {P.}~\bibnamefont {Vincent}},\ and\ \bibinfo {author}
  {\bibfnamefont {S.}~\bibnamefont {Seidelin}},\ }\bibfield  {title} {\bibinfo
  {title} {A single nitrogen-vacancy defect coupled to a nanomechanical
  oscillator},\ }\href {https://doi.org/10.1038/nphys2070} {\bibfield
  {journal} {\bibinfo  {journal} {Nat. Phys.}\ }\textbf {\bibinfo {volume}
  {7}},\ \bibinfo {pages} {879} (\bibinfo {year} {2011})}\BibitemShut {NoStop}%
\bibitem [{\citenamefont {Gieseler}\ \emph {et~al.}(2020)\citenamefont
  {Gieseler}, \citenamefont {Kabcenell}, \citenamefont {Rosenfeld},
  \citenamefont {Schaefer}, \citenamefont {Safira}, \citenamefont {Schuetz},
  \citenamefont {Gonzalez-Ballestero}, \citenamefont {Rusconi}, \citenamefont
  {Romero-Isart},\ and\ \citenamefont {Lukin}}]{PhysRevLett.124.163604}%
  \BibitemOpen
  \bibfield  {author} {\bibinfo {author} {\bibfnamefont {J.}~\bibnamefont
  {Gieseler}}, \bibinfo {author} {\bibfnamefont {A.}~\bibnamefont {Kabcenell}},
  \bibinfo {author} {\bibfnamefont {E.}~\bibnamefont {Rosenfeld}}, \bibinfo
  {author} {\bibfnamefont {J.~D.}\ \bibnamefont {Schaefer}}, \bibinfo {author}
  {\bibfnamefont {A.}~\bibnamefont {Safira}}, \bibinfo {author} {\bibfnamefont
  {M.~J.~A.}\ \bibnamefont {Schuetz}}, \bibinfo {author} {\bibfnamefont
  {C.}~\bibnamefont {Gonzalez-Ballestero}}, \bibinfo {author} {\bibfnamefont
  {C.~C.}\ \bibnamefont {Rusconi}}, \bibinfo {author} {\bibfnamefont
  {O.}~\bibnamefont {Romero-Isart}},\ and\ \bibinfo {author} {\bibfnamefont
  {M.~D.}\ \bibnamefont {Lukin}},\ }\bibfield  {title} {\bibinfo {title}
  {Single-spin magnetomechanics with levitated micromagnets},\ }\href
  {https://doi.org/10.1103/PhysRevLett.124.163604} {\bibfield  {journal}
  {\bibinfo  {journal} {Phys. Rev. Lett.}\ }\textbf {\bibinfo {volume} {124}},\
  \bibinfo {pages} {163604} (\bibinfo {year} {2020})}\BibitemShut {NoStop}%
\bibitem [{\citenamefont {Rusconi}\ \emph
  {et~al.}(2022{\natexlab{b}})\citenamefont {Rusconi}, \citenamefont
  {Perdriat}, \citenamefont {H{\'e}tet}, \citenamefont {Romero-Isart},\ and\
  \citenamefont {Stickler}}]{rusconi2022spin}%
  \BibitemOpen
  \bibfield  {author} {\bibinfo {author} {\bibfnamefont {C.~C.}\ \bibnamefont
  {Rusconi}}, \bibinfo {author} {\bibfnamefont {M.}~\bibnamefont {Perdriat}},
  \bibinfo {author} {\bibfnamefont {G.}~\bibnamefont {H{\'e}tet}}, \bibinfo
  {author} {\bibfnamefont {O.}~\bibnamefont {Romero-Isart}},\ and\ \bibinfo
  {author} {\bibfnamefont {B.~A.}\ \bibnamefont {Stickler}},\ }\bibfield
  {title} {\bibinfo {title} {Spin-controlled quantum interference of levitated
  nanorotors},\ }\href
  {https://journals.aps.org/prl/abstract/10.1103/PhysRevLett.129.093605}
  {\bibfield  {journal} {\bibinfo  {journal} {Phys. Rev. Lett.}\ }\textbf
  {\bibinfo {volume} {129}},\ \bibinfo {pages} {093605} (\bibinfo {year}
  {2022}{\natexlab{b}})}\BibitemShut {NoStop}%
\bibitem [{\citenamefont {Lee}\ \emph {et~al.}(2017)\citenamefont {Lee},
  \citenamefont {Lee}, \citenamefont {Cady}, \citenamefont {Ovartchaiyapong},\
  and\ \citenamefont {Jayich}}]{Lee_2017}%
  \BibitemOpen
  \bibfield  {author} {\bibinfo {author} {\bibfnamefont {D.}~\bibnamefont
  {Lee}}, \bibinfo {author} {\bibfnamefont {K.~W.}\ \bibnamefont {Lee}},
  \bibinfo {author} {\bibfnamefont {J.~V.}\ \bibnamefont {Cady}}, \bibinfo
  {author} {\bibfnamefont {P.}~\bibnamefont {Ovartchaiyapong}},\ and\ \bibinfo
  {author} {\bibfnamefont {A.~C.~B.}\ \bibnamefont {Jayich}},\ }\bibfield
  {title} {\bibinfo {title} {Topical review: spins and mechanics in diamond},\
  }\href {https://doi.org/10.1088/2040-8986/aa52cd} {\bibfield  {journal}
  {\bibinfo  {journal} {J. Opt.}\ }\textbf {\bibinfo {volume} {19}},\ \bibinfo
  {pages} {033001} (\bibinfo {year} {2017})}\BibitemShut {NoStop}%
\bibitem [{\citenamefont {Kolkowitz}\ \emph {et~al.}(2012)\citenamefont
  {Kolkowitz}, \citenamefont {Jayich}, \citenamefont {Unterreithmeier},
  \citenamefont {Bennett}, \citenamefont {Rabl}, \citenamefont {Harris},\ and\
  \citenamefont {Lukin}}]{doi:10.1126/science.1216821}%
  \BibitemOpen
  \bibfield  {author} {\bibinfo {author} {\bibfnamefont {S.}~\bibnamefont
  {Kolkowitz}}, \bibinfo {author} {\bibfnamefont {A.~C.~B.}\ \bibnamefont
  {Jayich}}, \bibinfo {author} {\bibfnamefont {Q.~P.}\ \bibnamefont
  {Unterreithmeier}}, \bibinfo {author} {\bibfnamefont {S.~D.}\ \bibnamefont
  {Bennett}}, \bibinfo {author} {\bibfnamefont {P.}~\bibnamefont {Rabl}},
  \bibinfo {author} {\bibfnamefont {J.~G.~E.}\ \bibnamefont {Harris}},\ and\
  \bibinfo {author} {\bibfnamefont {M.~D.}\ \bibnamefont {Lukin}},\ }\bibfield
  {title} {\bibinfo {title} {Coherent sensing of a mechanical resonator with a
  single-spin qubit},\ }\href {https://doi.org/10.1126/science.1216821}
  {\bibfield  {journal} {\bibinfo  {journal} {Science}\ }\textbf {\bibinfo
  {volume} {335}},\ \bibinfo {pages} {1603} (\bibinfo {year}
  {2012})}\BibitemShut {NoStop}%
\bibitem [{\citenamefont {Whiteley}\ \emph {et~al.}(2019)\citenamefont
  {Whiteley}, \citenamefont {Wolfowicz}, \citenamefont {Anderson},
  \citenamefont {Bourassa}, \citenamefont {Ma}, \citenamefont {Ye},
  \citenamefont {Koolstra}, \citenamefont {Satzinger}, \citenamefont {Holt},
  \citenamefont {Heremans}, \citenamefont {Cleland}, \citenamefont {Schuster},
  \citenamefont {Galli},\ and\ \citenamefont {Awschalom}}]{Whiteley2019}%
  \BibitemOpen
  \bibfield  {author} {\bibinfo {author} {\bibfnamefont {S.~J.}\ \bibnamefont
  {Whiteley}}, \bibinfo {author} {\bibfnamefont {G.}~\bibnamefont {Wolfowicz}},
  \bibinfo {author} {\bibfnamefont {C.~P.}\ \bibnamefont {Anderson}}, \bibinfo
  {author} {\bibfnamefont {A.}~\bibnamefont {Bourassa}}, \bibinfo {author}
  {\bibfnamefont {H.}~\bibnamefont {Ma}}, \bibinfo {author} {\bibfnamefont
  {M.}~\bibnamefont {Ye}}, \bibinfo {author} {\bibfnamefont {G.}~\bibnamefont
  {Koolstra}}, \bibinfo {author} {\bibfnamefont {K.~J.}\ \bibnamefont
  {Satzinger}}, \bibinfo {author} {\bibfnamefont {M.~V.}\ \bibnamefont {Holt}},
  \bibinfo {author} {\bibfnamefont {F.~J.}\ \bibnamefont {Heremans}}, \bibinfo
  {author} {\bibfnamefont {A.~N.}\ \bibnamefont {Cleland}}, \bibinfo {author}
  {\bibfnamefont {D.~I.}\ \bibnamefont {Schuster}}, \bibinfo {author}
  {\bibfnamefont {G.}~\bibnamefont {Galli}},\ and\ \bibinfo {author}
  {\bibfnamefont {D.~D.}\ \bibnamefont {Awschalom}},\ }\bibfield  {title}
  {\bibinfo {title} {Spin–phonon interactions in silicon carbide addressed by
  {Gaussian} acoustics},\ }\href {https://doi.org/10.1038/s41567-019-0420-0}
  {\bibfield  {journal} {\bibinfo  {journal} {Nat. Phys.}\ }\textbf {\bibinfo
  {volume} {15}},\ \bibinfo {pages} {490} (\bibinfo {year} {2019})}\BibitemShut
  {NoStop}%
\bibitem [{\citenamefont {Kuzyk}\ and\ \citenamefont
  {Wang}(2018)}]{PhysRevX.8.041027}%
  \BibitemOpen
  \bibfield  {author} {\bibinfo {author} {\bibfnamefont {M.~C.}\ \bibnamefont
  {Kuzyk}}\ and\ \bibinfo {author} {\bibfnamefont {H.}~\bibnamefont {Wang}},\
  }\bibfield  {title} {\bibinfo {title} {Scaling phononic quantum networks of
  solid-state spins with closed mechanical subsystems},\ }\href
  {https://doi.org/10.1103/PhysRevX.8.041027} {\bibfield  {journal} {\bibinfo
  {journal} {Phys. Rev. X}\ }\textbf {\bibinfo {volume} {8}},\ \bibinfo {pages}
  {041027} (\bibinfo {year} {2018})}\BibitemShut {NoStop}%
\bibitem [{\citenamefont {MacQuarrie}\ \emph {et~al.}(2013)\citenamefont
  {MacQuarrie}, \citenamefont {Gosavi}, \citenamefont {Jungwirth},
  \citenamefont {Bhave},\ and\ \citenamefont {Fuchs}}]{PhysRevLett.111.227602}%
  \BibitemOpen
  \bibfield  {author} {\bibinfo {author} {\bibfnamefont {E.~R.}\ \bibnamefont
  {MacQuarrie}}, \bibinfo {author} {\bibfnamefont {T.~A.}\ \bibnamefont
  {Gosavi}}, \bibinfo {author} {\bibfnamefont {N.~R.}\ \bibnamefont
  {Jungwirth}}, \bibinfo {author} {\bibfnamefont {S.~A.}\ \bibnamefont
  {Bhave}},\ and\ \bibinfo {author} {\bibfnamefont {G.~D.}\ \bibnamefont
  {Fuchs}},\ }\bibfield  {title} {\bibinfo {title} {Mechanical spin control of
  nitrogen-vacancy centers in diamond},\ }\href
  {https://doi.org/10.1103/PhysRevLett.111.227602} {\bibfield  {journal}
  {\bibinfo  {journal} {Phys. Rev. Lett.}\ }\textbf {\bibinfo {volume} {111}},\
  \bibinfo {pages} {227602} (\bibinfo {year} {2013})}\BibitemShut {NoStop}%
\bibitem [{\citenamefont {Meesala}\ \emph {et~al.}(2016)\citenamefont
  {Meesala}, \citenamefont {Sohn}, \citenamefont {Atikian}, \citenamefont
  {Kim}, \citenamefont {Burek}, \citenamefont {Choy},\ and\ \citenamefont
  {Lon\ifmmode~\check{c}\else \v{c}\fi{}ar}}]{PhysRevApplied.5.034010}%
  \BibitemOpen
  \bibfield  {author} {\bibinfo {author} {\bibfnamefont {S.}~\bibnamefont
  {Meesala}}, \bibinfo {author} {\bibfnamefont {Y.-I.}\ \bibnamefont {Sohn}},
  \bibinfo {author} {\bibfnamefont {H.~A.}\ \bibnamefont {Atikian}}, \bibinfo
  {author} {\bibfnamefont {S.}~\bibnamefont {Kim}}, \bibinfo {author}
  {\bibfnamefont {M.~J.}\ \bibnamefont {Burek}}, \bibinfo {author}
  {\bibfnamefont {J.~T.}\ \bibnamefont {Choy}},\ and\ \bibinfo {author}
  {\bibfnamefont {M.}~\bibnamefont {Lon\ifmmode~\check{c}\else \v{c}\fi{}ar}},\
  }\bibfield  {title} {\bibinfo {title} {Enhanced strain coupling of
  nitrogen-vacancy spins to nanoscale diamond cantilevers},\ }\href
  {https://doi.org/10.1103/PhysRevApplied.5.034010} {\bibfield  {journal}
  {\bibinfo  {journal} {Phys. Rev. Applied}\ }\textbf {\bibinfo {volume} {5}},\
  \bibinfo {pages} {034010} (\bibinfo {year} {2016})}\BibitemShut {NoStop}%
\bibitem [{\citenamefont {Rabl}\ \emph {et~al.}(2009)\citenamefont {Rabl},
  \citenamefont {Cappellaro}, \citenamefont {Dutt}, \citenamefont {Jiang},
  \citenamefont {Maze},\ and\ \citenamefont {Lukin}}]{PhysRevB.79.041302}%
  \BibitemOpen
  \bibfield  {author} {\bibinfo {author} {\bibfnamefont {P.}~\bibnamefont
  {Rabl}}, \bibinfo {author} {\bibfnamefont {P.}~\bibnamefont {Cappellaro}},
  \bibinfo {author} {\bibfnamefont {M.~V.~G.}\ \bibnamefont {Dutt}}, \bibinfo
  {author} {\bibfnamefont {L.}~\bibnamefont {Jiang}}, \bibinfo {author}
  {\bibfnamefont {J.~R.}\ \bibnamefont {Maze}},\ and\ \bibinfo {author}
  {\bibfnamefont {M.~D.}\ \bibnamefont {Lukin}},\ }\bibfield  {title} {\bibinfo
  {title} {Strong magnetic coupling between an electronic spin qubit and a
  mechanical resonator},\ }\href {https://doi.org/10.1103/PhysRevB.79.041302}
  {\bibfield  {journal} {\bibinfo  {journal} {Phys. Rev. B}\ }\textbf {\bibinfo
  {volume} {79}},\ \bibinfo {pages} {041302} (\bibinfo {year}
  {2009})}\BibitemShut {NoStop}%
\bibitem [{\citenamefont {Li}\ \emph {et~al.}(2015)\citenamefont {Li},
  \citenamefont {Liu}, \citenamefont {Gao}, \citenamefont {Xiang},
  \citenamefont {Rabl}, \citenamefont {Xiao},\ and\ \citenamefont
  {Li}}]{PhysRevApplied.4.044003}%
  \BibitemOpen
  \bibfield  {author} {\bibinfo {author} {\bibfnamefont {P.-B.}\ \bibnamefont
  {Li}}, \bibinfo {author} {\bibfnamefont {Y.-C.}\ \bibnamefont {Liu}},
  \bibinfo {author} {\bibfnamefont {S.-Y.}\ \bibnamefont {Gao}}, \bibinfo
  {author} {\bibfnamefont {Z.-L.}\ \bibnamefont {Xiang}}, \bibinfo {author}
  {\bibfnamefont {P.}~\bibnamefont {Rabl}}, \bibinfo {author} {\bibfnamefont
  {Y.-F.}\ \bibnamefont {Xiao}},\ and\ \bibinfo {author} {\bibfnamefont
  {F.-L.}\ \bibnamefont {Li}},\ }\bibfield  {title} {\bibinfo {title} {Hybrid
  quantum device based on {NV} centers in diamond nanomechanical resonators
  plus superconducting waveguide cavities},\ }\href
  {https://doi.org/10.1103/PhysRevApplied.4.044003} {\bibfield  {journal}
  {\bibinfo  {journal} {Phys. Rev. Applied}\ }\textbf {\bibinfo {volume} {4}},\
  \bibinfo {pages} {044003} (\bibinfo {year} {2015})}\BibitemShut {NoStop}%
\bibitem [{\citenamefont {Hong}\ \emph {et~al.}(2012)\citenamefont {Hong},
  \citenamefont {Grinolds}, \citenamefont {Maletinsky}, \citenamefont
  {Walsworth}, \citenamefont {Lukin},\ and\ \citenamefont
  {Yacoby}}]{doi:10.1021/nl300775c}%
  \BibitemOpen
  \bibfield  {author} {\bibinfo {author} {\bibfnamefont {S.}~\bibnamefont
  {Hong}}, \bibinfo {author} {\bibfnamefont {M.~S.}\ \bibnamefont {Grinolds}},
  \bibinfo {author} {\bibfnamefont {P.}~\bibnamefont {Maletinsky}}, \bibinfo
  {author} {\bibfnamefont {R.~L.}\ \bibnamefont {Walsworth}}, \bibinfo {author}
  {\bibfnamefont {M.~D.}\ \bibnamefont {Lukin}},\ and\ \bibinfo {author}
  {\bibfnamefont {A.}~\bibnamefont {Yacoby}},\ }\bibfield  {title} {\bibinfo
  {title} {Coherent, mechanical control of a single electronic spin},\ }\href
  {https://doi.org/10.1021/nl300775c} {\bibfield  {journal} {\bibinfo
  {journal} {Nano Lett.}\ }\textbf {\bibinfo {volume} {12}},\ \bibinfo {pages}
  {3920} (\bibinfo {year} {2012})}\BibitemShut {NoStop}%
\bibitem [{\citenamefont {Barfuss}\ \emph {et~al.}(2015)\citenamefont
  {Barfuss}, \citenamefont {Teissier}, \citenamefont {Neu}, \citenamefont
  {Nunnenkamp},\ and\ \citenamefont {Maletinsky}}]{Barfuss2015}%
  \BibitemOpen
  \bibfield  {author} {\bibinfo {author} {\bibfnamefont {A.}~\bibnamefont
  {Barfuss}}, \bibinfo {author} {\bibfnamefont {J.}~\bibnamefont {Teissier}},
  \bibinfo {author} {\bibfnamefont {E.}~\bibnamefont {Neu}}, \bibinfo {author}
  {\bibfnamefont {A.}~\bibnamefont {Nunnenkamp}},\ and\ \bibinfo {author}
  {\bibfnamefont {P.}~\bibnamefont {Maletinsky}},\ }\bibfield  {title}
  {\bibinfo {title} {Strong mechanical driving of a single electron spin},\
  }\href {https://doi.org/10.1038/nphys3411} {\bibfield  {journal} {\bibinfo
  {journal} {Nat. Phys.}\ }\textbf {\bibinfo {volume} {11}},\ \bibinfo {pages}
  {820} (\bibinfo {year} {2015})}\BibitemShut {NoStop}%
\bibitem [{\citenamefont {Cai}\ \emph {et~al.}(2014)\citenamefont {Cai},
  \citenamefont {Jelezko},\ and\ \citenamefont {Plenio}}]{Cai2014}%
  \BibitemOpen
  \bibfield  {author} {\bibinfo {author} {\bibfnamefont {J.}~\bibnamefont
  {Cai}}, \bibinfo {author} {\bibfnamefont {F.}~\bibnamefont {Jelezko}},\ and\
  \bibinfo {author} {\bibfnamefont {M.~B.}\ \bibnamefont {Plenio}},\ }\bibfield
   {title} {\bibinfo {title} {Hybrid sensors based on colour centres in diamond
  and piezoactive layers},\ }\href {https://doi.org/10.1038/ncomms5065}
  {\bibfield  {journal} {\bibinfo  {journal} {Nat. Commun.}\ }\textbf {\bibinfo
  {volume} {5}},\ \bibinfo {pages} {4065} (\bibinfo {year} {2014})}\BibitemShut
  {NoStop}%
\bibitem [{\citenamefont {Rabl}\ \emph {et~al.}(2010)\citenamefont {Rabl},
  \citenamefont {Kolkowitz}, \citenamefont {Koppens}, \citenamefont {Harris},
  \citenamefont {Zoller},\ and\ \citenamefont {Lukin}}]{Rabl2010}%
  \BibitemOpen
  \bibfield  {author} {\bibinfo {author} {\bibfnamefont {P.}~\bibnamefont
  {Rabl}}, \bibinfo {author} {\bibfnamefont {S.~J.}\ \bibnamefont {Kolkowitz}},
  \bibinfo {author} {\bibfnamefont {F.~H.~L.}\ \bibnamefont {Koppens}},
  \bibinfo {author} {\bibfnamefont {J.~G.~E.}\ \bibnamefont {Harris}}, \bibinfo
  {author} {\bibfnamefont {P.}~\bibnamefont {Zoller}},\ and\ \bibinfo {author}
  {\bibfnamefont {M.~D.}\ \bibnamefont {Lukin}},\ }\bibfield  {title} {\bibinfo
  {title} {A quantum spin transducer based on nanoelectromechanical resonator
  arrays},\ }\href {https://doi.org/10.1038/nphys1679} {\bibfield  {journal}
  {\bibinfo  {journal} {Nat. Phys.}\ }\textbf {\bibinfo {volume} {6}},\
  \bibinfo {pages} {602} (\bibinfo {year} {2010})}\BibitemShut {NoStop}%
\bibitem [{\citenamefont {MacQuarrie}\ \emph {et~al.}(2017)\citenamefont
  {MacQuarrie}, \citenamefont {Otten}, \citenamefont {Gray},\ and\
  \citenamefont {Fuchs}}]{MacQuarrie2017}%
  \BibitemOpen
  \bibfield  {author} {\bibinfo {author} {\bibfnamefont {E.~R.}\ \bibnamefont
  {MacQuarrie}}, \bibinfo {author} {\bibfnamefont {M.}~\bibnamefont {Otten}},
  \bibinfo {author} {\bibfnamefont {S.~K.}\ \bibnamefont {Gray}},\ and\
  \bibinfo {author} {\bibfnamefont {G.~D.}\ \bibnamefont {Fuchs}},\ }\bibfield
  {title} {\bibinfo {title} {Cooling a mechanical resonator with
  nitrogen-vacancy centres using a room temperature excited state spin–strain
  interaction},\ }\href {https://doi.org/10.1038/ncomms14358} {\bibfield
  {journal} {\bibinfo  {journal} {Nat. Commun.}\ }\textbf {\bibinfo {volume}
  {8}},\ \bibinfo {pages} {14358} (\bibinfo {year} {2017})}\BibitemShut
  {NoStop}%
\bibitem [{\citenamefont {Manenti}\ \emph {et~al.}(2017)\citenamefont
  {Manenti}, \citenamefont {Kockum}, \citenamefont {Patterson}, \citenamefont
  {Behrle}, \citenamefont {Rahamim}, \citenamefont {Tancredi}, \citenamefont
  {Nori},\ and\ \citenamefont {Leek}}]{RN1054}%
  \BibitemOpen
  \bibfield  {author} {\bibinfo {author} {\bibfnamefont {R.}~\bibnamefont
  {Manenti}}, \bibinfo {author} {\bibfnamefont {A.~F.}\ \bibnamefont {Kockum}},
  \bibinfo {author} {\bibfnamefont {A.}~\bibnamefont {Patterson}}, \bibinfo
  {author} {\bibfnamefont {T.}~\bibnamefont {Behrle}}, \bibinfo {author}
  {\bibfnamefont {J.}~\bibnamefont {Rahamim}}, \bibinfo {author} {\bibfnamefont
  {G.}~\bibnamefont {Tancredi}}, \bibinfo {author} {\bibfnamefont
  {F.}~\bibnamefont {Nori}},\ and\ \bibinfo {author} {\bibfnamefont {P.~J.}\
  \bibnamefont {Leek}},\ }\bibfield  {title} {\bibinfo {title} {Circuit quantum
  acoustodynamics with surface acoustic waves},\ }\href
  {https://doi.org/10.1038/s41467-017-01063-9} {\bibfield  {journal} {\bibinfo
  {journal} {Nat. Commun.}\ }\textbf {\bibinfo {volume} {8}},\ \bibinfo {pages}
  {2041} (\bibinfo {year} {2017})}\BibitemShut {NoStop}%
\bibitem [{\citenamefont {Neuman}\ \emph {et~al.}(2020)\citenamefont {Neuman},
  \citenamefont {Wang},\ and\ \citenamefont {Narang}}]{PhysRevLett.125.247702}%
  \BibitemOpen
  \bibfield  {author} {\bibinfo {author} {\bibfnamefont {T.}~\bibnamefont
  {Neuman}}, \bibinfo {author} {\bibfnamefont {D.~S.}\ \bibnamefont {Wang}},\
  and\ \bibinfo {author} {\bibfnamefont {P.}~\bibnamefont {Narang}},\
  }\bibfield  {title} {\bibinfo {title} {Nanomagnonic cavities for strong
  spin-magnon coupling and magnon-mediated spin-spin interactions},\ }\href
  {https://doi.org/10.1103/PhysRevLett.125.247702} {\bibfield  {journal}
  {\bibinfo  {journal} {Phys. Rev. Lett.}\ }\textbf {\bibinfo {volume} {125}},\
  \bibinfo {pages} {247702} (\bibinfo {year} {2020})}\BibitemShut {NoStop}%
\bibitem [{\citenamefont {Hei}\ \emph {et~al.}(2021)\citenamefont {Hei},
  \citenamefont {Dong}, \citenamefont {Chen}, \citenamefont {Shen},
  \citenamefont {Qiao},\ and\ \citenamefont {Li}}]{PhysRevA.103.043706}%
  \BibitemOpen
  \bibfield  {author} {\bibinfo {author} {\bibfnamefont {X.-L.}\ \bibnamefont
  {Hei}}, \bibinfo {author} {\bibfnamefont {X.-L.}\ \bibnamefont {Dong}},
  \bibinfo {author} {\bibfnamefont {J.-Q.}\ \bibnamefont {Chen}}, \bibinfo
  {author} {\bibfnamefont {C.-P.}\ \bibnamefont {Shen}}, \bibinfo {author}
  {\bibfnamefont {Y.-F.}\ \bibnamefont {Qiao}},\ and\ \bibinfo {author}
  {\bibfnamefont {P.-B.}\ \bibnamefont {Li}},\ }\bibfield  {title} {\bibinfo
  {title} {Enhancing spin-photon coupling with a micromagnet},\ }\href
  {https://doi.org/10.1103/PhysRevA.103.043706} {\bibfield  {journal} {\bibinfo
   {journal} {Phys. Rev. A}\ }\textbf {\bibinfo {volume} {103}},\ \bibinfo
  {pages} {043706} (\bibinfo {year} {2021})}\BibitemShut {NoStop}%
\bibitem [{\citenamefont {Candido}\ \emph {et~al.}(2020)\citenamefont
  {Candido}, \citenamefont {Fuchs}, \citenamefont {Johnston-Halperin},\ and\
  \citenamefont {Flatt{\'{e}}}}]{Candido_2020}%
  \BibitemOpen
  \bibfield  {author} {\bibinfo {author} {\bibfnamefont {D.~R.}\ \bibnamefont
  {Candido}}, \bibinfo {author} {\bibfnamefont {G.~D.}\ \bibnamefont {Fuchs}},
  \bibinfo {author} {\bibfnamefont {E.}~\bibnamefont {Johnston-Halperin}},\
  and\ \bibinfo {author} {\bibfnamefont {M.~E.}\ \bibnamefont {Flatt{\'{e}}}},\
  }\bibfield  {title} {\bibinfo {title} {Predicted strong coupling of
  solid-state spins via a single magnon mode},\ }\href
  {https://doi.org/10.1088/2633-4356/ab9a55} {\bibfield  {journal} {\bibinfo
  {journal} {Mat. Quantum Technol.}\ }\textbf {\bibinfo {volume} {1}},\
  \bibinfo {pages} {011001} (\bibinfo {year} {2020})}\BibitemShut {NoStop}%
\bibitem [{\citenamefont {Fukami}\ \emph {et~al.}(2021)\citenamefont {Fukami},
  \citenamefont {Candido}, \citenamefont {Awschalom},\ and\ \citenamefont
  {Flatt\'e}}]{PRXQuantum.2.040314}%
  \BibitemOpen
  \bibfield  {author} {\bibinfo {author} {\bibfnamefont {M.}~\bibnamefont
  {Fukami}}, \bibinfo {author} {\bibfnamefont {D.~R.}\ \bibnamefont {Candido}},
  \bibinfo {author} {\bibfnamefont {D.~D.}\ \bibnamefont {Awschalom}},\ and\
  \bibinfo {author} {\bibfnamefont {M.~E.}\ \bibnamefont {Flatt\'e}},\
  }\bibfield  {title} {\bibinfo {title} {Opportunities for long-range
  magnon-mediated entanglement of spin qubits via on- and off-resonant
  coupling},\ }\href {https://doi.org/10.1103/PRXQuantum.2.040314} {\bibfield
  {journal} {\bibinfo  {journal} {PRX Quantum}\ }\textbf {\bibinfo {volume}
  {2}},\ \bibinfo {pages} {040314} (\bibinfo {year} {2021})}\BibitemShut
  {NoStop}%
\bibitem [{\citenamefont {Xiong}\ \emph {et~al.}(2022)\citenamefont {Xiong},
  \citenamefont {Tian}, \citenamefont {Zhang},\ and\ \citenamefont
  {You}}]{RN382}%
  \BibitemOpen
  \bibfield  {author} {\bibinfo {author} {\bibfnamefont {W.}~\bibnamefont
  {Xiong}}, \bibinfo {author} {\bibfnamefont {M.}~\bibnamefont {Tian}},
  \bibinfo {author} {\bibfnamefont {G.-Q.}\ \bibnamefont {Zhang}},\ and\
  \bibinfo {author} {\bibfnamefont {J.~Q.}\ \bibnamefont {You}},\ }\bibfield
  {title} {\bibinfo {title} {Strong long-range spin-spin coupling via a kerr
  magnon interface},\ }\href {https://doi.org/10.1103/PhysRevB.105.245310}
  {\bibfield  {journal} {\bibinfo  {journal} {Phys. Rev. B}\ }\textbf {\bibinfo
  {volume} {105}},\ \bibinfo {pages} {245310} (\bibinfo {year}
  {2022})}\BibitemShut {NoStop}%
\bibitem [{\citenamefont {Bai}\ \emph {et~al.}(2015)\citenamefont {Bai},
  \citenamefont {Harder}, \citenamefont {Chen}, \citenamefont {Fan},
  \citenamefont {Xiao},\ and\ \citenamefont {Hu}}]{RN623}%
  \BibitemOpen
  \bibfield  {author} {\bibinfo {author} {\bibfnamefont {L.}~\bibnamefont
  {Bai}}, \bibinfo {author} {\bibfnamefont {M.}~\bibnamefont {Harder}},
  \bibinfo {author} {\bibfnamefont {Y.~P.}\ \bibnamefont {Chen}}, \bibinfo
  {author} {\bibfnamefont {X.}~\bibnamefont {Fan}}, \bibinfo {author}
  {\bibfnamefont {J.~Q.}\ \bibnamefont {Xiao}},\ and\ \bibinfo {author}
  {\bibfnamefont {C.~M.}\ \bibnamefont {Hu}},\ }\bibfield  {title} {\bibinfo
  {title} {Spin pumping in electrodynamically coupled magnon-photon systems},\
  }\href {https://doi.org/10.1103/PhysRevLett.114.227201} {\bibfield  {journal}
  {\bibinfo  {journal} {Phys. Rev. Lett.}\ }\textbf {\bibinfo {volume} {114}},\
  \bibinfo {pages} {227201} (\bibinfo {year} {2015})}\BibitemShut {NoStop}%
\bibitem [{\citenamefont {Berk}\ \emph {et~al.}(2019)\citenamefont {Berk},
  \citenamefont {Jaris}, \citenamefont {Yang}, \citenamefont {Dhuey},
  \citenamefont {Cabrini},\ and\ \citenamefont {Schmidt}}]{RN691}%
  \BibitemOpen
  \bibfield  {author} {\bibinfo {author} {\bibfnamefont {C.}~\bibnamefont
  {Berk}}, \bibinfo {author} {\bibfnamefont {M.}~\bibnamefont {Jaris}},
  \bibinfo {author} {\bibfnamefont {W.}~\bibnamefont {Yang}}, \bibinfo {author}
  {\bibfnamefont {S.}~\bibnamefont {Dhuey}}, \bibinfo {author} {\bibfnamefont
  {S.}~\bibnamefont {Cabrini}},\ and\ \bibinfo {author} {\bibfnamefont
  {H.}~\bibnamefont {Schmidt}},\ }\bibfield  {title} {\bibinfo {title}
  {Strongly coupled magnon-phonon dynamics in a single nanomagnet},\ }\href
  {https://doi.org/10.1038/s41467-019-10545-x} {\bibfield  {journal} {\bibinfo
  {journal} {Nat. Commun.}\ }\textbf {\bibinfo {volume} {10}},\ \bibinfo
  {pages} {2652} (\bibinfo {year} {2019})}\BibitemShut {NoStop}%
\bibitem [{\citenamefont {Bittencourt}\ \emph {et~al.}(2022)\citenamefont
  {Bittencourt}, \citenamefont {Liberal},\ and\ \citenamefont
  {Viola~Kusminskiy}}]{RN472}%
  \BibitemOpen
  \bibfield  {author} {\bibinfo {author} {\bibfnamefont {V.}~\bibnamefont
  {Bittencourt}}, \bibinfo {author} {\bibfnamefont {I.}~\bibnamefont
  {Liberal}},\ and\ \bibinfo {author} {\bibfnamefont {S.}~\bibnamefont
  {Viola~Kusminskiy}},\ }\bibfield  {title} {\bibinfo {title} {Optomagnonics in
  dispersive media: Magnon-photon coupling enhancement at the epsilon-near-zero
  frequency},\ }\href {https://doi.org/10.1103/PhysRevLett.128.183603}
  {\bibfield  {journal} {\bibinfo  {journal} {Phys. Rev. Lett.}\ }\textbf
  {\bibinfo {volume} {128}},\ \bibinfo {pages} {183603} (\bibinfo {year}
  {2022})}\BibitemShut {NoStop}%
\bibitem [{\citenamefont {Harder}\ \emph {et~al.}(2018)\citenamefont {Harder},
  \citenamefont {Yang}, \citenamefont {Yao}, \citenamefont {Yu}, \citenamefont
  {Rao}, \citenamefont {Gui}, \citenamefont {Stamps},\ and\ \citenamefont
  {Hu}}]{RN433}%
  \BibitemOpen
  \bibfield  {author} {\bibinfo {author} {\bibfnamefont {M.}~\bibnamefont
  {Harder}}, \bibinfo {author} {\bibfnamefont {Y.}~\bibnamefont {Yang}},
  \bibinfo {author} {\bibfnamefont {B.~M.}\ \bibnamefont {Yao}}, \bibinfo
  {author} {\bibfnamefont {C.~H.}\ \bibnamefont {Yu}}, \bibinfo {author}
  {\bibfnamefont {J.~W.}\ \bibnamefont {Rao}}, \bibinfo {author} {\bibfnamefont
  {Y.~S.}\ \bibnamefont {Gui}}, \bibinfo {author} {\bibfnamefont {R.~L.}\
  \bibnamefont {Stamps}},\ and\ \bibinfo {author} {\bibfnamefont {C.~M.}\
  \bibnamefont {Hu}},\ }\bibfield  {title} {\bibinfo {title} {Level attraction
  due to dissipative magnon-photon coupling},\ }\href
  {https://doi.org/10.1103/PhysRevLett.121.137203} {\bibfield  {journal}
  {\bibinfo  {journal} {Phys. Rev. Lett.}\ }\textbf {\bibinfo {volume} {121}},\
  \bibinfo {pages} {137203} (\bibinfo {year} {2018})}\BibitemShut {NoStop}%
\bibitem [{\citenamefont {Tabuchi}\ \emph {et~al.}(2014)\citenamefont
  {Tabuchi}, \citenamefont {Ishino}, \citenamefont {Ishikawa}, \citenamefont
  {Yamazaki}, \citenamefont {Usami},\ and\ \citenamefont
  {Nakamura}}]{PhysRevLett.113.083603}%
  \BibitemOpen
  \bibfield  {author} {\bibinfo {author} {\bibfnamefont {Y.}~\bibnamefont
  {Tabuchi}}, \bibinfo {author} {\bibfnamefont {S.}~\bibnamefont {Ishino}},
  \bibinfo {author} {\bibfnamefont {T.}~\bibnamefont {Ishikawa}}, \bibinfo
  {author} {\bibfnamefont {R.}~\bibnamefont {Yamazaki}}, \bibinfo {author}
  {\bibfnamefont {K.}~\bibnamefont {Usami}},\ and\ \bibinfo {author}
  {\bibfnamefont {Y.}~\bibnamefont {Nakamura}},\ }\bibfield  {title} {\bibinfo
  {title} {Hybridizing ferromagnetic magnons and microwave photons in the
  quantum limit},\ }\href {https://doi.org/10.1103/PhysRevLett.113.083603}
  {\bibfield  {journal} {\bibinfo  {journal} {Phys. Rev. Lett.}\ }\textbf
  {\bibinfo {volume} {113}},\ \bibinfo {pages} {083603} (\bibinfo {year}
  {2014})}\BibitemShut {NoStop}%
\bibitem [{\citenamefont {Xu}\ \emph {et~al.}(2020)\citenamefont {Xu},
  \citenamefont {Zhong}, \citenamefont {Han}, \citenamefont {Jin},
  \citenamefont {Jiang},\ and\ \citenamefont {Zhang}}]{RN463}%
  \BibitemOpen
  \bibfield  {author} {\bibinfo {author} {\bibfnamefont {J.}~\bibnamefont
  {Xu}}, \bibinfo {author} {\bibfnamefont {C.}~\bibnamefont {Zhong}}, \bibinfo
  {author} {\bibfnamefont {X.}~\bibnamefont {Han}}, \bibinfo {author}
  {\bibfnamefont {D.}~\bibnamefont {Jin}}, \bibinfo {author} {\bibfnamefont
  {L.}~\bibnamefont {Jiang}},\ and\ \bibinfo {author} {\bibfnamefont
  {X.}~\bibnamefont {Zhang}},\ }\bibfield  {title} {\bibinfo {title} {Floquet
  cavity electromagnonics},\ }\href
  {https://doi.org/10.1103/PhysRevLett.125.237201} {\bibfield  {journal}
  {\bibinfo  {journal} {Phys. Rev. Lett.}\ }\textbf {\bibinfo {volume} {125}},\
  \bibinfo {pages} {237201} (\bibinfo {year} {2020})}\BibitemShut {NoStop}%
\bibitem [{\citenamefont {Yang}\ \emph {et~al.}(2019)\citenamefont {Yang},
  \citenamefont {Rao}, \citenamefont {Gui}, \citenamefont {Yao}, \citenamefont
  {Lu},\ and\ \citenamefont {Hu}}]{RN343}%
  \BibitemOpen
  \bibfield  {author} {\bibinfo {author} {\bibfnamefont {Y.}~\bibnamefont
  {Yang}}, \bibinfo {author} {\bibfnamefont {J.~W.}\ \bibnamefont {Rao}},
  \bibinfo {author} {\bibfnamefont {Y.~S.}\ \bibnamefont {Gui}}, \bibinfo
  {author} {\bibfnamefont {B.~M.}\ \bibnamefont {Yao}}, \bibinfo {author}
  {\bibfnamefont {W.}~\bibnamefont {Lu}},\ and\ \bibinfo {author}
  {\bibfnamefont {C.~M.}\ \bibnamefont {Hu}},\ }\bibfield  {title} {\bibinfo
  {title} {Control of the magnon-photon level attraction in a planar cavity},\
  }\href {https://doi.org/10.1103/PhysRevApplied.11.054023} {\bibfield
  {journal} {\bibinfo  {journal} {Phys. Rev. Applied}\ }\textbf {\bibinfo
  {volume} {11}},\ \bibinfo {pages} {054023} (\bibinfo {year}
  {2019})}\BibitemShut {NoStop}%
\bibitem [{\citenamefont {Yuan}\ \emph {et~al.}(2020)\citenamefont {Yuan},
  \citenamefont {Yan}, \citenamefont {Zheng}, \citenamefont {He}, \citenamefont
  {Xia},\ and\ \citenamefont {Yung}}]{RN448}%
  \BibitemOpen
  \bibfield  {author} {\bibinfo {author} {\bibfnamefont {H.~Y.}\ \bibnamefont
  {Yuan}}, \bibinfo {author} {\bibfnamefont {P.}~\bibnamefont {Yan}}, \bibinfo
  {author} {\bibfnamefont {S.}~\bibnamefont {Zheng}}, \bibinfo {author}
  {\bibfnamefont {Q.~Y.}\ \bibnamefont {He}}, \bibinfo {author} {\bibfnamefont
  {K.}~\bibnamefont {Xia}},\ and\ \bibinfo {author} {\bibfnamefont {M.~H.}\
  \bibnamefont {Yung}},\ }\bibfield  {title} {\bibinfo {title} {Steady bell
  state generation via magnon-photon coupling},\ }\href
  {https://doi.org/10.1103/PhysRevLett.124.053602} {\bibfield  {journal}
  {\bibinfo  {journal} {Phys. Rev. Lett.}\ }\textbf {\bibinfo {volume} {124}},\
  \bibinfo {pages} {053602} (\bibinfo {year} {2020})}\BibitemShut {NoStop}%
\bibitem [{\citenamefont {Zhang}\ \emph {et~al.}(2014)\citenamefont {Zhang},
  \citenamefont {Zou}, \citenamefont {Jiang},\ and\ \citenamefont
  {Tang}}]{PhysRevLett.113.156401}%
  \BibitemOpen
  \bibfield  {author} {\bibinfo {author} {\bibfnamefont {X.}~\bibnamefont
  {Zhang}}, \bibinfo {author} {\bibfnamefont {C.-L.}\ \bibnamefont {Zou}},
  \bibinfo {author} {\bibfnamefont {L.}~\bibnamefont {Jiang}},\ and\ \bibinfo
  {author} {\bibfnamefont {H.~X.}\ \bibnamefont {Tang}},\ }\bibfield  {title}
  {\bibinfo {title} {Strongly coupled magnons and cavity microwave photons},\
  }\href {https://doi.org/10.1103/PhysRevLett.113.156401} {\bibfield  {journal}
  {\bibinfo  {journal} {Phys. Rev. Lett.}\ }\textbf {\bibinfo {volume} {113}},\
  \bibinfo {pages} {156401} (\bibinfo {year} {2014})}\BibitemShut {NoStop}%
\bibitem [{\citenamefont {Kani}\ \emph
  {et~al.}(2022{\natexlab{b}})\citenamefont {Kani}, \citenamefont {Sarma},\
  and\ \citenamefont {Twamley}}]{kani2022intensive}%
  \BibitemOpen
  \bibfield  {author} {\bibinfo {author} {\bibfnamefont {A.}~\bibnamefont
  {Kani}}, \bibinfo {author} {\bibfnamefont {B.}~\bibnamefont {Sarma}},\ and\
  \bibinfo {author} {\bibfnamefont {J.}~\bibnamefont {Twamley}},\ }\bibfield
  {title} {\bibinfo {title} {Intensive cavity-magnomechanical cooling of a
  levitated macromagnet},\ }\href
  {https://journals.aps.org/prl/abstract/10.1103/PhysRevLett.128.013602}
  {\bibfield  {journal} {\bibinfo  {journal} {Phys. Rev. Lett.}\ }\textbf
  {\bibinfo {volume} {128}},\ \bibinfo {pages} {013602} (\bibinfo {year}
  {2022}{\natexlab{b}})}\BibitemShut {NoStop}%
\bibitem [{\citenamefont {Rugar}\ and\ \citenamefont
  {Gr\"utter}(1991)}]{PhysRevLett.67.699}%
  \BibitemOpen
  \bibfield  {author} {\bibinfo {author} {\bibfnamefont {D.}~\bibnamefont
  {Rugar}}\ and\ \bibinfo {author} {\bibfnamefont {P.}~\bibnamefont
  {Gr\"utter}},\ }\bibfield  {title} {\bibinfo {title} {Mechanical parametric
  amplification and thermomechanical noise squeezing},\ }\href
  {https://doi.org/10.1103/PhysRevLett.67.699} {\bibfield  {journal} {\bibinfo
  {journal} {Phys. Rev. Lett.}\ }\textbf {\bibinfo {volume} {67}},\ \bibinfo
  {pages} {699} (\bibinfo {year} {1991})}\BibitemShut {NoStop}%
\bibitem [{\citenamefont {Szorkovszky}\ \emph {et~al.}(2011)\citenamefont
  {Szorkovszky}, \citenamefont {Doherty}, \citenamefont {Harris},\ and\
  \citenamefont {Bowen}}]{PhysRevLett.107.213603}%
  \BibitemOpen
  \bibfield  {author} {\bibinfo {author} {\bibfnamefont {A.}~\bibnamefont
  {Szorkovszky}}, \bibinfo {author} {\bibfnamefont {A.~C.}\ \bibnamefont
  {Doherty}}, \bibinfo {author} {\bibfnamefont {G.~I.}\ \bibnamefont
  {Harris}},\ and\ \bibinfo {author} {\bibfnamefont {W.~P.}\ \bibnamefont
  {Bowen}},\ }\bibfield  {title} {\bibinfo {title} {Mechanical squeezing via
  parametric amplification and weak measurement},\ }\href
  {https://doi.org/10.1103/PhysRevLett.107.213603} {\bibfield  {journal}
  {\bibinfo  {journal} {Phys. Rev. Lett.}\ }\textbf {\bibinfo {volume} {107}},\
  \bibinfo {pages} {213603} (\bibinfo {year} {2011})}\BibitemShut {NoStop}%
\bibitem [{\citenamefont {Li}\ \emph {et~al.}(2020{\natexlab{c}})\citenamefont
  {Li}, \citenamefont {Zhou}, \citenamefont {Gao},\ and\ \citenamefont
  {Nori}}]{PhysRevLett.125.153602}%
  \BibitemOpen
  \bibfield  {author} {\bibinfo {author} {\bibfnamefont {P.-B.}\ \bibnamefont
  {Li}}, \bibinfo {author} {\bibfnamefont {Y.}~\bibnamefont {Zhou}}, \bibinfo
  {author} {\bibfnamefont {W.-B.}\ \bibnamefont {Gao}},\ and\ \bibinfo {author}
  {\bibfnamefont {F.}~\bibnamefont {Nori}},\ }\bibfield  {title} {\bibinfo
  {title} {Enhancing spin-phonon and spin-spin interactions using linear
  resources in a hybrid quantum system},\ }\href
  {https://doi.org/10.1103/PhysRevLett.125.153602} {\bibfield  {journal}
  {\bibinfo  {journal} {Phys. Rev. Lett.}\ }\textbf {\bibinfo {volume} {125}},\
  \bibinfo {pages} {153602} (\bibinfo {year} {2020}{\natexlab{c}})}\BibitemShut
  {NoStop}%
\bibitem [{\citenamefont {Szorkovszky}\ \emph {et~al.}(2014)\citenamefont
  {Szorkovszky}, \citenamefont {Clerk}, \citenamefont {Doherty},\ and\
  \citenamefont {Bowen}}]{Szorkovszky_2014}%
  \BibitemOpen
  \bibfield  {author} {\bibinfo {author} {\bibfnamefont {A.}~\bibnamefont
  {Szorkovszky}}, \bibinfo {author} {\bibfnamefont {A.~A.}\ \bibnamefont
  {Clerk}}, \bibinfo {author} {\bibfnamefont {A.~C.}\ \bibnamefont {Doherty}},\
  and\ \bibinfo {author} {\bibfnamefont {W.~P.}\ \bibnamefont {Bowen}},\
  }\bibfield  {title} {\bibinfo {title} {Mechanical entanglement via detuned
  parametric amplification},\ }\href
  {https://doi.org/10.1088/1367-2630/16/6/063043} {\bibfield  {journal}
  {\bibinfo  {journal} {New J. Phys.}\ }\textbf {\bibinfo {volume} {16}},\
  \bibinfo {pages} {063043} (\bibinfo {year} {2014})}\BibitemShut {NoStop}%
\bibitem [{\citenamefont {Lemonde}\ \emph {et~al.}(2016)\citenamefont
  {Lemonde}, \citenamefont {Didier},\ and\ \citenamefont
  {Clerk}}]{Lemonde2016}%
  \BibitemOpen
  \bibfield  {author} {\bibinfo {author} {\bibfnamefont {M.-A.}\ \bibnamefont
  {Lemonde}}, \bibinfo {author} {\bibfnamefont {N.}~\bibnamefont {Didier}},\
  and\ \bibinfo {author} {\bibfnamefont {A.~A.}\ \bibnamefont {Clerk}},\
  }\bibfield  {title} {\bibinfo {title} {Enhanced nonlinear interactions in
  quantum optomechanics via mechanical amplification},\ }\href
  {https://doi.org/10.1038/ncomms11338} {\bibfield  {journal} {\bibinfo
  {journal} {Nat. Commun.}\ }\textbf {\bibinfo {volume} {7}},\ \bibinfo {pages}
  {11338} (\bibinfo {year} {2016})}\BibitemShut {NoStop}%
\bibitem [{\citenamefont {Yin}\ \emph {et~al.}(2017)\citenamefont {Yin},
  \citenamefont {L\"u}, \citenamefont {Zheng}, \citenamefont {Wang},
  \citenamefont {Li},\ and\ \citenamefont {Wu}}]{PhysRevA.95.053861}%
  \BibitemOpen
  \bibfield  {author} {\bibinfo {author} {\bibfnamefont {T.-S.}\ \bibnamefont
  {Yin}}, \bibinfo {author} {\bibfnamefont {X.-Y.}\ \bibnamefont {L\"u}},
  \bibinfo {author} {\bibfnamefont {L.-L.}\ \bibnamefont {Zheng}}, \bibinfo
  {author} {\bibfnamefont {M.}~\bibnamefont {Wang}}, \bibinfo {author}
  {\bibfnamefont {S.}~\bibnamefont {Li}},\ and\ \bibinfo {author}
  {\bibfnamefont {Y.}~\bibnamefont {Wu}},\ }\bibfield  {title} {\bibinfo
  {title} {Nonlinear effects in modulated quantum optomechanics},\ }\href
  {https://doi.org/10.1103/PhysRevA.95.053861} {\bibfield  {journal} {\bibinfo
  {journal} {Phys. Rev. A}\ }\textbf {\bibinfo {volume} {95}},\ \bibinfo
  {pages} {053861} (\bibinfo {year} {2017})}\BibitemShut {NoStop}%
\bibitem [{\citenamefont {Ge}\ \emph {et~al.}(2019)\citenamefont {Ge},
  \citenamefont {Sawyer}, \citenamefont {Britton}, \citenamefont {Jacobs},
  \citenamefont {Bollinger},\ and\ \citenamefont
  {Foss-Feig}}]{PhysRevLett.122.030501}%
  \BibitemOpen
  \bibfield  {author} {\bibinfo {author} {\bibfnamefont {W.}~\bibnamefont
  {Ge}}, \bibinfo {author} {\bibfnamefont {B.~C.}\ \bibnamefont {Sawyer}},
  \bibinfo {author} {\bibfnamefont {J.~W.}\ \bibnamefont {Britton}}, \bibinfo
  {author} {\bibfnamefont {K.}~\bibnamefont {Jacobs}}, \bibinfo {author}
  {\bibfnamefont {J.~J.}\ \bibnamefont {Bollinger}},\ and\ \bibinfo {author}
  {\bibfnamefont {M.}~\bibnamefont {Foss-Feig}},\ }\bibfield  {title} {\bibinfo
  {title} {Trapped ion quantum information processing with squeezed phonons},\
  }\href {https://doi.org/10.1103/PhysRevLett.122.030501} {\bibfield  {journal}
  {\bibinfo  {journal} {Phys. Rev. Lett.}\ }\textbf {\bibinfo {volume} {122}},\
  \bibinfo {pages} {030501} (\bibinfo {year} {2019})}\BibitemShut {NoStop}%
\bibitem [{\citenamefont {Burd}\ \emph {et~al.}(2021)\citenamefont {Burd},
  \citenamefont {Srinivas}, \citenamefont {Knaack}, \citenamefont {Ge},
  \citenamefont {Wilson}, \citenamefont {Wineland}, \citenamefont {Leibfried},
  \citenamefont {Bollinger}, \citenamefont {Allcock},\ and\ \citenamefont
  {Slichter}}]{Burd2021}%
  \BibitemOpen
  \bibfield  {author} {\bibinfo {author} {\bibfnamefont {S.~C.}\ \bibnamefont
  {Burd}}, \bibinfo {author} {\bibfnamefont {R.}~\bibnamefont {Srinivas}},
  \bibinfo {author} {\bibfnamefont {H.~M.}\ \bibnamefont {Knaack}}, \bibinfo
  {author} {\bibfnamefont {W.}~\bibnamefont {Ge}}, \bibinfo {author}
  {\bibfnamefont {A.~C.}\ \bibnamefont {Wilson}}, \bibinfo {author}
  {\bibfnamefont {D.~J.}\ \bibnamefont {Wineland}}, \bibinfo {author}
  {\bibfnamefont {D.}~\bibnamefont {Leibfried}}, \bibinfo {author}
  {\bibfnamefont {J.~J.}\ \bibnamefont {Bollinger}}, \bibinfo {author}
  {\bibfnamefont {D.~T.~C.}\ \bibnamefont {Allcock}},\ and\ \bibinfo {author}
  {\bibfnamefont {D.~H.}\ \bibnamefont {Slichter}},\ }\bibfield  {title}
  {\bibinfo {title} {Quantum amplification of boson-mediated interactions},\
  }\href {https://doi.org/10.1038/s41567-021-01237-9} {\bibfield  {journal}
  {\bibinfo  {journal} {Nat. Phys.}\ }\textbf {\bibinfo {volume} {17}},\
  \bibinfo {pages} {898} (\bibinfo {year} {2021})}\BibitemShut {NoStop}%
\bibitem [{\citenamefont {Paul}(1990)}]{RN742}%
  \BibitemOpen
  \bibfield  {author} {\bibinfo {author} {\bibfnamefont {W.}~\bibnamefont
  {Paul}},\ }\bibfield  {title} {\bibinfo {title} {Electromagnetic traps for
  charged and neutral particles},\ }\href
  {https://doi.org/10.1103/RevModPhys.62.531} {\bibfield  {journal} {\bibinfo
  {journal} {Rev. Mod. Phys.}\ }\textbf {\bibinfo {volume} {62}},\ \bibinfo
  {pages} {531} (\bibinfo {year} {1990})}\BibitemShut {NoStop}%
\bibitem [{\citenamefont {Delord}\ \emph {et~al.}(2018)\citenamefont {Delord},
  \citenamefont {Huillery}, \citenamefont {Schwab}, \citenamefont {Nicolas},
  \citenamefont {Lecordier},\ and\ \citenamefont {Hetet}}]{RN738}%
  \BibitemOpen
  \bibfield  {author} {\bibinfo {author} {\bibfnamefont {T.}~\bibnamefont
  {Delord}}, \bibinfo {author} {\bibfnamefont {P.}~\bibnamefont {Huillery}},
  \bibinfo {author} {\bibfnamefont {L.}~\bibnamefont {Schwab}}, \bibinfo
  {author} {\bibfnamefont {L.}~\bibnamefont {Nicolas}}, \bibinfo {author}
  {\bibfnamefont {L.}~\bibnamefont {Lecordier}},\ and\ \bibinfo {author}
  {\bibfnamefont {G.}~\bibnamefont {Hetet}},\ }\bibfield  {title} {\bibinfo
  {title} {Ramsey interferences and spin echoes from electron spins inside a
  levitating macroscopic particle},\ }\href
  {https://doi.org/10.1103/PhysRevLett.121.053602} {\bibfield  {journal}
  {\bibinfo  {journal} {Phys. Rev. Lett.}\ }\textbf {\bibinfo {volume} {121}},\
  \bibinfo {pages} {053602} (\bibinfo {year} {2018})}\BibitemShut {NoStop}%
\bibitem [{\citenamefont {Delord}\ \emph
  {et~al.}(2017{\natexlab{a}})\citenamefont {Delord}, \citenamefont {Nicolas},
  \citenamefont {Bodini},\ and\ \citenamefont {Hétet}}]{RN735}%
  \BibitemOpen
  \bibfield  {author} {\bibinfo {author} {\bibfnamefont {T.}~\bibnamefont
  {Delord}}, \bibinfo {author} {\bibfnamefont {L.}~\bibnamefont {Nicolas}},
  \bibinfo {author} {\bibfnamefont {M.}~\bibnamefont {Bodini}},\ and\ \bibinfo
  {author} {\bibfnamefont {G.}~\bibnamefont {Hétet}},\ }\bibfield  {title}
  {\bibinfo {title} {Diamonds levitating in a paul trap under vacuum:
  Measurements of laser-induced heating via nv center thermometry},\ }\href
  {https://doi.org/10.1063/1.4991670} {\bibfield  {journal} {\bibinfo
  {journal} {Appl. Phys. Lett.}\ }\textbf {\bibinfo {volume} {111}},\ \bibinfo
  {pages} {013101} (\bibinfo {year} {2017}{\natexlab{a}})}\BibitemShut
  {NoStop}%
\bibitem [{\citenamefont {Delord}\ \emph
  {et~al.}(2017{\natexlab{b}})\citenamefont {Delord}, \citenamefont {Nicolas},
  \citenamefont {Schwab},\ and\ \citenamefont {Hétet}}]{RN737}%
  \BibitemOpen
  \bibfield  {author} {\bibinfo {author} {\bibfnamefont {T.}~\bibnamefont
  {Delord}}, \bibinfo {author} {\bibfnamefont {L.}~\bibnamefont {Nicolas}},
  \bibinfo {author} {\bibfnamefont {L.}~\bibnamefont {Schwab}},\ and\ \bibinfo
  {author} {\bibfnamefont {G.}~\bibnamefont {Hétet}},\ }\bibfield  {title}
  {\bibinfo {title} {Electron spin resonance from nv centers in diamonds
  levitating in an ion trap},\ }\href
  {https://doi.org/10.1088/1367-2630/aa659c} {\bibfield  {journal} {\bibinfo
  {journal} {New J. Phys.}\ }\textbf {\bibinfo {volume} {19}},\ \bibinfo
  {pages} {033031} (\bibinfo {year} {2017}{\natexlab{b}})}\BibitemShut
  {NoStop}%
\bibitem [{\citenamefont {Alda}\ \emph {et~al.}(2016)\citenamefont {Alda},
  \citenamefont {Berthelot}, \citenamefont {Rica},\ and\ \citenamefont
  {Quidant}}]{RN752}%
  \BibitemOpen
  \bibfield  {author} {\bibinfo {author} {\bibfnamefont {I.}~\bibnamefont
  {Alda}}, \bibinfo {author} {\bibfnamefont {J.}~\bibnamefont {Berthelot}},
  \bibinfo {author} {\bibfnamefont {R.~A.}\ \bibnamefont {Rica}},\ and\
  \bibinfo {author} {\bibfnamefont {R.}~\bibnamefont {Quidant}},\ }\bibfield
  {title} {\bibinfo {title} {Trapping and manipulation of individual
  nanoparticles in a planar paul trap},\ }\href
  {https://doi.org/10.1063/1.4965859} {\bibfield  {journal} {\bibinfo
  {journal} {Appl. Phys. Lett.}\ }\textbf {\bibinfo {volume} {109}},\ \bibinfo
  {pages} {163105} (\bibinfo {year} {2016})}\BibitemShut {NoStop}%
\bibitem [{\citenamefont {Kuhlicke}\ \emph {et~al.}(2014)\citenamefont
  {Kuhlicke}, \citenamefont {Schell}, \citenamefont {Zoll},\ and\ \citenamefont
  {Benson}}]{doi:10.1063/1.4893575}%
  \BibitemOpen
  \bibfield  {author} {\bibinfo {author} {\bibfnamefont {A.}~\bibnamefont
  {Kuhlicke}}, \bibinfo {author} {\bibfnamefont {A.~W.}\ \bibnamefont
  {Schell}}, \bibinfo {author} {\bibfnamefont {J.}~\bibnamefont {Zoll}},\ and\
  \bibinfo {author} {\bibfnamefont {O.}~\bibnamefont {Benson}},\ }\bibfield
  {title} {\bibinfo {title} {Nitrogen vacancy center fluorescence from a
  submicron diamond cluster levitated in a linear quadrupole ion trap},\ }\href
  {https://doi.org/10.1063/1.4893575} {\bibfield  {journal} {\bibinfo
  {journal} {Appl. Phys. Lett.}\ }\textbf {\bibinfo {volume} {105}},\ \bibinfo
  {pages} {073101} (\bibinfo {year} {2014})}\BibitemShut {NoStop}%
\bibitem [{\citenamefont {Conangla}\ \emph {et~al.}(2018)\citenamefont
  {Conangla}, \citenamefont {Schell}, \citenamefont {Rica},\ and\ \citenamefont
  {Quidant}}]{RN734}%
  \BibitemOpen
  \bibfield  {author} {\bibinfo {author} {\bibfnamefont {G.~P.}\ \bibnamefont
  {Conangla}}, \bibinfo {author} {\bibfnamefont {A.~W.}\ \bibnamefont
  {Schell}}, \bibinfo {author} {\bibfnamefont {R.~A.}\ \bibnamefont {Rica}},\
  and\ \bibinfo {author} {\bibfnamefont {R.}~\bibnamefont {Quidant}},\
  }\bibfield  {title} {\bibinfo {title} {Motion control and optical
  interrogation of a levitating single nitrogen vacancy in vacuum},\ }\href
  {https://doi.org/10.1021/acs.nanolett.8b01414} {\bibfield  {journal}
  {\bibinfo  {journal} {Nano. Lett.}\ }\textbf {\bibinfo {volume} {18}},\
  \bibinfo {pages} {3956} (\bibinfo {year} {2018})}\BibitemShut {NoStop}%
\bibitem [{\citenamefont {Vinante}\ \emph {et~al.}(2020)\citenamefont
  {Vinante}, \citenamefont {Falferi}, \citenamefont {Gasbarri}, \citenamefont
  {Setter}, \citenamefont {Timberlake},\ and\ \citenamefont
  {Ulbricht}}]{PhysRevApplied.13.064027}%
  \BibitemOpen
  \bibfield  {author} {\bibinfo {author} {\bibfnamefont {A.}~\bibnamefont
  {Vinante}}, \bibinfo {author} {\bibfnamefont {P.}~\bibnamefont {Falferi}},
  \bibinfo {author} {\bibfnamefont {G.}~\bibnamefont {Gasbarri}}, \bibinfo
  {author} {\bibfnamefont {A.}~\bibnamefont {Setter}}, \bibinfo {author}
  {\bibfnamefont {C.}~\bibnamefont {Timberlake}},\ and\ \bibinfo {author}
  {\bibfnamefont {H.}~\bibnamefont {Ulbricht}},\ }\bibfield  {title} {\bibinfo
  {title} {Ultralow mechanical damping with {Meissner}-levitated ferromagnetic
  microparticles},\ }\href {https://doi.org/10.1103/PhysRevApplied.13.064027}
  {\bibfield  {journal} {\bibinfo  {journal} {Phys. Rev. Applied}\ }\textbf
  {\bibinfo {volume} {13}},\ \bibinfo {pages} {064027} (\bibinfo {year}
  {2020})}\BibitemShut {NoStop}%
\bibitem [{\citenamefont {Rusconi}\ \emph {et~al.}(2017)\citenamefont
  {Rusconi}, \citenamefont {Pochhacker}, \citenamefont {Kustura}, \citenamefont
  {Cirac},\ and\ \citenamefont {Romero-Isart}}]{RN427}%
  \BibitemOpen
  \bibfield  {author} {\bibinfo {author} {\bibfnamefont {C.~C.}\ \bibnamefont
  {Rusconi}}, \bibinfo {author} {\bibfnamefont {V.}~\bibnamefont {Pochhacker}},
  \bibinfo {author} {\bibfnamefont {K.}~\bibnamefont {Kustura}}, \bibinfo
  {author} {\bibfnamefont {J.~I.}\ \bibnamefont {Cirac}},\ and\ \bibinfo
  {author} {\bibfnamefont {O.}~\bibnamefont {Romero-Isart}},\ }\bibfield
  {title} {\bibinfo {title} {Quantum spin stabilized magnetic levitation},\
  }\href {https://doi.org/10.1103/PhysRevLett.119.167202} {\bibfield  {journal}
  {\bibinfo  {journal} {Phys. Rev. Lett.}\ }\textbf {\bibinfo {volume} {119}},\
  \bibinfo {pages} {167202} (\bibinfo {year} {2017})}\BibitemShut {NoStop}%
\bibitem [{\citenamefont {Timberlake}\ \emph {et~al.}(2019)\citenamefont
  {Timberlake}, \citenamefont {Gasbarri}, \citenamefont {Vinante},
  \citenamefont {Setter},\ and\ \citenamefont
  {Ulbricht}}]{doi:10.1063/1.5129145}%
  \BibitemOpen
  \bibfield  {author} {\bibinfo {author} {\bibfnamefont {C.}~\bibnamefont
  {Timberlake}}, \bibinfo {author} {\bibfnamefont {G.}~\bibnamefont
  {Gasbarri}}, \bibinfo {author} {\bibfnamefont {A.}~\bibnamefont {Vinante}},
  \bibinfo {author} {\bibfnamefont {A.}~\bibnamefont {Setter}},\ and\ \bibinfo
  {author} {\bibfnamefont {H.}~\bibnamefont {Ulbricht}},\ }\bibfield  {title}
  {\bibinfo {title} {Acceleration sensing with magnetically levitated
  oscillators above a superconductor},\ }\href
  {https://doi.org/10.1063/1.5129145} {\bibfield  {journal} {\bibinfo
  {journal} {Appl. Phys. Lett.}\ }\textbf {\bibinfo {volume} {115}},\ \bibinfo
  {pages} {224101} (\bibinfo {year} {2019})}\BibitemShut {NoStop}%
\bibitem [{Sup()}]{SupplementalMaterial}%
  \BibitemOpen
  \href@noop {} {}\bibinfo {note} {See Supplemental Material at https://xxx for
  more details, which includes Refs. [124-136].}\BibitemShut {Stop}%
\bibitem [{\citenamefont {Leibfried}\ \emph {et~al.}(2003)\citenamefont
	{Leibfried}, \citenamefont {Blatt}, \citenamefont {Monroe},\ and\
	\citenamefont {Wineland}}]{RN666}%
\BibitemOpen
\bibfield  {author} {\bibinfo {author} {\bibfnamefont {D.}~\bibnamefont
		{Leibfried}}, \bibinfo {author} {\bibfnamefont {R.}~\bibnamefont {Blatt}},
	\bibinfo {author} {\bibfnamefont {C.}~\bibnamefont {Monroe}},\ and\ \bibinfo
	{author} {\bibfnamefont {D.}~\bibnamefont {Wineland}},\ }\bibfield  {title}
{\bibinfo {title} {Quantum dynamics of single trapped ions},\ }\href
{https://doi.org/10.1103/RevModPhys.75.281} {\bibfield  {journal} {\bibinfo
		{journal} {Rev. Mod. Phys.}\ }\textbf {\bibinfo {volume} {75}},\ \bibinfo
	{pages} {281} (\bibinfo {year} {2003})}\BibitemShut {NoStop}%
\bibitem [{\citenamefont {Stancil}\ and\ \citenamefont
	{Prabhakar}(2009)}]{stancil2009spin}%
\BibitemOpen
\bibfield  {author} {\bibinfo {author} {\bibfnamefont {D.~D.}\ \bibnamefont
		{Stancil}}\ and\ \bibinfo {author} {\bibfnamefont {A.}~\bibnamefont
		{Prabhakar}},\ }\href@noop {} {\emph {\bibinfo {title} {Spin Waves: Theory
			and Applications}}}\ (\bibinfo  {publisher} {Springer, New York},\ \bibinfo
{year} {2009})\BibitemShut {NoStop}%
\bibitem [{\citenamefont {Aharoni}(2000)}]{aharoni2000introduction}%
\BibitemOpen
\bibfield  {author} {\bibinfo {author} {\bibfnamefont {A.}~\bibnamefont
		{Aharoni}},\ }\href@noop {} {\emph {\bibinfo {title} {Introduction to the
			Theory of Ferromagnetism}}}\ (\bibinfo  {publisher} {Clarendon Press},\
\bibinfo {year} {2000})\BibitemShut {NoStop}%
\bibitem [{\citenamefont {Gurevich}\ and\ \citenamefont
	{Melkov}(1996)}]{gurevich1996magnetization}%
\BibitemOpen
\bibfield  {author} {\bibinfo {author} {\bibfnamefont {A.~G.}\ \bibnamefont
		{Gurevich}}\ and\ \bibinfo {author} {\bibfnamefont {G.~A.}\ \bibnamefont
		{Melkov}},\ }\href@noop {} {\emph {\bibinfo {title} {Magnetization
			oscillations and waves}}}\ (\bibinfo  {publisher} {CRC press},\ \bibinfo
{year} {1996})\BibitemShut {NoStop}%
\bibitem [{\citenamefont {Kostylev}\ \emph {et~al.}(2016)\citenamefont
	{Kostylev}, \citenamefont {Goryachev},\ and\ \citenamefont
	{Tobar}}]{doi:10.1063/1.4941730}%
\BibitemOpen
\bibfield  {author} {\bibinfo {author} {\bibfnamefont {N.}~\bibnamefont
		{Kostylev}}, \bibinfo {author} {\bibfnamefont {M.}~\bibnamefont
		{Goryachev}},\ and\ \bibinfo {author} {\bibfnamefont {M.~E.}\ \bibnamefont
		{Tobar}},\ }\bibfield  {title} {\bibinfo {title} {Superstrong coupling of a
		microwave cavity to yttrium iron garnet magnons},\ }\href
{https://doi.org/10.1063/1.4941730} {\bibfield  {journal} {\bibinfo
		{journal} {Appl. Phys. Lett.}\ }\textbf {\bibinfo {volume} {108}},\ \bibinfo
	{pages} {062402} (\bibinfo {year} {2016})}\BibitemShut {NoStop}%
\bibitem [{\citenamefont {Maier-Flaig}\ \emph {et~al.}(2017)\citenamefont
	{Maier-Flaig}, \citenamefont {Klingler}, \citenamefont {Dubs}, \citenamefont
	{Surzhenko}, \citenamefont {Gross}, \citenamefont {Weiler}, \citenamefont
	{Huebl},\ and\ \citenamefont {Goennenwein}}]{PhysRevB.95.214423}%
\BibitemOpen
\bibfield  {author} {\bibinfo {author} {\bibfnamefont {H.}~\bibnamefont
		{Maier-Flaig}}, \bibinfo {author} {\bibfnamefont {S.}~\bibnamefont
		{Klingler}}, \bibinfo {author} {\bibfnamefont {C.}~\bibnamefont {Dubs}},
	\bibinfo {author} {\bibfnamefont {O.}~\bibnamefont {Surzhenko}}, \bibinfo
	{author} {\bibfnamefont {R.}~\bibnamefont {Gross}}, \bibinfo {author}
	{\bibfnamefont {M.}~\bibnamefont {Weiler}}, \bibinfo {author} {\bibfnamefont
		{H.}~\bibnamefont {Huebl}},\ and\ \bibinfo {author} {\bibfnamefont
		{S.~T.~B.}\ \bibnamefont {Goennenwein}},\ }\bibfield  {title} {\bibinfo
	{title} {Temperature-dependent magnetic damping of yttrium iron garnet
		spheres},\ }\href {https://doi.org/10.1103/PhysRevB.95.214423} {\bibfield
	{journal} {\bibinfo  {journal} {Phys. Rev. B}\ }\textbf {\bibinfo {volume}
		{95}},\ \bibinfo {pages} {214423} (\bibinfo {year} {2017})}\BibitemShut
{NoStop}%
\bibitem [{\citenamefont {Jackson}(2007)}]{jackson2007classical}%
\BibitemOpen
\bibfield  {author} {\bibinfo {author} {\bibfnamefont {J.~D.}\ \bibnamefont
		{Jackson}},\ }\href@noop {} {\emph {\bibinfo {title} {Classical
			electrodynamics}}}\ (\bibinfo  {publisher} {John Wiley \& Sons},\ \bibinfo
{year} {2007})\BibitemShut {NoStop}%
\bibitem [{\citenamefont {Fletcher}\ and\ \citenamefont
	{Bell}(1959)}]{doi:10.1063/1.1735216}%
\BibitemOpen
\bibfield  {author} {\bibinfo {author} {\bibfnamefont {P.~C.}\ \bibnamefont
		{Fletcher}}\ and\ \bibinfo {author} {\bibfnamefont {R.~O.}\ \bibnamefont
		{Bell}},\ }\bibfield  {title} {\bibinfo {title} {Ferrimagnetic resonance
		modes in spheres},\ }\href {https://doi.org/10.1063/1.1735216} {\bibfield
	{journal} {\bibinfo  {journal} {J. Appl. Phys.}\ }\textbf {\bibinfo {volume}
		{30}},\ \bibinfo {pages} {687} (\bibinfo {year} {1959})}\BibitemShut
{NoStop}%
\bibitem [{\citenamefont {Röschmann}\ and\ \citenamefont
	{Dötsch}(1977)}]{doi:10.1002/pssb.2220820102}%
\BibitemOpen
\bibfield  {author} {\bibinfo {author} {\bibfnamefont {P.}~\bibnamefont
		{Röschmann}}\ and\ \bibinfo {author} {\bibfnamefont {H.}~\bibnamefont
		{Dötsch}},\ }\bibfield  {title} {\bibinfo {title} {Properties of
		magnetostatic modes in ferrimagnetic spheroids},\ }\href
{https://doi.org/10.1002/pssb.2220820102} {\bibfield  {journal} {\bibinfo
		{journal} {Phys. Status Solidi B}\ }\textbf {\bibinfo {volume} {82}},\
	\bibinfo {pages} {11} (\bibinfo {year} {1977})}\BibitemShut {NoStop}%
\bibitem [{\citenamefont {Mills}(2006)}]{MILLS200616}%
\BibitemOpen
\bibfield  {author} {\bibinfo {author} {\bibfnamefont {D.}~\bibnamefont
		{Mills}},\ }\bibfield  {title} {\bibinfo {title} {Quantum theory of spin
		waves in finite samples},\ }\href
{https://doi.org/https://doi.org/10.1016/j.jmmm.2006.02.267} {\bibfield
	{journal} {\bibinfo  {journal} {J. Magn. Magn. Mater.}\ }\textbf {\bibinfo
		{volume} {306}},\ \bibinfo {pages} {16} (\bibinfo {year} {2006})}\BibitemShut
{NoStop}%
\bibitem [{\citenamefont {Walker}(1957)}]{PhysRev.105.390}%
\BibitemOpen
\bibfield  {author} {\bibinfo {author} {\bibfnamefont {L.~R.}\ \bibnamefont
		{Walker}},\ }\bibfield  {title} {\bibinfo {title} {Magnetostatic modes in
		ferromagnetic resonance},\ }\href {https://doi.org/10.1103/PhysRev.105.390}
{\bibfield  {journal} {\bibinfo  {journal} {Phys. Rev.}\ }\textbf {\bibinfo
		{volume} {105}},\ \bibinfo {pages} {390} (\bibinfo {year}
	{1957})}\BibitemShut {NoStop}%
\bibitem [{\citenamefont {Ding}\ \emph {et~al.}(2018)\citenamefont {Ding},
	\citenamefont {Maslennikov}, \citenamefont {Habl\"utzel},\ and\ \citenamefont
	{Matsukevich}}]{PhysRevLett.121.130502}%
\BibitemOpen
\bibfield  {author} {\bibinfo {author} {\bibfnamefont {S.}~\bibnamefont
		{Ding}}, \bibinfo {author} {\bibfnamefont {G.}~\bibnamefont {Maslennikov}},
	\bibinfo {author} {\bibfnamefont {R.}~\bibnamefont {Habl\"utzel}},\ and\
	\bibinfo {author} {\bibfnamefont {D.}~\bibnamefont {Matsukevich}},\
}\bibfield  {title} {\bibinfo {title} {Quantum simulation with a trilinear
		{Hamiltonian}},\ }\href {https://doi.org/10.1103/PhysRevLett.121.130502}
{\bibfield  {journal} {\bibinfo  {journal} {Phys. Rev. Lett.}\ }\textbf
	{\bibinfo {volume} {121}},\ \bibinfo {pages} {130502} (\bibinfo {year}
	{2018})}\BibitemShut {NoStop}%
\bibitem [{\citenamefont {Gieseler}\ \emph {et~al.}(2012)\citenamefont
	{Gieseler}, \citenamefont {Deutsch}, \citenamefont {Quidant},\ and\
	\citenamefont {Novotny}}]{RN744}%
\BibitemOpen
\bibfield  {author} {\bibinfo {author} {\bibfnamefont {J.}~\bibnamefont
		{Gieseler}}, \bibinfo {author} {\bibfnamefont {B.}~\bibnamefont {Deutsch}},
	\bibinfo {author} {\bibfnamefont {R.}~\bibnamefont {Quidant}},\ and\ \bibinfo
	{author} {\bibfnamefont {L.}~\bibnamefont {Novotny}},\ }\bibfield  {title}
{\bibinfo {title} {Subkelvin parametric feedback cooling of a laser-trapped
		nanoparticle},\ }\href {https://doi.org/10.1103/PhysRevLett.109.103603}
{\bibfield  {journal} {\bibinfo  {journal} {Phys. Rev. Lett.}\ }\textbf
	{\bibinfo {volume} {109}},\ \bibinfo {pages} {103603} (\bibinfo {year}
	{2012})}\BibitemShut {NoStop}%
\bibitem [{\citenamefont {Prat-Camps}\ \emph {et~al.}(2017)\citenamefont
  {Prat-Camps}, \citenamefont {Teo}, \citenamefont {Rusconi}, \citenamefont
  {Wieczorek},\ and\ \citenamefont {Romero-Isart}}]{PhysRevApplied.8.034002}%
  \BibitemOpen
  \bibfield  {author} {\bibinfo {author} {\bibfnamefont {J.}~\bibnamefont
  {Prat-Camps}}, \bibinfo {author} {\bibfnamefont {C.}~\bibnamefont {Teo}},
  \bibinfo {author} {\bibfnamefont {C.~C.}\ \bibnamefont {Rusconi}}, \bibinfo
  {author} {\bibfnamefont {W.}~\bibnamefont {Wieczorek}},\ and\ \bibinfo
  {author} {\bibfnamefont {O.}~\bibnamefont {Romero-Isart}},\ }\bibfield
  {title} {\bibinfo {title} {Ultrasensitive inertial and force sensors with
  diamagnetically levitated magnets},\ }\href
  {https://doi.org/10.1103/PhysRevApplied.8.034002} {\bibfield  {journal}
  {\bibinfo  {journal} {Phys. Rev. Applied}\ }\textbf {\bibinfo {volume} {8}},\
  \bibinfo {pages} {034002} (\bibinfo {year} {2017})}\BibitemShut {NoStop}%
\bibitem [{\citenamefont {Heinzen}\ and\ \citenamefont
  {Wineland}(1990)}]{PhysRevA.42.2977}%
  \BibitemOpen
  \bibfield  {author} {\bibinfo {author} {\bibfnamefont {D.~J.}\ \bibnamefont
  {Heinzen}}\ and\ \bibinfo {author} {\bibfnamefont {D.~J.}\ \bibnamefont
  {Wineland}},\ }\bibfield  {title} {\bibinfo {title} {Quantum-limited cooling
  and detection of radio-frequency oscillations by laser-cooled ions},\ }\href
  {https://doi.org/10.1103/PhysRevA.42.2977} {\bibfield  {journal} {\bibinfo
  {journal} {Phys. Rev. A}\ }\textbf {\bibinfo {volume} {42}},\ \bibinfo
  {pages} {2977} (\bibinfo {year} {1990})}\BibitemShut {NoStop}%
\bibitem [{\citenamefont {Delord}\ \emph
  {et~al.}(2017{\natexlab{c}})\citenamefont {Delord}, \citenamefont {Nicolas},
  \citenamefont {Chassagneux},\ and\ \citenamefont
  {H\'etet}}]{PhysRevA.96.063810}%
  \BibitemOpen
  \bibfield  {author} {\bibinfo {author} {\bibfnamefont {T.}~\bibnamefont
  {Delord}}, \bibinfo {author} {\bibfnamefont {L.}~\bibnamefont {Nicolas}},
  \bibinfo {author} {\bibfnamefont {Y.}~\bibnamefont {Chassagneux}},\ and\
  \bibinfo {author} {\bibfnamefont {G.}~\bibnamefont {H\'etet}},\ }\bibfield
  {title} {\bibinfo {title} {Strong coupling between a single nitrogen-vacancy
  spin and the rotational mode of diamonds levitating in an ion trap},\ }\href
  {https://doi.org/10.1103/PhysRevA.96.063810} {\bibfield  {journal} {\bibinfo
  {journal} {Phys. Rev. A}\ }\textbf {\bibinfo {volume} {96}},\ \bibinfo
  {pages} {063810} (\bibinfo {year} {2017}{\natexlab{c}})}\BibitemShut
  {NoStop}%
\bibitem [{\citenamefont {Adesso}\ and\ \citenamefont
  {Illuminati}(2006)}]{doi:10.1142/S0219749906001852}%
  \BibitemOpen
  \bibfield  {author} {\bibinfo {author} {\bibfnamefont {G.}~\bibnamefont
  {Adesso}}\ and\ \bibinfo {author} {\bibfnamefont {F.}~\bibnamefont
  {Illuminati}},\ }\bibfield  {title} {\bibinfo {title} {Entanglement sharing:
  From qubits to {Gaussian} states},\ }\href
  {https://doi.org/10.1142/S0219749906001852} {\bibfield  {journal} {\bibinfo
  {journal} {Int. J. Quantum Inform.}\ }\textbf {\bibinfo {volume} {04}},\
  \bibinfo {pages} {383} (\bibinfo {year} {2006})}\BibitemShut {NoStop}%
\bibitem [{\citenamefont {Coffman}\ \emph {et~al.}(2000)\citenamefont
  {Coffman}, \citenamefont {Kundu},\ and\ \citenamefont
  {Wootters}}]{PhysRevA.61.052306}%
  \BibitemOpen
  \bibfield  {author} {\bibinfo {author} {\bibfnamefont {V.}~\bibnamefont
  {Coffman}}, \bibinfo {author} {\bibfnamefont {J.}~\bibnamefont {Kundu}},\
  and\ \bibinfo {author} {\bibfnamefont {W.~K.}\ \bibnamefont {Wootters}},\
  }\bibfield  {title} {\bibinfo {title} {Distributed entanglement},\ }\href
  {https://doi.org/10.1103/PhysRevA.61.052306} {\bibfield  {journal} {\bibinfo
  {journal} {Phys. Rev. A}\ }\textbf {\bibinfo {volume} {61}},\ \bibinfo
  {pages} {052306} (\bibinfo {year} {2000})}\BibitemShut {NoStop}%
\bibitem [{\citenamefont {Karlsson}\ and\ \citenamefont
  {Bourennane}(1998)}]{1998KarlssonP43944400}%
  \BibitemOpen
  \bibfield  {author} {\bibinfo {author} {\bibfnamefont {A.}~\bibnamefont
  {Karlsson}}\ and\ \bibinfo {author} {\bibfnamefont {M.}~\bibnamefont
  {Bourennane}},\ }\bibfield  {title} {\bibinfo {title} {Quantum teleportation
  using three-particle entanglement},\ }\href
  {https://doi.org/10.1103/PhysRevA.58.4394} {\bibfield  {journal} {\bibinfo
  {journal} {Phys. Rev. A}\ }\textbf {\bibinfo {volume} {58}},\ \bibinfo
  {pages} {4394} (\bibinfo {year} {1998})}\BibitemShut {NoStop}%
\bibitem [{\citenamefont {Yonezawa}\ \emph {et~al.}(2004)\citenamefont
  {Yonezawa}, \citenamefont {Aoki},\ and\ \citenamefont
  {Furusawa}}]{2004YonezawaP430433}%
  \BibitemOpen
  \bibfield  {author} {\bibinfo {author} {\bibfnamefont {H.}~\bibnamefont
  {Yonezawa}}, \bibinfo {author} {\bibfnamefont {T.}~\bibnamefont {Aoki}},\
  and\ \bibinfo {author} {\bibfnamefont {A.}~\bibnamefont {Furusawa}},\
  }\bibfield  {title} {\bibinfo {title} {Demonstration of a quantum
  teleportation network for continuous variables},\ }\href
  {https://doi.org/10.1038/nature02858} {\bibfield  {journal} {\bibinfo
  {journal} {Nature}\ }\textbf {\bibinfo {volume} {431}},\ \bibinfo {pages}
  {430} (\bibinfo {year} {2004})}\BibitemShut {NoStop}%
\bibitem [{\citenamefont {Yeo}\ and\ \citenamefont
  {Chua}(2006)}]{2006YeoP6050260502}%
  \BibitemOpen
  \bibfield  {author} {\bibinfo {author} {\bibfnamefont {Y.}~\bibnamefont
  {Yeo}}\ and\ \bibinfo {author} {\bibfnamefont {W.~K.}\ \bibnamefont {Chua}},\
  }\bibfield  {title} {\bibinfo {title} {Teleportation and dense coding with
  genuine multipartite entanglement},\ }\href
  {https://doi.org/10.1103/PhysRevLett.96.060502} {\bibfield  {journal}
  {\bibinfo  {journal} {Phys. Rev. Lett.}\ }\textbf {\bibinfo {volume} {96}},\
  \bibinfo {pages} {060502} (\bibinfo {year} {2006})}\BibitemShut {NoStop}%
\bibitem [{\citenamefont {Briegel}\ \emph {et~al.}(2009)\citenamefont
  {Briegel}, \citenamefont {Browne}, \citenamefont {Dür}, \citenamefont
  {Raussendorf},\ and\ \citenamefont {Van~den Nest}}]{2009BriegelP1926}%
  \BibitemOpen
  \bibfield  {author} {\bibinfo {author} {\bibfnamefont {H.~J.}\ \bibnamefont
  {Briegel}}, \bibinfo {author} {\bibfnamefont {D.~E.}\ \bibnamefont {Browne}},
  \bibinfo {author} {\bibfnamefont {W.}~\bibnamefont {Dür}}, \bibinfo {author}
  {\bibfnamefont {R.}~\bibnamefont {Raussendorf}},\ and\ \bibinfo {author}
  {\bibfnamefont {M.}~\bibnamefont {Van~den Nest}},\ }\bibfield  {title}
  {\bibinfo {title} {Measurement-based quantum computation},\ }\href
  {https://doi.org/10.1038/nphys1157} {\bibfield  {journal} {\bibinfo
  {journal} {Nat. Phys.}\ }\textbf {\bibinfo {volume} {5}},\ \bibinfo {pages}
  {19} (\bibinfo {year} {2009})}\BibitemShut {NoStop}%
\bibitem [{\citenamefont {Cleve}\ \emph {et~al.}(1999)\citenamefont {Cleve},
  \citenamefont {Gottesman},\ and\ \citenamefont {Lo}}]{1999CleveP648651}%
  \BibitemOpen
  \bibfield  {author} {\bibinfo {author} {\bibfnamefont {R.}~\bibnamefont
  {Cleve}}, \bibinfo {author} {\bibfnamefont {D.}~\bibnamefont {Gottesman}},\
  and\ \bibinfo {author} {\bibfnamefont {H.-K.}\ \bibnamefont {Lo}},\
  }\bibfield  {title} {\bibinfo {title} {How to share a quantum secret},\
  }\href {https://doi.org/10.1103/PhysRevLett.83.648} {\bibfield  {journal}
  {\bibinfo  {journal} {Phys. Rev. Lett.}\ }\textbf {\bibinfo {volume} {83}},\
  \bibinfo {pages} {648} (\bibinfo {year} {1999})}\BibitemShut {NoStop}%
\bibitem [{\citenamefont {Lance}\ \emph {et~al.}(2004)\citenamefont {Lance},
  \citenamefont {Symul}, \citenamefont {Bowen}, \citenamefont {Sanders},\ and\
  \citenamefont {Lam}}]{2004LanceP177903177903}%
  \BibitemOpen
  \bibfield  {author} {\bibinfo {author} {\bibfnamefont {A.~M.}\ \bibnamefont
  {Lance}}, \bibinfo {author} {\bibfnamefont {T.}~\bibnamefont {Symul}},
  \bibinfo {author} {\bibfnamefont {W.~P.}\ \bibnamefont {Bowen}}, \bibinfo
  {author} {\bibfnamefont {B.~C.}\ \bibnamefont {Sanders}},\ and\ \bibinfo
  {author} {\bibfnamefont {P.~K.}\ \bibnamefont {Lam}},\ }\bibfield  {title}
  {\bibinfo {title} {Tripartite quantum state sharing},\ }\href
  {https://doi.org/10.1103/PhysRevLett.92.177903} {\bibfield  {journal}
  {\bibinfo  {journal} {Phys. Rev. Lett.}\ }\textbf {\bibinfo {volume} {92}},\
  \bibinfo {pages} {177903} (\bibinfo {year} {2004})}\BibitemShut {NoStop}%
\bibitem [{\citenamefont {Resch}\ \emph {et~al.}(2005)\citenamefont {Resch},
  \citenamefont {Walther},\ and\ \citenamefont
  {Zeilinger}}]{PhysRevLett.94.070402}%
  \BibitemOpen
  \bibfield  {author} {\bibinfo {author} {\bibfnamefont {K.~J.}\ \bibnamefont
  {Resch}}, \bibinfo {author} {\bibfnamefont {P.}~\bibnamefont {Walther}},\
  and\ \bibinfo {author} {\bibfnamefont {A.}~\bibnamefont {Zeilinger}},\
  }\bibfield  {title} {\bibinfo {title} {Full characterization of a
  three-photon greenberger-horne-zeilinger state using quantum state
  tomography},\ }\href {https://doi.org/10.1103/PhysRevLett.94.070402}
  {\bibfield  {journal} {\bibinfo  {journal} {Phys. Rev. Lett.}\ }\textbf
  {\bibinfo {volume} {94}},\ \bibinfo {pages} {070402} (\bibinfo {year}
  {2005})}\BibitemShut {NoStop}%
\bibitem [{\citenamefont {Christandl}\ and\ \citenamefont
  {Renner}(2012)}]{PhysRevLett.109.120403}%
  \BibitemOpen
  \bibfield  {author} {\bibinfo {author} {\bibfnamefont {M.}~\bibnamefont
  {Christandl}}\ and\ \bibinfo {author} {\bibfnamefont {R.}~\bibnamefont
  {Renner}},\ }\bibfield  {title} {\bibinfo {title} {Reliable quantum state
  tomography},\ }\href {https://doi.org/10.1103/PhysRevLett.109.120403}
  {\bibfield  {journal} {\bibinfo  {journal} {Phys. Rev. Lett.}\ }\textbf
  {\bibinfo {volume} {109}},\ \bibinfo {pages} {120403} (\bibinfo {year}
  {2012})}\BibitemShut {NoStop}%
\bibitem [{\citenamefont {Bent}\ \emph {et~al.}(2015)\citenamefont {Bent},
  \citenamefont {Qassim}, \citenamefont {Tahir}, \citenamefont {Sych},
  \citenamefont {Leuchs}, \citenamefont {S\'anchez-Soto}, \citenamefont
  {Karimi},\ and\ \citenamefont {Boyd}}]{PhysRevX.5.041006}%
  \BibitemOpen
  \bibfield  {author} {\bibinfo {author} {\bibfnamefont {N.}~\bibnamefont
  {Bent}}, \bibinfo {author} {\bibfnamefont {H.}~\bibnamefont {Qassim}},
  \bibinfo {author} {\bibfnamefont {A.~A.}\ \bibnamefont {Tahir}}, \bibinfo
  {author} {\bibfnamefont {D.}~\bibnamefont {Sych}}, \bibinfo {author}
  {\bibfnamefont {G.}~\bibnamefont {Leuchs}}, \bibinfo {author} {\bibfnamefont
  {L.~L.}\ \bibnamefont {S\'anchez-Soto}}, \bibinfo {author} {\bibfnamefont
  {E.}~\bibnamefont {Karimi}},\ and\ \bibinfo {author} {\bibfnamefont {R.~W.}\
  \bibnamefont {Boyd}},\ }\bibfield  {title} {\bibinfo {title} {Experimental
  realization of quantum tomography of photonic qudits via symmetric
  informationally complete positive operator-valued measures},\ }\href
  {https://doi.org/10.1103/PhysRevX.5.041006} {\bibfield  {journal} {\bibinfo
  {journal} {Phys. Rev. X}\ }\textbf {\bibinfo {volume} {5}},\ \bibinfo {pages}
  {041006} (\bibinfo {year} {2015})}\BibitemShut {NoStop}%
\bibitem [{\citenamefont {Lundeen}\ \emph {et~al.}(2011)\citenamefont
  {Lundeen}, \citenamefont {Sutherland}, \citenamefont {Patel}, \citenamefont
  {Stewart},\ and\ \citenamefont {Bamber}}]{RN758}%
  \BibitemOpen
  \bibfield  {author} {\bibinfo {author} {\bibfnamefont {J.~S.}\ \bibnamefont
  {Lundeen}}, \bibinfo {author} {\bibfnamefont {B.}~\bibnamefont {Sutherland}},
  \bibinfo {author} {\bibfnamefont {A.}~\bibnamefont {Patel}}, \bibinfo
  {author} {\bibfnamefont {C.}~\bibnamefont {Stewart}},\ and\ \bibinfo {author}
  {\bibfnamefont {C.}~\bibnamefont {Bamber}},\ }\bibfield  {title} {\bibinfo
  {title} {Direct measurement of the quantum wavefunction},\ }\href
  {https://doi.org/10.1038/nature10120} {\bibfield  {journal} {\bibinfo
  {journal} {Nature}\ }\textbf {\bibinfo {volume} {474}},\ \bibinfo {pages}
  {188} (\bibinfo {year} {2011})}\BibitemShut {NoStop}%
\bibitem [{\citenamefont {Zhou}\ \emph {et~al.}(2021)\citenamefont {Zhou},
  \citenamefont {Zhao}, \citenamefont {Hay}, \citenamefont {McGonagle},
  \citenamefont {Boyd},\ and\ \citenamefont {Shi}}]{RN755}%
  \BibitemOpen
  \bibfield  {author} {\bibinfo {author} {\bibfnamefont {Y.}~\bibnamefont
  {Zhou}}, \bibinfo {author} {\bibfnamefont {J.}~\bibnamefont {Zhao}}, \bibinfo
  {author} {\bibfnamefont {D.}~\bibnamefont {Hay}}, \bibinfo {author}
  {\bibfnamefont {K.}~\bibnamefont {McGonagle}}, \bibinfo {author}
  {\bibfnamefont {R.~W.}\ \bibnamefont {Boyd}},\ and\ \bibinfo {author}
  {\bibfnamefont {Z.}~\bibnamefont {Shi}},\ }\bibfield  {title} {\bibinfo
  {title} {Direct tomography of high-dimensional density matrices for general
  quantum states of photons},\ }\href
  {https://doi.org/10.1103/PhysRevLett.127.040402} {\bibfield  {journal}
  {\bibinfo  {journal} {Phys. Rev. Lett.}\ }\textbf {\bibinfo {volume} {127}},\
  \bibinfo {pages} {040402} (\bibinfo {year} {2021})}\BibitemShut {NoStop}%
\bibitem [{\citenamefont {Thekkadath}\ \emph {et~al.}(2016)\citenamefont
  {Thekkadath}, \citenamefont {Giner}, \citenamefont {Chalich}, \citenamefont
  {Horton}, \citenamefont {Banker},\ and\ \citenamefont {Lundeen}}]{RN751}%
  \BibitemOpen
  \bibfield  {author} {\bibinfo {author} {\bibfnamefont {G.~S.}\ \bibnamefont
  {Thekkadath}}, \bibinfo {author} {\bibfnamefont {L.}~\bibnamefont {Giner}},
  \bibinfo {author} {\bibfnamefont {Y.}~\bibnamefont {Chalich}}, \bibinfo
  {author} {\bibfnamefont {M.~J.}\ \bibnamefont {Horton}}, \bibinfo {author}
  {\bibfnamefont {J.}~\bibnamefont {Banker}},\ and\ \bibinfo {author}
  {\bibfnamefont {J.~S.}\ \bibnamefont {Lundeen}},\ }\bibfield  {title}
  {\bibinfo {title} {Direct measurement of the density matrix of a quantum
  system},\ }\href {https://doi.org/10.1103/PhysRevLett.117.120401} {\bibfield
  {journal} {\bibinfo  {journal} {Phys. Rev. Lett.}\ }\textbf {\bibinfo
  {volume} {117}},\ \bibinfo {pages} {120401} (\bibinfo {year}
  {2016})}\BibitemShut {NoStop}%
\bibitem [{\citenamefont {Lachance-Quirion}\ \emph {et~al.}(2017)\citenamefont
  {Lachance-Quirion}, \citenamefont {Tabuchi}, \citenamefont {Ishino},
  \citenamefont {Noguchi}, \citenamefont {Ishikawa}, \citenamefont {Yamazaki},\
  and\ \citenamefont {Nakamura}}]{2017LachanceQuirionP16031501603150}%
  \BibitemOpen
  \bibfield  {author} {\bibinfo {author} {\bibfnamefont {D.}~\bibnamefont
  {Lachance-Quirion}}, \bibinfo {author} {\bibfnamefont {Y.}~\bibnamefont
  {Tabuchi}}, \bibinfo {author} {\bibfnamefont {S.}~\bibnamefont {Ishino}},
  \bibinfo {author} {\bibfnamefont {A.}~\bibnamefont {Noguchi}}, \bibinfo
  {author} {\bibfnamefont {T.}~\bibnamefont {Ishikawa}}, \bibinfo {author}
  {\bibfnamefont {R.}~\bibnamefont {Yamazaki}},\ and\ \bibinfo {author}
  {\bibfnamefont {Y.}~\bibnamefont {Nakamura}},\ }\bibfield  {title} {\bibinfo
  {title} {Resolving quanta of collective spin excitations in a
  millimeter-sized ferromagnet},\ }\href
  {https://doi.org/10.1126/sciadv.1603150} {\bibfield  {journal} {\bibinfo
  {journal} {Sci. Adv.}\ }\textbf {\bibinfo {volume} {3}},\ \bibinfo {pages}
  {e1603150} (\bibinfo {year} {2017})}\BibitemShut {NoStop}%
\bibitem [{\citenamefont {Togan}\ \emph {et~al.}(2010)\citenamefont {Togan},
  \citenamefont {Chu}, \citenamefont {Trifonov}, \citenamefont {Jiang},
  \citenamefont {Maze}, \citenamefont {Childress}, \citenamefont {Dutt},
  \citenamefont {Sorensen}, \citenamefont {Hemmer}, \citenamefont {Zibrov},\
  and\ \citenamefont {Lukin}}]{RN591}%
  \BibitemOpen
  \bibfield  {author} {\bibinfo {author} {\bibfnamefont {E.}~\bibnamefont
  {Togan}}, \bibinfo {author} {\bibfnamefont {Y.}~\bibnamefont {Chu}}, \bibinfo
  {author} {\bibfnamefont {A.~S.}\ \bibnamefont {Trifonov}}, \bibinfo {author}
  {\bibfnamefont {L.}~\bibnamefont {Jiang}}, \bibinfo {author} {\bibfnamefont
  {J.}~\bibnamefont {Maze}}, \bibinfo {author} {\bibfnamefont {L.}~\bibnamefont
  {Childress}}, \bibinfo {author} {\bibfnamefont {M.~V.}\ \bibnamefont {Dutt}},
  \bibinfo {author} {\bibfnamefont {A.~S.}\ \bibnamefont {Sorensen}}, \bibinfo
  {author} {\bibfnamefont {P.~R.}\ \bibnamefont {Hemmer}}, \bibinfo {author}
  {\bibfnamefont {A.~S.}\ \bibnamefont {Zibrov}},\ and\ \bibinfo {author}
  {\bibfnamefont {M.~D.}\ \bibnamefont {Lukin}},\ }\bibfield  {title} {\bibinfo
  {title} {Quantum entanglement between an optical photon and a solid-state
  spin qubit},\ }\href {https://doi.org/10.1038/nature09256} {\bibfield
  {journal} {\bibinfo  {journal} {Nature}\ }\textbf {\bibinfo {volume} {466}},\
  \bibinfo {pages} {730} (\bibinfo {year} {2010})}\BibitemShut {NoStop}%
\bibitem [{\citenamefont {Siyushev}\ \emph {et~al.}(2019)\citenamefont
  {Siyushev}, \citenamefont {Nesladek}, \citenamefont {Bourgeois},
  \citenamefont {Gulka}, \citenamefont {Hruby}, \citenamefont {Yamamoto},
  \citenamefont {Trupke}, \citenamefont {Teraji}, \citenamefont {Isoya},\ and\
  \citenamefont {Jelezko}}]{RN566}%
  \BibitemOpen
  \bibfield  {author} {\bibinfo {author} {\bibfnamefont {P.}~\bibnamefont
  {Siyushev}}, \bibinfo {author} {\bibfnamefont {M.}~\bibnamefont {Nesladek}},
  \bibinfo {author} {\bibfnamefont {E.}~\bibnamefont {Bourgeois}}, \bibinfo
  {author} {\bibfnamefont {M.}~\bibnamefont {Gulka}}, \bibinfo {author}
  {\bibfnamefont {J.}~\bibnamefont {Hruby}}, \bibinfo {author} {\bibfnamefont
  {T.}~\bibnamefont {Yamamoto}}, \bibinfo {author} {\bibfnamefont
  {M.}~\bibnamefont {Trupke}}, \bibinfo {author} {\bibfnamefont
  {T.}~\bibnamefont {Teraji}}, \bibinfo {author} {\bibfnamefont
  {J.}~\bibnamefont {Isoya}},\ and\ \bibinfo {author} {\bibfnamefont
  {F.}~\bibnamefont {Jelezko}},\ }\bibfield  {title} {\bibinfo {title}
  {Photoelectrical imaging and coherent spin-state readout of single
  nitrogen-vacancy centers in diamond},\ }\href
  {https://doi.org/10.1126/science.aav2789} {\bibfield  {journal} {\bibinfo
  {journal} {Science}\ }\textbf {\bibinfo {volume} {363}},\ \bibinfo {pages}
  {728} (\bibinfo {year} {2019})}\BibitemShut {NoStop}%
\bibitem [{\citenamefont {Delord}\ \emph {et~al.}(2020)\citenamefont {Delord},
  \citenamefont {Huillery}, \citenamefont {Nicolas},\ and\ \citenamefont
  {Hetet}}]{RN714}%
  \BibitemOpen
  \bibfield  {author} {\bibinfo {author} {\bibfnamefont {T.}~\bibnamefont
  {Delord}}, \bibinfo {author} {\bibfnamefont {P.}~\bibnamefont {Huillery}},
  \bibinfo {author} {\bibfnamefont {L.}~\bibnamefont {Nicolas}},\ and\ \bibinfo
  {author} {\bibfnamefont {G.}~\bibnamefont {Hetet}},\ }\bibfield  {title}
  {\bibinfo {title} {Spin-cooling of the motion of a trapped diamond},\ }\href
  {https://doi.org/10.1038/s41586-020-2133-z} {\bibfield  {journal} {\bibinfo
  {journal} {Nature}\ }\textbf {\bibinfo {volume} {580}},\ \bibinfo {pages}
  {56} (\bibinfo {year} {2020})}\BibitemShut {NoStop}%
\bibitem [{\citenamefont {Gieseler}\ \emph {et~al.}(2013)\citenamefont
  {Gieseler}, \citenamefont {Novotny},\ and\ \citenamefont {Quidant}}]{RN743}%
  \BibitemOpen
  \bibfield  {author} {\bibinfo {author} {\bibfnamefont {J.}~\bibnamefont
  {Gieseler}}, \bibinfo {author} {\bibfnamefont {L.}~\bibnamefont {Novotny}},\
  and\ \bibinfo {author} {\bibfnamefont {R.}~\bibnamefont {Quidant}},\
  }\bibfield  {title} {\bibinfo {title} {Thermal nonlinearities in a
  nanomechanical oscillator},\ }\href {https://doi.org/10.1038/nphys2798}
  {\bibfield  {journal} {\bibinfo  {journal} {Nat. Phys.}\ }\textbf {\bibinfo
  {volume} {9}},\ \bibinfo {pages} {806} (\bibinfo {year} {2013})}\BibitemShut
  {NoStop}%
\bibitem [{\citenamefont {Beresnev}\ \emph {et~al.}(2006)\citenamefont
  {Beresnev}, \citenamefont {Chernyak},\ and\ \citenamefont
  {Fomyagin}}]{RN246}%
  \BibitemOpen
  \bibfield  {author} {\bibinfo {author} {\bibfnamefont {S.~A.}\ \bibnamefont
  {Beresnev}}, \bibinfo {author} {\bibfnamefont {V.~G.}\ \bibnamefont
  {Chernyak}},\ and\ \bibinfo {author} {\bibfnamefont {G.~A.}\ \bibnamefont
  {Fomyagin}},\ }\bibfield  {title} {\bibinfo {title} {Motion of a spherical
  particle in a rarefied gas. {Part 2. Drag} and thermal polarization},\ }\href
  {https://doi.org/10.1017/s0022112090003007} {\bibfield  {journal} {\bibinfo
  {journal} {J. Fluid Mech.}\ }\textbf {\bibinfo {volume} {219}},\ \bibinfo
  {pages} {405} (\bibinfo {year} {2006})}\BibitemShut {NoStop}%
\bibitem [{\citenamefont {Johansson}\ \emph {et~al.}(2012)\citenamefont
  {Johansson}, \citenamefont {Nation},\ and\ \citenamefont
  {Nori}}]{JOHANSSON20121760}%
  \BibitemOpen
  \bibfield  {author} {\bibinfo {author} {\bibfnamefont {J.~R.}\ \bibnamefont
  {Johansson}}, \bibinfo {author} {\bibfnamefont {P.~D.}\ \bibnamefont
  {Nation}},\ and\ \bibinfo {author} {\bibfnamefont {F.}~\bibnamefont {Nori}},\
  }\bibfield  {title} {\bibinfo {title} {Qutip: An open-source python framework
  for the dynamics of open quantum systems},\ }\href
  {https://doi.org/https://doi.org/10.1016/j.cpc.2012.02.021} {\bibfield
  {journal} {\bibinfo  {journal} {Comput. Phys. Commun.}\ }\textbf {\bibinfo
  {volume} {183}},\ \bibinfo {pages} {1760} (\bibinfo {year}
  {2012})}\BibitemShut {NoStop}%
\bibitem [{\citenamefont {Johansson}\ \emph {et~al.}(2013)\citenamefont
  {Johansson}, \citenamefont {Nation},\ and\ \citenamefont
  {F.}}]{JOHANSSON20131234}%
  \BibitemOpen
  \bibfield  {author} {\bibinfo {author} {\bibfnamefont {J.~R.}\ \bibnamefont
  {Johansson}}, \bibinfo {author} {\bibfnamefont {P.~D.}\ \bibnamefont
  {Nation}},\ and\ \bibinfo {author} {\bibfnamefont {F.}~\bibnamefont {Nori}},\
  }\bibfield  {title} {\bibinfo {title} {Qutip 2: A python framework for the
  dynamics of open quantum systems},\ }\href
  {https://doi.org/https://doi.org/10.1016/j.cpc.2012.11.019} {\bibfield
  {journal} {\bibinfo  {journal} {Comput. Phys. Commun.}\ }\textbf {\bibinfo
  {volume} {184}},\ \bibinfo {pages} {1234} (\bibinfo {year}
  {2013})}\BibitemShut {NoStop}%
\end{thebibliography}
\providecommand{\noopsort}[1]{}\providecommand{\singleletter}[1]{#1}%

\end{document}